\pdfoutput=1
\documentclass[12pt,reqno]{article}
\RequirePackage{etex}

\usepackage{etex}
\usepackage{graphicx}
\usepackage{xcolor}

\newcommand{\bslfigure}[2]{%
  \IfFileExists{#2}{%
    \includegraphics[width=#1]{#2}%
  }{%
    \begingroup
    \setlength{\fboxsep}{4pt}%
    \fbox{\parbox[c][0.09\textheight][c]{\dimexpr#1-2\fboxsep-2\fboxrule\relax}{%
      \centering\scriptsize Figure file not supplied\\[-0.15em]
      \texttt{\detokenize{#2}}}}%
    \endgroup
  }%
}
\usepackage{amsmath}
\usepackage{amssymb}
\usepackage{amsfonts}
\usepackage{titlesec}
\usepackage{pictex}
\usepackage{comment}
\usepackage{amsthm}
\usepackage{graphics,epsfig,verbatim,bm,latexsym,url,amsbsy}
\usepackage{rotating}
\usepackage[authoryear,round]{natbib}
\IfFileExists{ecta.bst}{\bibliographystyle{ecta}}{\bibliographystyle{plainnat}}
\usepackage{float}
\usepackage[position=bottom]{subfig}
\usepackage{mathrsfs}
\usepackage{multirow}
\usepackage{array}
\usepackage{bigints}
\usepackage{mathtools}
\usepackage{enumitem}
\usepackage{multirow,array}
\usepackage{blkarray}
\allowdisplaybreaks

\newtheorem{manualtheoreminner}{Theorem}
\newenvironment{manualtheorem}[1]{%
  \renewcommand\themanualtheoreminner{#1}%
  \manualtheoreminner
}{\endmanualtheoreminner}

\usepackage[hypertexnames=false,hyperfootnotes=false]{hyperref}
\hypersetup{
	colorlinks=true,
	linkcolor=red!70!black,
	citecolor=blue!70!black,
	filecolor=magenta,      
	urlcolor=cyan,
}
\usepackage[nameinlink,noabbrev,sort,capitalise]{cleveref}
\crefname{manualtheoreminner}{Theorem}{Theorems}
\crefname{assumption}{Assumption}{Assumptions}
\Crefname{manualtheoreminner}{Theorem}{Theorems}
\DeclarePairedDelimiterX{\inp}[2]{\langle}{\rangle}{#1, #2}

\newcommand{\Ex}{\mathbb{E}}
\newcommand{\KL}{\mathsf{KL}}

\newcommand{\V}{\mathsf{Var}}

\renewcommand{\Pr}{\mathbb{P}}

\theoremstyle{definition} \newtheorem{example}{Example}
\theoremstyle{definition} 
\theoremstyle{definition} 

\theoremstyle{definition} 
\theoremstyle{definition} \newtheorem{definition}{Definition}
\theoremstyle{definition} 
\theoremstyle{definition} 
\theoremstyle{definition} 
\theoremstyle{definition} \newtheorem{lemma}{Lemma}
\theoremstyle{plain} \newtheorem{theorem}{Theorem}
\theoremstyle{definition} \newtheorem{assumption}{Assumption}
\theoremstyle{definition} 
\theoremstyle{definition} 
\theoremstyle{definition}\newtheorem{proposition}{Proposition}
\theoremstyle{definition} 
\theoremstyle{definition} 
\theoremstyle{definition} 
\theoremstyle{definition} 
\theoremstyle{definition} 

\theoremstyle{definition} 

\usepackage{mathtools}
\usepackage{geometry}
\usepackage{setspace}
\theoremstyle{definition}

\titleformat{\section}
		{\bfseries\center \MakeUppercase}
         {\thesection}
        {0.5em}
        {}
        []

\titleformat{\subsection}[runin]
        {\normalfont\bfseries}
        {\thesubsection}
        {0.5em}
        {\addperiod}
        []
\newcommand{\addperiod}[1]{#1.}
\allowdisplaybreaks

\titlespacing*{\subsection}{0pt}{0.5ex plus 0.2ex minus 0.1ex}{0.25em}

\title{Balanced Social Learning}

\author{\makebox[.25\linewidth]{{Andrew Koh}\thanks{Columbia Department of Economics; email: \protect\texttt{ajk2316@columbia.edu}}}\\{Columbia} 
\and 
\makebox[.25\linewidth]{Ricky Li\thanks{MIT Department of Economics; email: \protect\texttt{rickyli@mit.edu} \\~\\
We are especially grateful to Drew Fudenberg and Stephen Morris for guidance and helpful suggestions. We also thank Daron Acemoglu, Isaiah Andrews, Ian Ball, Yeon-Koo Che, In-Koo Cho, Krishna Dasaratha, Kevin He, Navin Kartik, Jan Knoepfle, Tianhao Liu, Anna Mikusheva, Roi Orzach, Sivakorn Sanguanmoo, Alex Wolitzky, Julian Wright, Weijie Zhong, as well as audiences at Wharton and Cornell for helpful comments. First posted version: November 2025; this paper supersedes a previous paper circulated under the title `Competition and Social Learning' (February 2023).
}} \\
{MIT}}

\date{This version: October 19, 2026
}

\begin{document}
\maketitle
\thispagestyle{empty}
\begin{abstract} 
    When is social learning efficient if agents have heterogeneous preferences over a set of actions? When myopic agents generate information only about the action they choose, efficiency requires \emph{balance} such that all high-value actions are discovered. We characterize the value of policies that control the flow of past information to future agents. Under \emph{full transparency} of past information, the probability of efficient outcomes is approximately one if and only if the prior and preferences are balanced and the log-likelihood ratio distribution satisfies a set of intuitive moment conditions. Noisy bad news is a canonical example. By contrast, approximate efficiency is obtainable on the set of arbitrary \emph{adaptive information policies} if and only if the prior alone is balanced. Moreover, it suffices to use \emph{adaptive evidence policies} which permanently and fully disclose subsets of past information. When the prior is imbalanced and approximate efficiency is unattainable, we characterize the value of adaptive information and show that it is uniformly approximated by the value of adaptive evidence as the impact of each individual signal on beliefs vanishes.
\end{abstract}

\clearpage

\setcounter{page}{1}

\section{Introduction}

Many social learning environments feature a tension between \emph{speed} in aggregating dispersed information and \emph{balance} in the composition of actions that society learns about. When social learning is fast, society quickly grows confident in the value of \emph{an} action. When social learning is balanced, enough actions are sampled that society eventually learns about the value of \emph{all} actions. Speed is preferred to balance when preferences are identical across society: identifying a single high-value action suffices for efficiency. However, balance may be preferred to speed when preferences are heterogeneous: to serve diverse individual tastes, society must learn about a broad set of actions. When and how is balanced social learning achieved?

To answer this question, we develop a binary-action, binary-state model of multidimensional social learning.\footnote{Analogs of our results hold in a model with more than two actions. The restriction to binary states carries more weight, and we leave extensions to richer state spaces for future work.} Each action has an unknown value that is common across agents. Agents are uncertain about these values and have heterogeneous preferences over the two actions. An \emph{information policy} controls the flow of information from past to future. At each time $t$, an agent arrives and chooses an action given their idiosyncratic preference and the public belief about each action's value. The choice generates a signal from a statistical experiment about the chosen action. The information policy then determines what future agents learn from that signal.

Two running interpretations illustrate the model:
\begin{enumerate}
    \item \textbf{Product markets}: consumers choose between two products that differ in quality and in fit with individual taste. An information policy is a review system that determines what future consumers learn about past consumers' experiences.
    \item \textbf{Technology adoption}: adopters choose between two technologies that differ in quality and in fit. An information policy determines how early outcomes are communicated to later adopters.
\end{enumerate}

A starting observation (\cref{lem:sources}) is that, under any informational environment in which infinitely sampled actions are correctly learned, long-run inefficiencies arise \emph{only from} insufficient exploration in which \emph{too few} high-value actions are learnt about. When society abandons a high-value action, agents with an idiosyncratic taste for that action forgo the payoff from a better horizontal match. This is the sense in which \emph{balance}---high-value actions being learned about in tandem---is required for efficiency.

We study three nested \emph{information regimes}---sets of feasible information policies---that differ in how precisely a designer (e.g., a platform or a government) can control the flow of information from past to future:
\begin{enumerate}[leftmargin = *]
    \item \textbf{Full transparency.} The designer has \emph{no control}: future agents observe all information generated by past agents. In our product-market example, transparency is common practice and is often regarded as a desideratum by policymakers and consumers.\footnote{For instance, a recent proposal in the U.S. Congress, the ``Filter Bubble Transparency Act'' (S.2024), would require platforms to show users ``unmanipulated content''.} Moreover, third-party websites such as Trustpilot allow users to review businesses across a wide variety of industries, which limits the ability of any single platform or designer to disclose reviews selectively.
    \item \textbf{Adaptive evidence design.} An \emph{adaptive evidence policy} specifies, for each history of past evidence, a subset of these signals to permanently and fully disclose as a function of the public history. The designer cannot alter the content of a signal: in this sense, disclosure is `all or nothing'.\footnote{Equivalently, the designer chooses a network that specifies, for each time-$t$ agent, the subset of past signals that the agent observes, subject to the constraint that later agents observe weakly larger sets of signals than earlier agents. This formulation is similar to the networked observational learning model of \cite{acemoglu2011bayesian}, except that agents here observe signals about outcomes rather than past actions.}
    \item \textbf{Adaptive information design.} An \emph{adaptive information policy} specifies, as a function of the public history, a statistical experiment about the stock of past evidence. This regime places no restriction on how the designer reveals information about past signals.
\end{enumerate}

\begin{figure}[H]
\centering
\caption{The form and value of information regimes}
    {\bslfigure{0.7\textwidth}{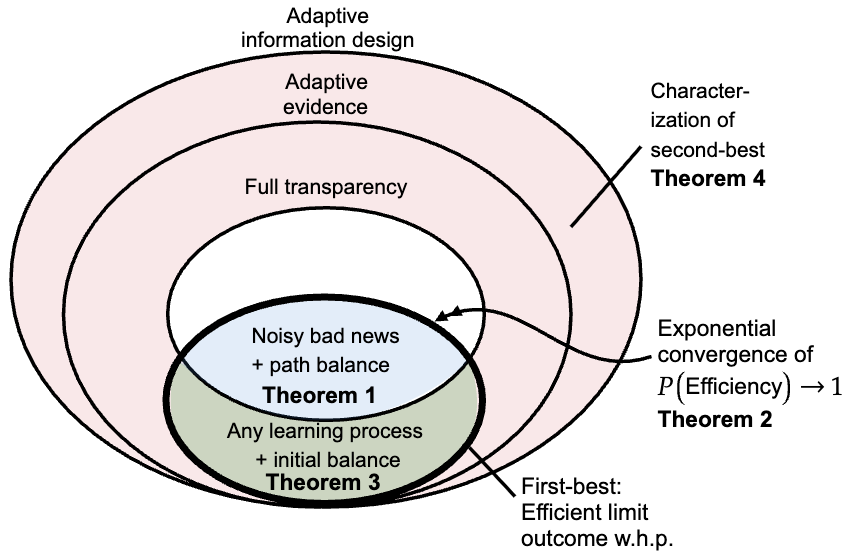}}
    \label{fig:outline}
\end{figure}

\paragraph{Efficiency under full transparency.} We begin our analysis with the regime of full transparency. \cref{lem:sources} establishes a key benchmark: under full transparency, the probability of efficient outcomes is strictly less than $1$ for \emph{any} fixed experiment. We therefore ask how closely almost-sure efficiency can be approximated. \cref{thrm:fulltrans} shows that, along any sequence of experiments whose high-state log-likelihood ratios are uniformly bounded above, the efficiency probability approaches $1$ if and only if two conditions hold.\footnote{To obtain a tight characterization, the ``only if'' direction of this result additionally assumes a mild non-tangency condition on the deterministic learning path (\cref{as:no-tangential-pb}).} First, the prior and the preference distribution satisfy \emph{path balance}. Second, conditional on a high-value action, the expected movement in the log-likelihood ratio (the \emph{drift}) vanishes and the variance of this movement (the \emph{noise}) is of smaller order than the drift. The blue region in \cref{fig:outline} describes this asymptotic result.

Intuitively, slowing learning down generates endogenous exploration. When each signal moves public beliefs only slightly, the dispersion in the horizontal preferences of arriving agents is large relative to the movement in beliefs, so both actions continue to be chosen. Using diffusion asymptotics, we show that in the noisy limit, where each agent's signal has a vanishing impact on public beliefs, beliefs about high-value actions can be made to increase in tandem along a \emph{balanced learning path}. Efficiency is then restored because no high-value action is abandoned. Notably, this exploration arises even though agents are myopic and never choose a myopically suboptimal action.

A key insight of our analysis is that noise alone does not deliver asymptotic efficiency: the restriction that the high-state noise be of smaller order than the drift is essential. \emph{Noisy bad news}, in which conclusive evidence that an action is low value arrives only rarely, is a canonical example: conditional on a high-value action, the upward drift in beliefs dominates the signal variance, and posterior beliefs deterministically rise in small steps. By contrast, whenever the signal variance remains of the same order as the drift, there always remains a possibility for beliefs to diverge from the balanced learning path, which bounds the efficiency probability away from $1$.

In product markets, \cref{thrm:fulltrans} implies that approximately efficient limit outcomes require that society learn from reviews slowly, and in a specific way. By contrast, in single-product environments faster learning from reviews always improves welfare \citep{acemoglu2022learning}. In technology adoption, increasingly rare `breakdowns' \citep{keller2015breakdowns} are a canonical sufficient route to balanced social learning, whereas \cref{prop:goodnews} shows that perfect-good-news `breakthrough' experiments \citep{keller2005strategic}, in the sense of \cref{def:perfect-good-news}, generate maximal imbalance.

We briefly discuss the technical contribution of \cref{thrm:fulltrans}. In our setting, beliefs evolve as a multidimensional stochastic process with two sources of randomness: (i) the realization of idiosyncratic preferences among arriving agents; and (ii) the realization of signals about action values. Both sources of randomness are important features of the economic environments we study, but together they make the evolution of beliefs difficult to analyze, because past beliefs determine which action is sampled and hence which belief is updated. We show that, for sufficiently smooth sequences of experiments in the noisy information limit, the law of the suitably rescaled belief process converges weakly to the law of the solution to a stochastic differential equation.\footnote{We establish convergence by verifying a diffusion-approximation criterion based on the martingale problem associated with the limiting stochastic differential equation. Such methods were pioneered by Stroock and Varadhan; see \cite{jacod2013limit} for a modern treatment.} A separate argument establishing the arrival of rare signals handles the remaining ``non-smooth'' sequences. Together, the two arguments yield the tight conditions for asymptotic efficiency in \cref{thrm:fulltrans}.

\noindent \textbf{How quickly do efficiency gains obtain as learning slows?} \cref{thrm:fulltrans} gives tight conditions under which the efficiency probability approaches $1$, but it requires signals to become arbitrarily noisy. We show
(\cref{thrm:exp_convergence}) that, when perfect bad news arrives with probability $p$, the efficiency probability under path balance converges to $1$ at least \emph{exponentially} fast as $p \to 0$. We establish this by bounding the probability that the belief process escapes from the balanced learning path. The exponential bound suggests that, in practice, perfect bad news signals may not need to be very noisy: moderate noise can yield a high efficiency probability without sacrificing much speed.

\noindent \textbf{What is the value of design?} Can a designer approximately obtain efficiency under less restrictive conditions than those of \cref{thrm:fulltrans} by controlling the flow of information from past to future? \cref{thrm:evidence_iff} shows that adaptive evidence policies approximately achieve efficiency if and only if the prior satisfies \emph{static balance}, which merely requires that both actions are chosen with positive probability under the prior. This is depicted by the green region in \cref{fig:outline}. Approximately efficient adaptive evidence policies take the following two-stage form. The designer first commits to silence for a long deterministic accumulation period, during which both actions continue to be sampled. She then discloses the accumulated evidence and thereafter fully discloses all future evidence.

When static balance fails, \cref{thrm:adaptiveinfo} characterizes the second-best value of all adaptive information policies. The key insight is that Bayes plausibility bounds the probability of lowering the belief about the favored action into the balanced region. A simple two-message garbling of a sufficiently informative designer posterior approximates the bound. Finally, \cref{thrm:adaptiveinfoB} shows that the efficiency gap between adaptive evidence and adaptive information uniformly disappears as the impact of individual signals on beliefs vanishes.

\subsection{Related Literature} Our work contributes to the literatures on (i) learning from bandits; (ii) social learning from others' outcomes; and (iii) dynamic or adaptive information design. We discuss each in turn.

\noindent \textbf{Myopic and contextual bandits.} 
We contribute to a broad literature on multi-agent learning from bandits (in our model, actions with uncertain values). Our model shares with this literature the property that information externalities may cause insufficient exploration.\footnote{A classic strand of this literature (\cite*{bolton1999strategic}, \cite*{keller2005strategic}, \cite*{keller2010strategic}, \cite*{keller2015breakdowns}) studies strategic bandits, in which multiple forward-looking agents decide how much to explore. The friction in those papers is informational free-riding among strategic agents. The friction we study is that myopic agents have no incentive to explore beyond their myopically optimal action.} Within this literature, our paper relates most closely to myopic bandits (since our decision makers are short-run) and to contextual bandits, in which dispersion in horizontal preferences plays the role of ``contexts.'' One strand of this literature (\cite{kannan2018smoothed}, \cite*{bastani2021mostly}, \cite{raghavan2023greedy}, \cite*{slivkins2025greedy}) provides conditions under which myopic maximization (a so-called ``greedy'' algorithm) guarantees efficient learning outcomes in a contextual bandit setting.\footnote{An important exception is \cite*{banihashem2023bandit}, who document conditions under which learning succeeds and fails in a non-contextual setting.} Many of these papers fix the signal structure and study conditions which ensure that contexts are sufficiently diverse.\footnote{\cite*{goeree2006social}'s main result may also be interpreted in this light.} By contrast, we impose an upper bound on context diversity and provide a tight characterization of the signal structures for which myopic subjective expected utility maximization (the Bayesian ``greedy'' algorithm) achieves approximate long-run efficiency.\footnote{We make progress by assuming a stylized payoff structure: each action (`arm') yields the per-period deterministic payoff distribution $\delta_0$ or $\delta_1$, but there is ex-ante uncertainty about which one. In this sense, the payoff structure is consistent with Bernoulli bandits under a particular two-point prior. However, we do not require payoffs and signals to be coupled, which allows us to vary the signal structure as a primitive of our model while holding payoffs fixed.} Our results therefore complement the insight that sufficiently diverse contexts can induce efficient outcomes under greedy behavior: sufficiently noisy and asymmetric information can do so as well.

\noindent\textbf{Optimal decision rules and diffusion approximation of bandits.} A recent strand of work in econometrics and statistics studies the optimality properties of bandit algorithms in a local asymptotic regime where the decision-maker has $n$ periods to sample arms whose unknown expected rewards differ on the same order $n^{-1/2}$ of sampling uncertainty. Building on the limits-of-experiments approach of \cite{le2012asymptotic}, many of these papers use diffusion asymptotics to approximate discrete-time bandits. Their focus is on evaluating classic algorithms such as Thompson sampling (\cite*{fan2021diffusion}, \cite*{kuang2024weak}) or providing a general analysis of locally asymptotically Bayes- and minimax-optimal policy rules \citep*{adusumilli2022sample,adusumilli2025risk}. By contrast, we take agents' myopic Bayesian behavior as given and ask which information policies produce efficient outcomes. Our characterization draws on diffusion asymptotics for sequences of informational environments that are sufficiently smooth, but also requires a direct rare-signal construction for the remaining sequences.

\noindent \textbf{Social learning from others' outcomes.} A closely related paper is \cite*{acemoglu2022learning} (henceforth, AMMO), who study learning about the quality of a monopolist's product. In AMMO's single-product environment, learning always succeeds under their maintained conditions, and their focus is on how the `selection effect'\footnote{That is, the composition of agents leaving reviews is endogenous to the public belief at which purchases took place.} and the granularity of reviews affect the speed of learning. In our multi-action case, agents always choose \emph{some} action, but this does not imply learning about \emph{all} actions. Our focus is therefore on the composition of the actions that are chosen and learned about. We show that, in contrast to AMMO, fast learning can exacerbate imbalance and balanced learning can be restored by slowing learning down in a specific way.\footnote{In AMMO, finer review structures (partitions) always speed up learning, which is unambiguously good in their monopolist environment. By contrast, faster learning can reduce the efficiency probability in our multi-action environment.}

Several other papers model social learning from others' outcomes in environments with some degree of competition or product selection \citep*{vaccari2016social,wolitzky2018learning,vellodi2018ratings,kakhbod2021heterogeneous,banihashem2023bandit}.\footnote{Of note is \cite{vellodi2018ratings}, who studies a stationary model of how ratings (modeled as a drift-diffusion) shape entry. Our environment is non-stationary with a fixed set of actions. This allows us to analyze how the experiment, horizontal preferences, and the information policy shape the joint evolution of multidimensional beliefs, and when this evolution leads to inefficiencies.} Our full-transparency regime is closest to the environments in these papers. The main difference is our focus on horizontal preference dispersion, which is absent in these papers: we study how this dispersion interacts with the experiment to shape multidimensional limit learning outcomes.\footnote{\cite*{banihashem2023bandit} study a two-arm Bernoulli bandit with no horizontal preference dispersion; by contrast, our setting allows a general class of statistical experiments and a general distribution of horizontal preferences. Farther afield, \cite{kartik2024beyond} also study the interplay between information and preferences, but in the different context of observational social learning.}

\noindent \textbf{Dynamic or adaptive information design.} Our work is also related to the literature on recommendation systems and information design with myopic agents \citep*{kremer2014implementing,che2018recommender}. In these papers, a principal observes the actions and/or outcomes of a sequence of myopic agents and chooses an (incentive-compatible) recommendation for the next agent who arrives. In this paper, the designer is constrained to only reveal information about \emph{past evidence}, and we focus on three key regimes which are motivated by the economic environments we study. Our analysis of adaptive evidence design is similar in spirit but formally distinct from \cite*{immorlica2018incentivizing,immorlica2019bayesian}'s ``selective data disclosure.'' An insight of our framework is that adaptive evidence and adaptive information are closely linked in terms of welfare outcomes, since the value of adaptive evidence is maximal if and only if the value of adaptive information is maximal. Moreover, when static balance fails, the two values coincide in the vanishing-impact limit established in \cref{thrm:adaptiveinfoB}.

Our work also relates to more recent work on adaptive and/or dynamic information design. Our notion of adaptive evidence design is similar to that of \cite{knoepfle2024dynamic}, who study an environment with single-dimensional uncertainty and fully revealing evidence (conclusive good or bad news).\footnote{See also \cite{board2025braeking}, who study how information should be released when agents learn via a network. Our notion of adaptive evidence design is also related to \cite{acemoglu2011bayesian}, since our designer in effect chooses the subset of past signals that each agent observes perfectly.} Our notion of adaptive information design is related to recent work on dynamic information design with sender learning \citep{fainmesser2023consumer,koh2024informational,arieli2024persuading}. Unlike these papers, we focus on balanced learning rather than on other desiderata such as speed, persuasion, coordination, or price discrimination. Some of our results draw on tools from this literature: for example, our characterization of the second-best value in \cref{thrm:adaptiveinfo} combines a Bayes-plausibility upper bound with a one-step posterior-splitting construction under endogenous designer learning.

\section{Model}\label{sec:model}

\begin{figure}[H]
    \centering
      \caption{Taxonomy of social learning under horizontal dispersion} 
      \vspace{-0.8em}
{\bslfigure{0.8\textwidth}{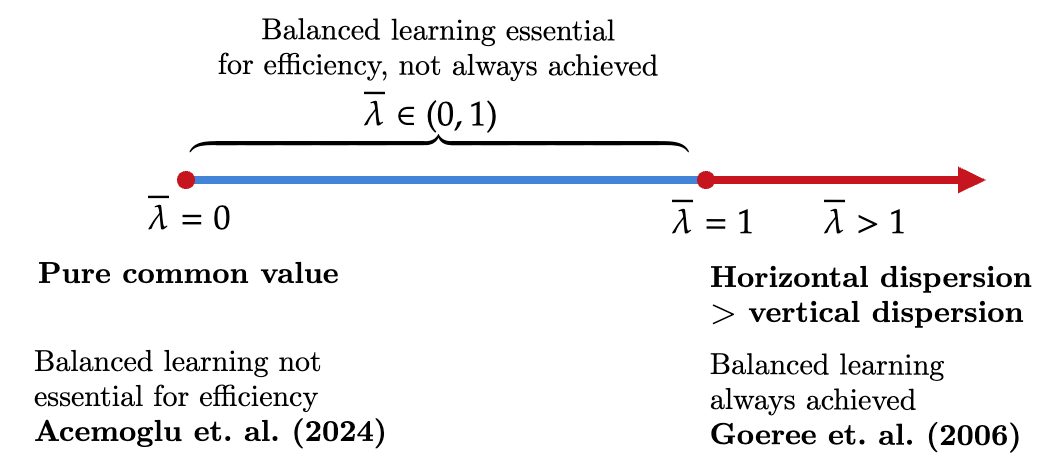}} 
    \label{fig:dispersion_diagram}
\end{figure}

\paragraph{Actions, preferences, payoffs.} There are two actions $\mathcal{I} = \{1,2\}$.\footnote{In product markets, the actions correspond to different products; in technology adoption, the actions correspond to different technologies.} Each action $i$ has an unknown binary value $\theta_i \in \{0,1\}$. Action $i$ is \emph{high value} if $\theta_i=1$, and \emph{low value} otherwise. The \emph{state} $\theta=(\theta_1,\theta_2)$ collects the value of each action. Let $\Theta=\{0,1\}^{2}$ be the \emph{state space}. Time is discrete: $\mathcal{T} = \{0,1,2,\ldots\}$. At each time $t \in \mathcal{T}$, an agent arrives and decides which action to take. The agent at $t$ observes their idiosyncratic preference (or \emph{type}) $\lambda_t \in \Lambda := [0, \bar \lambda]^2$, drawn i.i.d.\ across time from the distribution $\nu \in \Delta(\Lambda)$. Actions are \emph{vertically} and \emph{horizontally differentiated}: an agent of type $\lambda_t \in \Lambda$ who chooses action $i \in \mathcal{I}$ under state $\theta \in \Theta$ obtains payoff
\[
U(i,\theta,\lambda_t) := \underbrace{\lambda_{it}}_{\substack{\text{Horizontal}\\ \text{component}}} + \underbrace{\theta_i}_{\substack{\text{Common} \\ \text{vertical}\\ \text{component}}} 
\]
We impose a regularity condition on the type distribution. Say that $\nu$ satisfies \emph{regular horizontal demand} if $\nu$ has a density $f_\nu$ with respect to Lebesgue measure on $\Lambda=[0,\bar\lambda]^2$, and there exist $\underline f,\overline f$ such that $0<\underline f\leq f_\nu\leq\overline f<\infty$ a.e. We also assume that $\bar \lambda  \in (0,1)$, so that horizontal dispersion in idiosyncratic preferences is non-trivial but smaller than the common component of values. This case is depicted by the blue region in \cref{fig:dispersion_diagram}. For the case of full transparency, it is immediate to show that when $\bar \lambda = 0$, there is no heterogeneity and limit outcomes are always efficient, as in \cite{acemoglu2022learning};\footnote{Our model neither nests nor is nested by that of \cite{acemoglu2022learning}, who focus on the `selection effect', in which a type-specific confounder affects the signals. They give conditions ensuring complete learning despite the selection effect and characterize its speed in a single-product model. We abstract from the selection effect, after which it is immediate that limit outcomes are always efficient when $\bar \lambda = 0$.} when $\bar \lambda \geq 1$, limit outcomes are again always efficient.\footnote{When $\bar \lambda> 1$, every action is chosen with probability at least $\min_{i=1,2}\nu(\{\lambda:\lambda_i-\lambda_{-i}\geq1\})>0$ at every history, so every action is sampled infinitely often and every value is learned. An analogous point has been made in the context of observational social learning by \cite{goeree2006social}.} 

\paragraph{Choices.}
Agents are uncertain about the state $\theta$. When the time-$t$ agent arrives, they observe the \emph{public history} $H_t$ and form the \emph{public posterior} $\pi_t\in\Delta(\Theta)$,\footnote{As we discuss below, \emph{adaptive information policies} may correlate the two state coordinates in $\pi_t$, even though the prior is independent.} which induces marginal \emph{public beliefs} about each action $i$: $\mu_{it}:=\pi_t(\theta_i=1)$ and $\mu_t:=(\mu_{1t},\mu_{2t})$. At time $t$, the agent with preference $\lambda_t$ chooses an action to maximize their expected utility:\footnote{We break ties uniformly at random, but all results in this paper hold under any other tie-breaking rule.} 
\[i_t \in \operatorname*{arg\,max}_{i \in \mathcal{I}}\, \Big(\lambda_{it}+\mu_{it} \Big).
\]

\paragraph{Signals.} 
The value of each action is drawn independently from the time-$0$ common
prior $\mu_0=(\mu_{10},\mu_{20})$. To rule out uninteresting cases, we
assume throughout that $\mu_0\in(0,1)^2$ and
$\max_{i\in\mathcal I}\mu_{i0}<1-\bar\lambda$. At the end of each time $t$, agent $t$'s action $i_t$ is \emph{publicly revealed} and information is generated from a statistical experiment about the value $\theta_{i_t}$ of the time-$t$ action. Formally, each action $i$ is endowed with the same experiment $\phi: \{0,1\} \rightarrow \Delta([0,1])$, which produces a collection of signals $\{s_{ik}\}_{k\geq 1} \stackrel{\mathrm{iid}}{\sim} \phi^{\theta_i}$. Conditional on the state $\theta$, the signals $\{s_{ik}\}_{i\in \mathcal{I}, k\geq 1}$ are independent. For each action $i\in\mathcal I$, let $N_i(t):=\sum_{s<t}\mathbf 1\{i_s=i\}$ be the number of times action $i$ has been chosen before
date $t$. The signal generated at date $t$ is the next unused signal about the chosen action: $s_t:=s_{i_t,N_{i_t}(t)+1}$. The \emph{evidence} (or \emph{record}) generated at date $t$ is
\[
x_t:=\bigl(t,i_t,s_t\bigr)
\in \mathsf X:=\mathcal T\times\mathcal I\times[0,1].
\]
Each item of evidence contains its date, the action taken at that date (about which it is informative), and the realized signal. We maintain the non-triviality assumption that $\phi^{1} \neq \phi^0$, so values are learnable. Unless otherwise stated, we also impose the technical assumptions that $\phi^1$ is absolutely continuous with respect to $\phi^0$ (denoted $\phi^1 \ll \phi^0$), and that the log-likelihood ratio $\log \frac{d\phi^1}{d \phi^0}(s)$ is $\phi^1$-a.s. bounded. These assumptions rule out signals that perfectly reveal that the action is high value, and they simplify the statement of Theorem \ref{thrm:fulltrans}. 

\paragraph{Information regimes.} We introduce a general framework for modeling how a \emph{designer} can shape the flow of information from the past to the future. By imposing constraints on the designer's choice set, we recover the \emph{information regimes} described in the introduction. The designer is uncertain about the state $\theta$, but observes each item of evidence $x_t$ at the end of each date $t$. The \emph{time-$t$ designer history} collects the evidence generated up to and including date $t$:
\[
D_t=(x_0,\ldots,x_t)\in\mathcal D_t:=\mathsf X^{t+1}.
\] 

Evidence is not directly observed by agents. Instead, agents receive
\emph{public messages} about the designer history. Let $Y$ be the public message space.\footnote{Formally, let $\mathcal Y_{\mathsf{EV}}$ be the space of all finite sequences of elements of $\mathsf X$ having distinct, strictly increasing date coordinates, together with the empty sequence $\varnothing$. Endow this space with the Borel
$\sigma$-algebra inherited from the countable disjoint union
$\bigsqcup_{m\geq0}\mathsf X^m$. Take the public-message space to be the
standard Borel space $Y:=\mathcal Y_{\mathsf{EV}}\sqcup[0,1]$
where the second component allows for direct posterior reports.} The \emph{time-$t$ public history}
\[
H_t=(i_s,y_s)_{s<t}\in\mathsf H_t:=(\mathcal I\times Y)^t
\]
collects the actions and public messages observed before date $t$.\footnote{Define $H_0:=\varnothing$.} Let
$\mathscr H_t$ be the $\sigma$-algebra generated by $H_t$, which encodes the public information at the beginning of time $t$, and let $(\mathscr H_t)_{t\in\mathcal T}$ denote the public filtration. Let $\mathscr H_\infty$ be the $\sigma$-algebra generated by $\bigcup_{t}\mathscr H_t$. A \emph{time-$t$ information policy} is a Borel measurable Markov kernel
\[
\sigma_t:\mathcal D_t\times\mathsf H_t\to\Delta(Y),
\]
and an \emph{information policy} is a sequence $\sigma=(\sigma_t)_{t\in\mathcal T}$. After $x_t$ is generated, $y_t$ is drawn according to
$\sigma_t(\,\cdot\mid D_t,H_t)$ and appended to the public history: $H_{t+1}=(H_t;(i_t,y_t))$.

For $A\subseteq\{0,\ldots,t\}$ and
$D_t=(x_0,\ldots,x_t)$, let $D_t|_A:=(x_s)_{s\in A}$ listed in increasing order of dates, with
$D_t|_\varnothing=\varnothing$. For any two time-ordered evidence histories
$R$ and $R'$, write $R\preceq R'$ if $R$ is a time-ordered subhistory of $R'$.
In particular, $R\preceq D_t$ means $R=D_t|_A$ for some $A$. We
consider the following nested regimes.

\begin{enumerate}
    \item \textbf{Full transparency.}
    The full-transparency policy reveals the entire available stock of evidence:
    \[
    \sigma_{\mathsf{TR},t}(\,\cdot\mid D_t,H_t)=\delta_{D_t} \quad \forall t\in\mathcal{T}.
    \]
    \item \textbf{Adaptive evidence design.}
    Under an adaptive evidence policy, each public history records the cumulative stock of evidence disclosed. To make this precise, we need some additional definitions. For
    $D_t=(x_0,\ldots,x_t)$ and $H_t=(i_s,y_s)_{s<t}$, call the pair
    $(D_t,H_t)$ \emph{evidence-consistent} if, for each $s<t$, the action
    coordinate of $x_s$ is $i_s$,
    $y_s\preceq D_s:=(x_0,\ldots,x_s)$, and $y_{s-1}\preceq y_s$ whenever $1\leq s<t$. At an evidence-consistent
    pair, let $R(H_0):=\varnothing$ and, for $t\geq1$, let
    $R(H_t):=y_{t-1}$. A policy $\sigma$ is an \emph{adaptive evidence
    policy} if, at every evidence-consistent pair $(D_t,H_t)$,
    \[
    \sigma_t\!\left(
       \left\{R: R(H_t)\preceq R\preceq D_t\right\} \,\middle|\,D_t,H_t\right)=1.
    \]
    In words, the designer may condition disclosure on her realized history and
    may disclose any collection of previously undisclosed signals, but she may neither retract nor garble disclosed evidence. Off-path kernels are specified arbitrarily. Let $\Sigma_{\mathsf{EV}}$ be the set of such
    policies.

    \item \textbf{Adaptive information design.} Any information policy is allowed. Let $\Sigma_{\mathsf{ID}}$ denote the set of all information policies.
\end{enumerate}

Note that $\sigma_{\mathsf{TR}}\in\Sigma_{\mathsf{EV}}
\subseteq\Sigma_{\mathsf{ID}}$.

\paragraph{Summary of model timing.} 
We summarize the timing of the model. Before $t=0$, the state $\theta$ is drawn from the common independent prior $\mu_0$, the collection of signals $\{s_{ik}\}_{i\in \mathcal{I},k\geq 1}$ is drawn (conditionally i.i.d.\ across $k$ for each $i$ and independent across $i$) from $(\phi^{\theta_1},\phi^{\theta_2})$, and the designer commits to a policy $\sigma$. Then, at each time $t\in \mathcal{T}$, the following occur in order.\footnote{Let $D_{-1}=\varnothing$.} 
\begin{itemize}
    \item[(i)] Agent $t$ observes the public history $H_t$ and forms the
    public posterior $\pi_t$, with marginals
    $(\mu_{1t},\mu_{2t})$, by Bayes' rule.
    \item[(ii)] Agent $t$'s type $\lambda_t$ is drawn independently from
    $\nu$.
    \item[(iii)] Agent $t$ chooses $i_t$ to maximize their myopic expected
    utility.
    \item[(iv)] The evidence
    $x_t=(t,i_t,s_t)$ is observed by the designer and appended to their history: $D_t=(D_{t-1};x_t)$.
    \item[(v)] The public message $y_t$ is drawn from
    $\sigma_t(\,\cdot\mid D_t,H_t)$.
    \item[(vi)] The current action and message are appended to the public
    history: $H_{t+1}=(H_t;(i_t,y_t))$.
\end{itemize}

\paragraph{Asymptotic efficiency.} The main goal of this paper is to understand when and how social learning leads to inefficient limit outcomes. Say that \emph{learning is asymptotically efficient} if the following event occurs: 
\[
E:=\bigg\{ \lim_{t \to \infty}\Big(\underbrace{\max_{i \in \mathcal{I}} U(i,\theta,\lambda_t)}_{\text{Best possible payoff at $t$}} -\underbrace{U(i_t,\theta,\lambda_t)}_{\text{Actual payoff at $t$}}
\Big)=0 \bigg\}.
\]
Following the experimentation literature (e.g., \cite{aghion1991optimal}), our notion of efficiency requires that per-period regret vanish in the limit. This occurs whenever beliefs converge to the truth. However, as we show below, convergence to the truth is not necessary: not every value needs to be learned for regret to vanish. We emphasize that in our model, payoffs and information jointly determine the path of learning. In much of the bandit literature, these two forces are mechanically linked: agents learn from their consumption. By contrast, we decouple payoffs from signals to isolate how the experiment alone shapes learning outcomes.  

\paragraph{Learning environments.} Learning environments collect the primitives of our model. A \emph{learning environment} (henceforth \emph{environment}) is a tuple $(\sigma,\phi,\mu_0,\nu)$ consisting of (i) the information policy $\sigma$; (ii) the experiment $\phi$ about values, which is common across actions; (iii) the prior $\mu_0$; and (iv) the distribution $\nu$ of idiosyncratic preferences. Each environment $(\sigma,\phi,\mu_0,\nu)$ induces a probability measure $\mathbb{P}_{\sigma,\phi,\mu_0,\nu}$ over outcomes, as well as a probability measure over outcomes conditional on the state: $\mathbb{P}_{\sigma,\phi,\mu_0,\nu}(\cdot|\theta)$. Our main object of interest is the \emph{efficiency probability}, the probability that learning is asymptotically efficient, as a function of the environment:
\[
V(\sigma,\phi,\mu_0,\nu) := \mathbb{P}_{\sigma,\phi,\mu_0,\nu}(E).
\]
When an environment is fixed, we often omit the dependence on some or all of $\sigma,\phi,\mu_0,\nu$. 

\paragraph{Summary of modeling assumptions.} To summarize, we maintain the following assumptions for the rest of the paper unless otherwise stated. Priors are independent, interior, and satisfy $\max_i\mu_{i0}<1-\bar\lambda$, where $0<\bar\lambda<1$.
Types are i.i.d.\ and the type distribution satisfies regular horizontal demand, which ensures that both actions are chosen with positive probability whenever the public-belief gap lies strictly inside $(-\bar\lambda,\bar\lambda)$. Conditional on the state, each action's signal streams are independent across actions and i.i.d. across time, with a common
nontrivial experiment $\phi^1\ne\phi^0$.
Except where explicitly relaxed, $\phi^1\ll\phi^0$ and the
high-state log-likelihood ratio is essentially bounded for each experiment. The model also maintains two independence assumptions: (i) conditional on $(H_t,i_t)$, the signal $s_t$ is independent of the idiosyncratic preference $\lambda_t$ of the time-$t$ agent; and (ii) conditional on the state, signals about action $i$ are not informative about the other action $j \neq i$.\footnote{These assumptions are standard in the bandit literature and fit a variety of economic settings. For instance, in product markets, reviews are about particular products and typically contain enough information to `debias' information about the common component of quality from past consumers' individual tastes. This `selection effect' is the focus of \cite{acemoglu2022learning}.} 

\section{Balanced Learning under Transparency}\label{sec:transparency}
\paragraph{When do inefficiencies arise?} This section studies the regime of \emph{full transparency}. We begin with a benchmark that identifies when inefficiencies arise in full-transparency environments. This benchmark delimits how closely the efficiency probability $V$ can approach one. 

\begin{lemma}[Sources of inefficiency]\label{lem:sources} Fix any environment with $\sigma=\sigma_{\mathsf{TR}}$. In every state
$\theta\neq(1,1)$, $E$ occurs almost surely. In state $\theta=(1,1)$, $E^c$ occurs with positive probability. Consequently,
\[
V=1-\mu_{10}\mu_{20}
\Pr\bigl(E^c\mid\theta=(1,1)\bigr)<1.
\]
\end{lemma}

The proof is in \cref{app:prelim-trap}. The first claim and the formula for $V$ extend to every environment in which the value of any action that is sampled infinitely often is eventually learned (\cref{lem:sources-general}). We use this extension in \cref{sec:design} for adaptive evidence policies that eventually disclose every record.

\begin{figure}[h]
\centering
\caption{Illustration of \cref{lem:sources}}
    \begin{quote}
    \vspace{-1em}
    \centering 
    \scriptsize Experiment: raw signals are $\widetilde s_{ik}=\theta_i+\zeta_{ik}$, where $\zeta_{ik}\stackrel{\mathrm{iid}}{\sim}\mathcal N(0,5)$ across $(i,k)$ (Borel-encoded in $[0,1]$); preferences satisfy $\lambda_t\sim U([0,\bar\lambda]^2)$ with $\bar\lambda=0.5$. 
    \end{quote} 
    \subfloat[$\theta = (0,0)$]{\bslfigure{0.32\textwidth}{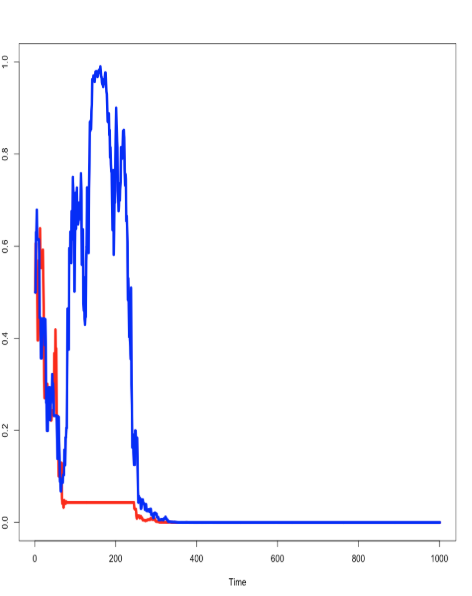}}
    \subfloat[$\theta = (1,1)$, succeed]{\bslfigure{0.32\textwidth}{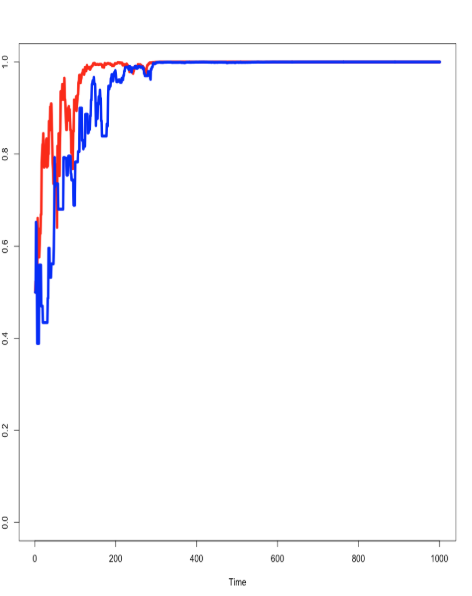}}
    \subfloat[$\theta = (1,1)$, fail]{\bslfigure{0.32\textwidth}{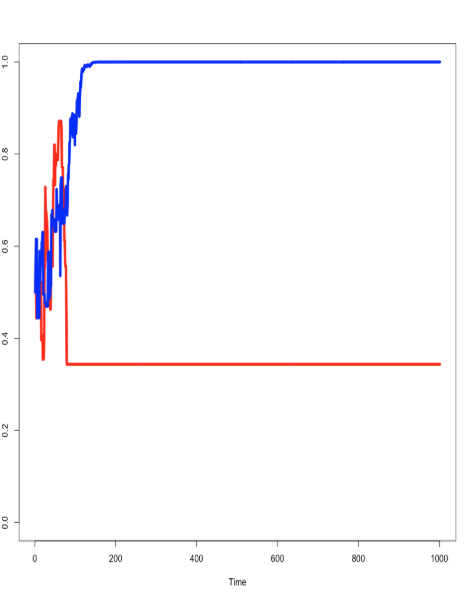}}
    \label{fig:sources}
\end{figure}
\cref{fig:sources} illustrates the lemma. In panel (a), the streak of mistaken optimism about the blue action is self-correcting, and both low-value actions are eventually learned about. In panel (b), both high-value actions are sampled and learned. In panel (c), society abandons one high-value action and never corrects its pessimism about it. Lemma \ref{lem:sources} establishes that inefficient belief paths such as the one depicted in panel (c) occur with positive probability in every full-transparency environment.

\paragraph{Efficiency under full transparency.} Lemma \ref{lem:sources} shows that no \emph{fixed} full-transparency environment achieves efficiency with probability one. How closely can full-transparency environments \emph{approximate} efficiency? To answer this question, we seek necessary and sufficient conditions on a \emph{sequence} of full-transparency environments indexed by $n$ under which efficiency holds approximately, i.e., $V_n\to 1$, where $V_n:=V(\sigma_{\mathsf{TR}},\phi_n,\mu_0,\nu)$ denotes the efficiency probability under the $n$th experiment. This criterion is a natural generalization of our efficiency criterion when the first-best of exact efficiency is unattainable. Informally, for efficiency to obtain with high probability, beliefs about both actions must become `more optimistic together' with high probability. To state when this occurs, we need a few more definitions. 

First, it is convenient to take a log-odds transformation of beliefs. Define
\[
\mathsf L(\mu):=\log\frac{\mu}{1-\mu} \quad \text{and its inverse} \quad \ell(x):=\frac{e^x}{1+e^x}
\]
with the conventions $\mathsf L(0)=-\infty$ and
$\mathsf L(1)=+\infty$. We apply $\mathsf L$ componentwise to
vectors. For $\bm x=(x_1,x_2)\in\mathbb R^2$, let
\[
\Gamma_1(\bm x)
:=
\nu\!\left(
 \left\{\lambda\in\Lambda:
 \lambda_1+\ell(x_1)\geq\lambda_2+\ell(x_2)
 \right\}
\right)
\qquad\text{and}\qquad
\Gamma_2(\bm x):=1-\Gamma_1(\bm x)
\]
be the probabilities that actions $1$ and $2$ are chosen at log-odds $\bm x$,\footnote{Since $\nu$ admits a density, the indifference set has
$\nu$-measure zero, so this definition does not depend on the
tie-breaking rule.} and
collect them in the vector
\[
\bm\Gamma(\bm x)
:=
\begin{pmatrix}
\Gamma_1(\bm x)\\
\Gamma_2(\bm x)
\end{pmatrix}
\]

Let $\bm X^{\mathrm{PB}}$ denote the unique global solution to the initial-value problem
\begin{equation}
\dot{\bm X}(t)
=
\bm\Gamma\!\left(\bm X(t)\right),
\qquad
\bm X(0)
=
\begin{pmatrix}
\mathsf L(\mu_{10})\\
\mathsf L(\mu_{20})
\end{pmatrix}
\tag{PATH}\label{eqn:path}
\end{equation}
where such a unique global solution exists by \cref{app:existence_ode}.

\begin{definition}[Path balance condition]
The prior $\mu_0$ and preferences $\nu$ satisfy \emph{path balance} if
\[
\lim_{t\to\infty}\bm X^{\mathrm{PB}}(t)
=
(+\infty,+\infty).
\]
\end{definition}

To interpret path balance, consider an alternative environment in which the state is $(1,1)$, public beliefs evolve deterministically in continuous time, and the log-odds of each action $i$ rise at a rate equal to the probability that action $i$ is chosen. Along this path, an `advantage' (a difference in beliefs) can reinforce itself and lead to herding on one action. However, an advantage does not necessarily imply imbalance, since beliefs are bounded: if the advantage does not accumulate sufficiently quickly, the belief about the disadvantaged action can nonetheless also converge to $1$.

\begin{figure}[h]
\centering
\caption{Examples of the path balance condition}
    \subfloat[Equal prior, symm. $\nu$]{\bslfigure{0.32\textwidth}{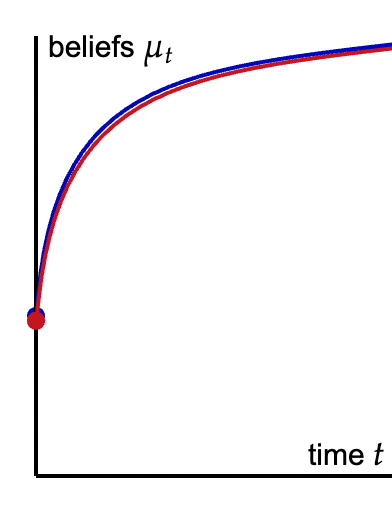}}
    \subfloat[Unequal prior, asymm. $\nu$]{\bslfigure{0.32\textwidth}{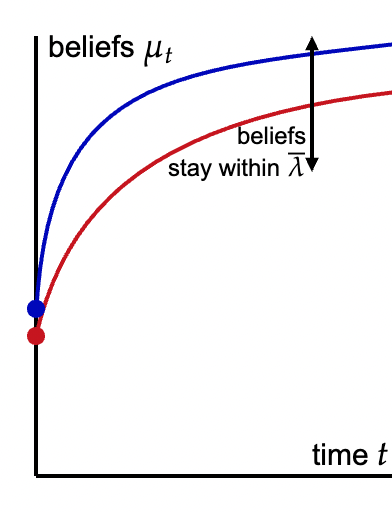}}
    \subfloat[Unequal prior, asymm. $\nu$]{\bslfigure{0.32\textwidth}{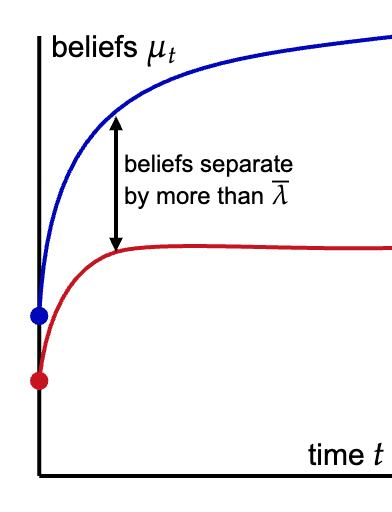}}
    \label{fig:path_balance}
\end{figure}

Path balance holds whenever the prior is equal across actions,
i.e., $\mu_{10}=\mu_{20}$, and the distribution of horizontal preferences is
symmetric across actions. Moreover, symmetry is not necessary: path balance is \emph{locally robust} in the sense that it holds for \emph{any} environment with sufficiently small asymmetry in the prior and the distribution of horizontal preferences (\cref{prop:path-balance-robustness}).\footnote{A concrete example is \(\bar\lambda=1/2\), \(\mu_0=(0.46,0.44)\), and \(f_\nu(\lambda_1,\lambda_2)=2+8\lambda_1\) on \([0,1/2]^2\). This environment satisfies path balance, although both the prior and horizontal preferences favor action \(1\).} Path balance might fail, however, when the prior beliefs are so far apart that one action is not chosen at time $0$, or when the distribution of preferences $\nu$ is such that the difference in beliefs grows so large that agents cease to choose one action. \cref{fig:path_balance} illustrates these cases.

Our main result for the full-transparency regime gives tight conditions under which outcomes are approximately efficient. Besides path balance, the conditions are stated in terms of two statistics of the experiment.

\begin{definition}[Drift and noise of experiments]
For experiment $\phi$, the \emph{drift} and \emph{noise} are: 
\[
    \mathsf{KL}[\phi]:= \Ex_{\phi^1} \left[ \log \frac{d \phi^1}{d\phi^0} \right] 
    \quad 
    \text{and} 
    \quad
    \mathsf{Var}[\phi] := \operatorname{Var}_{\phi^1} \left[ \log \frac{d \phi^1}{d\phi^0} \right].
\]
\end{definition}

To build intuition for these objects, note that the distribution of the log-likelihood ratio under $\phi^1$ governs how beliefs update after observing a signal from $\phi^1$. Our conditions apply to the first and second moments of this distribution. 

Finally, to obtain a tight characterization, we rule out the edge case in which path balance fails but $\bm X^{\mathrm{PB}}$ only tangentially approaches the abandonment boundary (e.g., where the belief gap equals $\bar\lambda$ and one action ceases to be chosen). Once the belief gap along $\bm X^{\mathrm{PB}}$ reaches $\bar\lambda$, the log-odds of the action with the lower belief stop moving and the gap grows strictly. The only case excluded by the following assumption is therefore a gap that converges to $\bar\lambda$ without ever reaching it.\footnote{This is because the coordinates of $\bm X^{\mathrm{PB}}$ are nondecreasing and their sum grows linearly. If path balance fails and the gap never reaches $\bar\lambda$, then one coordinate diverges while the other converges, so the gap converges to a limit of at most $\bar\lambda$. A limit strictly below $\bar\lambda$ would keep the choice probability of the disfavored action bounded away from zero, and its log-odds would diverge.}

\begin{assumption}[No tangential path-balance failure]
\label{as:no-tangential-pb}
If $(\mu_0,\nu)$ fails path balance, then there exists
$T_0<\infty$ such that
\[
\left|
\ell\!\left(X^{\mathrm{PB}}_1(T_0)\right)
-
\ell\!\left(X^{\mathrm{PB}}_2(T_0)\right)
\right|
>
\bar\lambda.
\]
\end{assumption}

\begin{theorem} \label{thrm:fulltrans}
Suppose Assumption \ref{as:no-tangential-pb} holds. Let $(\phi_n)_n$ be any sequence of experiments satisfying the maintained assumptions of \cref{sec:model} whose log-likelihood ratios are uniformly bounded above under $\phi_n^1$:\footnote{The `if' direction requires neither this bound nor Assumption \ref{as:no-tangential-pb}.}
    \[
      \sup_n\ \operatorname*{ess\,sup}_{s\sim\phi_n^1}\ \log\frac{d\phi_n^1}{d\phi_n^0}(s)<\infty.
    \]
Then $\lim_{n\to\infty}V(\sigma_{\mathsf{TR}},\phi_n,\mu_0,\nu)=1$ if and only if
    \begin{itemize}
        \item[(i)] \textbf{Prior and preferences}: $(\mu_0,\nu)$ satisfy path balance; and  
        \item[(ii)] \textbf{Information:} $(\phi_n)_n$ satisfies 
        \begin{itemize}
            \item[A.] {Drift vanishes}: $\mathsf{KL}[\phi_n] \to 0$.
            \item[B.] {Drift dominates noise}: $\mathsf{Var}[\phi_n]=o(\mathsf{KL}[\phi_n])$.
        \end{itemize}
    \end{itemize}

    \end{theorem}

\cref{thrm:fulltrans} characterizes asymptotic efficiency under the maintained assumptions. The uniform upper bound rules out sequences along which good news becomes arbitrarily close to conclusive. This assumption is imposed mainly for technical convenience: Proposition \ref{prop:goodnews} separately handles the case of perfect good news. Condition (i) is path balance. Condition (ii) imposes two properties on the sequence of experiments, which play complementary roles in ensuring balanced learning: 
\begin{itemize}[leftmargin = 2em]
    \item[A.] Vanishing drift controls \emph{randomness in the sampling share across actions}. Conditional on the sampled action being high value, beliefs about it grow only slightly more optimistic in expectation. This ensures that both actions are sampled at a `balanced' rate: over a fixed interval of time, the randomness in the share of periods in which each action is chosen is small relative to the movement in beliefs.
    \item[B.] Drift dominating noise controls \emph{randomness in signals for a given action}. Conditional on the sampled action being high value, the expected movement in beliefs must dominate the variance of this movement. This ensures that, conditional on action $i$ being chosen along a subsequence of times, public beliefs about $i$ along this subsequence concentrate around a fixed path.
\end{itemize}
The canonical experiments satisfying these conditions are `noisy bad news' experiments, in which conclusive `bad news' arrives only when the action is low value, and then only with low probability. To verify the conditions, observe that, conditional on the action being high value, bad news never arrives and beliefs increase deterministically, so the noise is identically zero: $\V[\phi_n]=0$. As the probability of bad news vanishes, $\KL[\phi_n] \to 0$, and hence conditions A and B hold. This example is stark, but both conditions can hold even when bad news is not conclusive: as \cref{ex:pbn} shows, they can be achieved by coarsening Gaussian signals.

A distinctive feature of many such experiments (including the binary bad news and coarsened Gaussian examples discussed later) is a Type-I/II asymmetry: false positives are much more likely than false negatives. Conditional on a high-value action, favorable but weakly informative signals are therefore common. Bayes' rule tempers the response to each such signal, so both beliefs track the balanced learning path with probability approaching one.

\begin{proof}[Proof sketch of \cref{thrm:fulltrans}] 
The proof of \cref{thrm:fulltrans} is involved and deferred to Appendix \ref{appendix:proofs1}; here we sketch the main ideas. Throughout the sketch, let $\KL_n:=\KL[\phi_n]$ and $\V_n:=\V[\phi_n]$. By \cref{lem:sources}, it suffices to study efficiency conditional on $\theta=(1,1)$, so all processes and probabilities below are taken with respect to this conditional law.

\underline{Drift does not vanish.} If $\KL_n\not\to0$, we may choose a subsequence along which, with positive probability uniformly across the subsequence, a finite set of signal realizations separates beliefs by more than $\bar \lambda+\varepsilon$. Conditional on that escape, applying Ville's maximal inequality to the nonnegative belief supermartingale $1-\mu_i$ gives a positive probability, uniformly along the subsequence, that the beliefs remain separated. Efficiency therefore fails with probability bounded away from zero in state $(1,1)$.
\\~\\
\underline{Drift vanishes.} Next, suppose that $\KL_n\to0$ and, along a subsequence, $\sqrt{\V_n/\KL_n}\to C\geq0$. If path balance holds, $C=0$ yields asymptotic efficiency. If $C>0$ is the limit along the retained subsequence, we require different tools to analyze the limit behavior of the belief process. Because the expected movement in beliefs each time an action is taken vanishes, a suitably rescaled version of the log-odds process, with shrinking time steps, converges to a continuous-time stochastic process that is more tractable. We show this in several steps. 

\noindent \underline{\smash{Step 1: Shrinking time steps.}} Let $(\bm{X}^{(n)}_t)_{t\in\mathcal T}$ denote the log-odds process under experiment $\phi_n$, which collects the log-odds of the two public beliefs at date $t$. For these processes, subscripts index discrete dates and parentheses index rescaled, continuous time. Define the \emph{rescaled} process
\[
\bm{X}^{(n)}(t) := 
    \bm{X}^{(n)}_{\lfloor t/\KL_n \rfloor} 
    \quad \text{for each $ t \in [0,\infty)$,}
\]
where $\lfloor x \rfloor$ denotes the largest integer not exceeding $x$, recalling that we are in the regime in which $\KL_n \to 0$.\footnote{The rescaling is usually taken with respect to an exogenous grid, as in Donsker's theorem on the convergence of random walks to Brownian motion. Because our sequence of experiments $(\phi_n)_n$ is arbitrary, we instead choose the rescaling to offset the drift $\KL_n$.} For fixed $n$, the rescaled process $(\bm{X}^{(n)}(t))_{t\geq0}$ is indexed by continuous time $[0,\infty)$ and takes values in the same log-odds space $\mathbb{R}^2$. Crucially, it has the same $t\to\infty$ behavior as the original process, and in particular the probability of learning is identical under both processes.

\noindent\underline{\smash{Step 2: Define the limit process.}}
Fix a subsequence, retained throughout this argument, along which
\[
\sqrt{\frac{\V[\phi_n]}{\KL[\phi_n]}}\to C
\qquad\text{for some }C\geq0.
\]
Let $(\bm X_t^*)_{t\geq0}$ solve
\begin{align}
d\bm X_t^*
&=
\bm\Gamma(\bm X_t^*)\,dt
+C\operatorname{diag}\!\bigl(\bm\Gamma(\bm X_t^*)\bigr)^{1/2}\,d\bm B_t,
\tag{SDE}\label{eqn:sde}\\
\bm X_0^*
&=
\begin{pmatrix}
\mathsf L(\mu_{10})\\
\mathsf L(\mu_{20})
\end{pmatrix}
\quad\text{a.s.},
\nonumber
\end{align}
where $\bm B$ is a standard two-dimensional Brownian motion and
$\operatorname{diag}(\bm\Gamma(\bm x))^{1/2}$ is the diagonal matrix with
entries $\sqrt{\Gamma_1(\bm x)}$ and $\sqrt{\Gamma_2(\bm x)}$. Each
coordinate of the log-odds thus drifts at the rate at which its action is
chosen and, when $C>0$, exhibits diffusion noise whose variance is
proportional to that rate. By \cref{lem:ex-unique-SDE}, the SDE admits a unique strong solution. Here and below, $\bm X^*$ denotes this SDE solution, while $\bm X^{\mathrm{PB}}$ denotes the deterministic solution of
\eqref{eqn:path}.

\noindent\underline{\smash{Step 3: Weak convergence via the martingale problem.}}
We identify the limit by convergence of local characteristics. For continuous coefficients $b:\mathbb R^2\to\mathbb R^2$ and
$a:\mathbb R^2\to\mathbb R^{2\times2}$, with $a(\bm x)$ symmetric and
positive semidefinite for every $\bm x$, define the second-order
operator
\[
\mathcal A f(\bm x)
:=
\sum_{i=1}^2 b_i(\bm x)\,\partial_i f(\bm x)
+\frac12\sum_{i,j=1}^2 a_{ij}(\bm x)\,\partial_{ij}f(\bm x),
\qquad f\in C_c^\infty(\mathbb R^2),
\]
where $C_c^\infty(\mathbb R^2)$ denotes the smooth functions on $\mathbb R^2$ with compact support.

\begin{definition}[Martingale problem]
\label{def:main-martingale-problem}
Let $\bm\omega=(\bm\omega_t)_{t\geq0}$ be the coordinate process on the path space
$C([0,\infty);\mathbb R^2)$ with its Borel
$\sigma$-algebra, and let
$(\mathscr C_t)_{t\geq0}$ be the natural filtration of $\bm\omega$.
A probability law $Q_{\bm x}$ solves the martingale problem for
$\mathcal A$ started from $\bm x$ if
$Q_{\bm x}(\bm\omega_0=\bm x)=1$ and, for each $f\in C_c^\infty(\mathbb R^2)$,
\[
f(\bm\omega_t)-f(\bm\omega_0)-\int_0^t\mathcal A f(\bm\omega_s)\,ds
\]
is a $(Q_{\bm x},\mathscr C_t)$-martingale.
We also call this the martingale problem for the coefficients
$(a,b)$ started from $\bm x$.
The martingale problem is \emph{well posed} if, for every
$\bm x\in\mathbb R^2$, such a law exists and is unique.\footnote{This is the definition in
\cite[Section~6.0]{stroock2007multidimensional}; see
\cite[Chapter~4, Sections~3--4]{ethier2009markov}
for the general formulation.}
\end{definition}

For \eqref{eqn:sde}, $b=\bm\Gamma$ and $a=C^2\operatorname{diag}(\bm\Gamma)$. It\^o's
formula shows that every weak solution law solves this martingale problem, and the converse SDE--martingale-problem correspondence and uniqueness in law
identify its unique solution. \cref{app:sde_existence} proves the
needed well-posedness. We then show in \cref{app:weakconvergence} that the finite-$n$ chains have local
characteristics converging to $(b,a)$. A diffusion-approximation criterion of
Stroock and Varadhan then identifies every subsequential limit as a solution of
the well-posed limiting martingale problem, which must equal the law of
$\bm X^*$.
\vspace{0.5em}

Say that $(\phi_n)_n$ is \emph{stochastically smooth} if it satisfies a technical condition, stated as \eqref{eqn:stochsmoothness} in Appendix \ref{app:weakconvergence} and described after the following lemma.

\begin{lemma}[Informal statement of weak convergence]
\label{lem:informal_weakconvergence}
Suppose
\[
\KL_n\to0
\qquad\text{and}\qquad
\sqrt{\frac{\V_n}{\KL_n}}\to C\in[0,\infty).
\]
If $(\phi_n)_n$ is stochastically smooth (which is automatically the case when $C=0$) then, for each
$T<\infty$,
\[
\bm X^{(n)}
\Rightarrow
\bm X^*
\qquad\text{in }\mathbb D([0,T];\mathbb R^2),
\]
where $\Rightarrow$ denotes weak convergence, $\mathbb D([0,T];\mathbb R^2)$ is the space of c\`adl\`ag paths from $[0,T]$ to $\mathbb R^2$ with the Skorokhod $J_1$ topology, and $\bm X^*$ is the unique solution of \eqref{eqn:sde} with diffusion
coefficient $C$.
\end{lemma}

A more precise statement of \cref{lem:informal_weakconvergence} is given in Appendix \ref{app:weakconvergence}. The stochastic smoothness condition is a Lindeberg-type tail restriction on the one-period log-likelihood-ratio increment $Z_n$ under $\phi_n^1$. Relative to the $\KL_n$ time scale, both the probability of an increment whose magnitude exceeds any fixed $\beta>0$ and the second-moment contribution of increments with $|Z_n|\geq1$ must vanish. 

\noindent \underline{\smash{Step 4: Analyze the limit process.}} The SDE characterizes the $n\to\infty$ law of the rescaled process. When $C>0$, a change of measure (Girsanov's theorem) shows that the diffusion enters a strict abandonment region with positive probability, and a posterior-supermartingale bound gives a positive probability of remaining there forever. When $C=0$, the limit is the path-balance ODE.

When stochastic smoothness fails, the upper bound on likelihood ratios implies that fixed-size signals occur with probability of order $\KL_n$. In a separate argument, we isolate finite blocks of such signals and control every remaining selected signal uniformly, even though sampling is endogenous. A rare tail of sufficiently positive or negative signals implies that learning fails with strictly positive probability.

Finally, when drift dominates noise, i.e., $\V_n/\KL_n\to0$, stochastic smoothness holds automatically. By \cref{lem:informal_weakconvergence}, the rescaled log-odds process converges to the special case of \eqref{eqn:sde} with $C=0$, which is the ODE \eqref{eqn:path}. If path balance holds, finite-horizon convergence ensures that beliefs enter into a high-belief region with high probability. Ville's maximal inequality for the supermartingales $1-\mu_i$ then keeps both actions active with probability approaching one. Conversely, if path balance fails, \cref{as:no-tangential-pb} yields a strict abandonment region and a positive probability that this abandonment is permanent.
\end{proof}

\begin{example}[Illustration of \cref{thrm:fulltrans}]
\label{ex:pbn}
Let $\lambda_t$ be uniformly distributed on $[0,\bar\lambda]^2$ and let $\mu_{10}=\mu_{20}$. First, consider the following experiment in which signals are noisy observations of values where the noise is Gaussian:
\[
s_{ik}=\theta_i+\zeta_{ik},
\qquad
\zeta_{ik}\stackrel{\mathrm{iid}}{\sim}\mathcal N(0,\sigma_\zeta^2),
\]
Panel (a) of \cref{fig:example_speed} illustrates a sample path. Between times $100$ and $200$, agents choosing action $1$ receive a string of positive signals that separates its belief from that of action $2$ by more than $\bar\lambda=0.5$. Subsequent agents then abandon action $2$ even though it is also high value.
\begin{figure}[h!]
    \centering
      \caption{Normal signals vs coarsened signals ($\approx$ perfect bad news)} 
      \begin{quote}
    \vspace{-1em}
    \centering 
    \footnotesize Note: $\theta = (1,1)$, $\sigma_\zeta = 5$, $\mu_0 = (0.1,0.1)$, $\bar \lambda = 0.5$.
    \end{quote} 
    \vspace{-1em}
    \subfloat[Normal signals]{\bslfigure{0.43\textwidth}{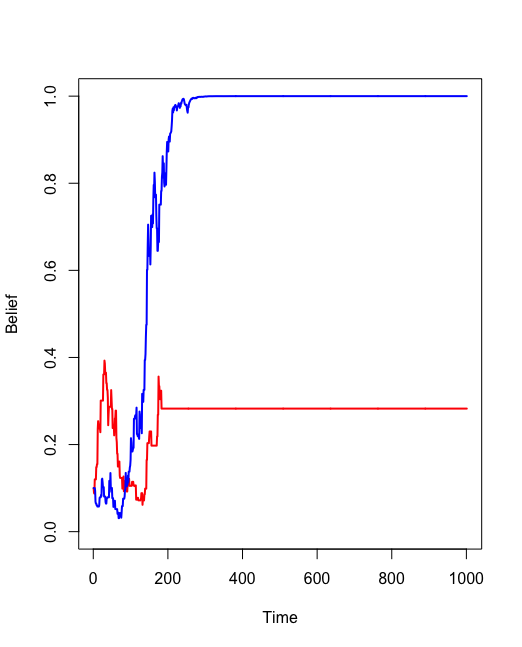}} 
    \subfloat[Coarsened signals $\approx$ PBN]{\bslfigure{0.43\textwidth}{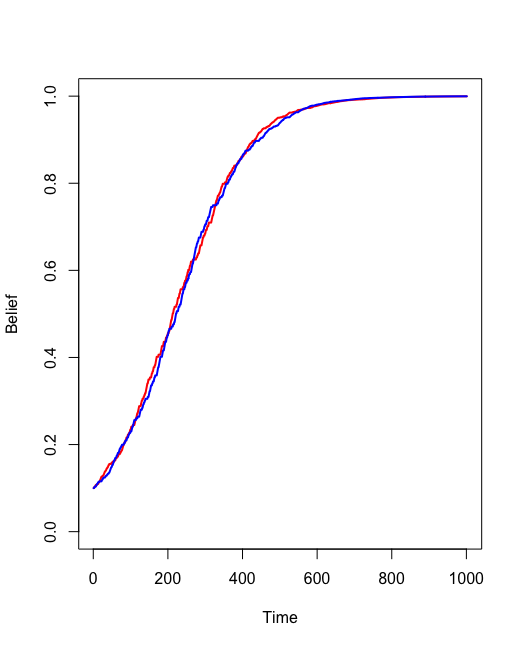}} 
    \label{fig:example_speed}
\end{figure}

Next, for each $K>0$, coarsen each signal into the binary partition
$\{s<-K\},\{s\geq-K\}$, and denote the induced experiment by $\phi^K$.
Let
\[
p_\vartheta^K
:=
\Pr(s<-K\mid\theta_i=\vartheta)
=
\Phi\!\left(-\frac{K+\vartheta}{\sigma_\zeta}\right),
\qquad \vartheta\in\{0,1\}.
\]
The two high-state log-likelihood-ratio increments are
\[
z_B^K:=\log\frac{p_1^K}{p_0^K},
\qquad
z_G^K:=\log\frac{1-p_1^K}{1-p_0^K}.
\]
By standard Mills ratio bounds,
\[
\frac{p_1^K}{p_0^K}=O\!\left(e^{-K/\sigma_\zeta^2}\right),
\qquad
z_B^K=-\frac{K}{\sigma_\zeta^2}+O(1),
\qquad
z_G^K\sim p_0^K.
\]
Consequently,
\[
\KL[\phi^K]\sim p_0^K\to0,
\qquad
\frac{\V[\phi^K]}{\KL[\phi^K]}
=
O\!\left(K^2\frac{p_1^K}{p_0^K}\right)
\to0.
\]
The rare bad-signal increment $z_B^K$ is unbounded below as $K\to\infty$, but
the log-likelihood ratios are uniformly bounded above, because $z_B^K<0<z_G^K\leq-\log(1-p_0^K)\leq\log2$ for every $K\geq0$, so
the sequence satisfies the hypothesis of \cref{thrm:fulltrans}. Path balance
holds by symmetry. The theorem therefore gives
\[
V(\sigma_{\mathsf{TR}},\phi^K,\mu_0,\nu)\to1.
\]
Panel~(b) of \cref{fig:example_speed}
illustrates the tandem movement of beliefs for a representative coarsened
experiment. In this example, note that because
$\V[\phi^K]/\KL[\phi^K]$ is of order $K^2e^{-K/\sigma_\zeta^2}/\sigma_\zeta^4$,
the ratio is small only when $K$ is large relative to $\sigma_\zeta^2$. $\hfill \diamondsuit$
\end{example}

\noindent \textbf{Maximally imbalanced experiments.} The sufficient conditions in \cref{thrm:fulltrans} identify noisy bad news as one canonical route to restoring efficiency. We now study the opposite extreme: every perfect-good-news experiment leads to imbalanced learning, and hence to inefficient limit outcomes whenever both actions are high value. To accommodate such experiments, we relax the assumptions that $\phi^1\ll\phi^0$ and that the high-state log-likelihood ratio is bounded.

\begin{definition}[Perfect good news]\label{def:perfect-good-news}
Let $\rho:=\phi^0+\phi^1$ and let
$g_\vartheta:=d\phi^\vartheta/d\rho$ for $\vartheta\in\{0,1\}$. The
experiment $\phi$ is \emph{perfect good news} if there exists a measurable
set $G$ such that
\[
\phi^0(G)=0,
\qquad
q:=\phi^1(G)>0,
\]
and
\[
g_1(s)\leq g_0(s)
\qquad
\text{for $\rho$-almost every $s\in G^c$.}
\]
In words, a signal in $G$ conclusively reveals that the sampled action is high
value, whereas every nonconclusive signal weakly lowers its posterior
probability of being high value.
\end{definition}

The first condition says that conclusive news that the value is high arrives with positive probability. The second condition says that, in the absence of such news, public beliefs become weakly more pessimistic. A canonical example is a `breakthrough' technology \citep{keller2005strategic}: a conclusive signal arrives only if the technology is high value, and in its absence public beliefs become more pessimistic. The following result shows that, under perfect good news, limit outcomes are never efficient when both actions are high value.
\vspace{0.5em}

\begin{proposition}[Good news generate imbalance]\label{prop:goodnews}
Under the maintained assumptions on $(\mu_0,\nu)$, if $\phi$ is perfect good
news, then, under full transparency,
\[
\Pr_{\sigma_{\mathsf{TR}},\phi,\mu_0,\nu}
\bigl(E\mid\theta=(1,1)\bigr)
=
0.
\]
\end{proposition}

\begin{figure}[H]
\begin{minipage}[t]{0.5\linewidth}
\cref{fig:diagram_outcomes} illustrates a partial characterization of the speed--efficiency pairs that can arise as we vary the experiment $\phi$. \\~\\ 
\cref{prop:goodnews} identifies perfect-good-news experiments under which limit outcomes are never efficient when both actions are high value. \\~\\
\cref{thrm:fulltrans} characterizes the limiting point at which the efficiency probability approaches one. Heuristically, this corresponds to a `noisy bad news' limit.  
\end{minipage}%
\hfill%
\begin{minipage}[t]{0.45\textwidth}\vspace{0pt}
\vspace{-1em}
\centering
{\bslfigure{\textwidth}{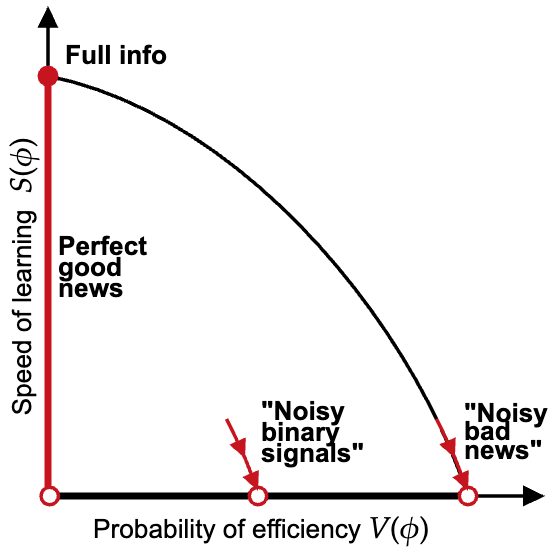}}
     \caption{Speed vs balance} 
      \begin{quote}
    \vspace{-1em}
    \centering 
    \scriptsize Note: Fix $\mu_0,\nu$ s.t. path balance holds, $\sigma = \sigma_{\mathsf{TR}}$ (full transparency)
    \end{quote} 
    \label{fig:diagram_outcomes}
\end{minipage}
\end{figure}

Combining \cref{thrm:fulltrans} and \cref{prop:goodnews}, \cref{fig:diagram_outcomes} gives a heuristic illustration of how different experiments map to different learning outcomes. Fix some $(\mu_0,\nu)$ under which the path balance condition holds. Fix $(\mu_0,\nu)$ satisfying path balance and a discount factor
$\delta\in(0,1)$. Define
\[
S_\delta(\phi):=
(1-\delta)
\Ex_{\sigma_{\mathsf{TR}},\phi,\mu_0,\nu}
\left[
\sum_{t=0}^{\infty}\delta^t
\mathbf 1\{\theta_{i_t}=\max_{i\in\mathcal I}\theta_i\}
\right].
\]
This is the expected discounted frequency with which agents choose a best-available action. Holding $\delta$ fixed, $S_\delta$ assigns greater weight to earlier corrections of mistakenly choosing a low-value action when a high-value action is available. In this sense, it measures the component of efficiency attributable to the \emph{speed of learning}.

\noindent \textbf{How noisy does bad news have to be?} \cref{thrm:fulltrans} shows that `noisy bad news' suffices for the efficiency probability to approach one under path balance. But how noisy must bad news be (and hence, how slow must learning be)? To study this question, we focus on \emph{binary bad news} $\phi^p$, under which bad news arrives with probability $p$ when an action is low value and never when it is high value. Conditional on $\theta_i = 1$, each time action $i$ is chosen no bad news arrives and its log-odds increase deterministically by $\log \frac{1}{1-p} > 0$. Our next result, which is \emph{non-asymptotic in $p$}, shows that whenever the path balance condition holds, the efficiency probability converges to $1$ at least exponentially fast as the bad news becomes noisier ($p \to 0$). \vspace{0.5em}

\begin{theorem}\label{thrm:exp_convergence}
Assume that $\nu$ satisfies regular horizontal demand, and suppose
$(\mu_0,\nu)$ satisfies path balance. Then there exist constants $K>0$ and
$p_0>0$, depending only on $(\mu_0,\bar\lambda,\nu)$, such that, for each
$p\in(0,p_0)$, full transparency under binary bad news $\phi^p$ satisfies
\[
V(\sigma_{\mathsf{TR}},\phi^p,\mu_0,\nu)
\geq
1-2\exp\!\left(-\frac Kp\right).
\]
\end{theorem}

The proof of \cref{thrm:exp_convergence} is in \cref{appendix:proof2}. Here, we briefly sketch the idea. Recall (\cref{lem:sources}) that efficiency can fail only in the state $\theta = (1,1)$, on which we condition throughout. In this state, binary bad news shuts down one of the two sources of randomness. Since bad news never arrives, the log-odds of a chosen action move up by the same fixed amount each time. Public beliefs are therefore a deterministic function of the \emph{counts} of how often each action has been chosen, and the only remaining randomness is in the idiosyncratic types that arrive. The crux of our argument is to bound how quickly the stochastic log-odds process concentrates around the deterministic path $\bm X^{\mathrm{PB}}$. Because one action is chosen each period, the sum of the two log-odds is deterministic, so the deviation of the process from $\bm X^{\mathrm{PB}}$ reduces to a single scalar.

\begin{figure}[H]
\begin{minipage}[t]{0.5\linewidth}
The main step is to observe that this deviation is a martingale plus a Lipschitz drift.  \\~\\
We then use a maximal version of the Azuma--Hoeffding inequality together with a finite-time Gr\"onwall factor to show that, with probability at least $1 - 2\exp(-K/p)$, the log-odds process stays within a fixed distance of $\bm X^{\mathrm{PB}}$ over $O(1/p)$ periods. 
This is enough time to lift both beliefs into an absorbing `robust region', in which beliefs are high enough that the belief gap can never again exceed $\bar\lambda$ (see \cref{fig:concentration}).
\end{minipage}%
\hfill%
\begin{minipage}[t]{0.48\textwidth}\vspace{0pt}
{\bslfigure{\textwidth}{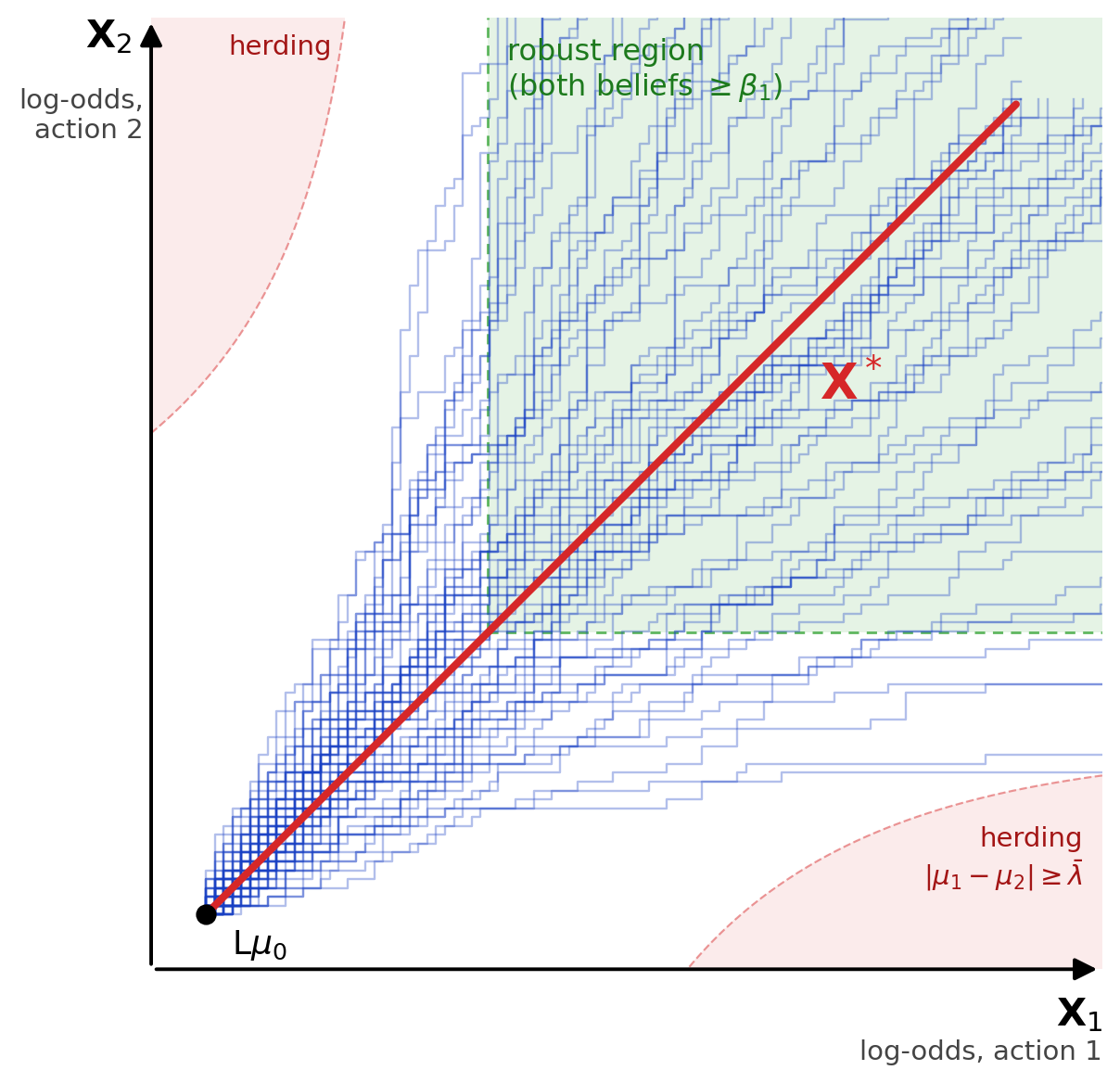}}
\caption{Path concentration}\label{fig:concentration}
\end{minipage}
\end{figure}

Once in this region, full support of preferences keeps both actions chosen with probability bounded away from zero, so both are sampled infinitely often and both beliefs converge to $1$.
\cref{thrm:exp_convergence} suggests that binary bad news that is `not too noisy' can achieve a high efficiency probability while sacrificing little speed.\footnote{We conjecture that \emph{every} point on the speed--balance frontier can be achieved with binary bad news experiments, i.e., that varying $p > 0$ traces out the frontier. However, we expect that a useful upper bound on the efficiency probability will require a more subtle combinatorial argument, since (i) the event of escape comprises all paths that can lead to escape; and (ii) the probabilities of these paths are path-dependent.}

\section{The value of design}\label{sec:design}

\cref{thrm:fulltrans} gives tight conditions under which the efficiency probability approaches one under full transparency. We now ask whether---and if so, by how much---a designer who controls the information policy can do better. In product markets, the designer is a platform that designs a review system, e.g., by coarsening or censoring reviews to encourage balanced exploration and thereby cater to idiosyncratic tastes (e.g., for color or shape). In technology adoption, the designer is a government that decides how to disclose information about the efficacy of new technologies.

\paragraph{When is efficiency attainable under design?} We begin by stating a tight condition under which adaptive evidence design and adaptive information design can bring the efficiency probability arbitrarily close to one. 

\begin{definition}[Static balance condition]
    A belief $\mu=(\mu_1,\mu_2)$ satisfies the \emph{static balance condition} if $|\mu_1 - \mu_2| < \bar \lambda$. 
\end{definition}

The static balance condition is implied by, but much weaker than, path balance.\footnote{Note that static balance can fail only when $\bar\lambda<\tfrac12$: the maintained restriction $\max_i\mu_{i0}<1-\bar\lambda$ gives $|\mu_{10}-\mu_{20}|<1-\bar\lambda$, which is at most $\bar\lambda$ whenever $\bar\lambda\geq\tfrac12$.} Under regular horizontal demand, it says that both actions are chosen with positive probability at the prior. Under the generic-prior assumption below, static balance characterizes when efficiency is attainable with adaptive evidence and adaptive information policies.

\begin{assumption}[Generic prior for design]\label{as:generic-design}
The prior does not lie on the static-balance boundary:
$|\mu_{10}-\mu_{20}|\neq\bar\lambda$.
\end{assumption}

\begin{theorem}\label{thrm:evidence_iff}
Suppose \cref{as:generic-design} holds, and fix any (nontrivial) experiment $\phi$.
\begin{itemize}
    \item[(i)] \textbf{Sufficiency.}
    If $\mu_0$ satisfies static balance, then for each
    $\varepsilon>0$ there is a finite $T$ and an accumulation policy
    $\sigma^{\mathsf{ACC},T}\in\Sigma_{\mathsf{EV}}$ such that
    \[
    V(\sigma^{\mathsf{ACC},T},\phi,\mu_0,\nu)
    \geq1-\varepsilon.
    \]
    Consequently,
    \[
    \sup_{\sigma\in\Sigma_{\mathsf{EV}}}
    V(\sigma,\phi,\mu_0,\nu)=1.
    \]

    \item[(ii)] \textbf{Necessity.}
    If $\mu_0$ does not satisfy static balance, then
    \[
    \sup_{\sigma\in\Sigma_{\mathsf{EV}}}
    V(\sigma,\phi,\mu_0,\nu)
    \leq
    \sup_{\sigma\in\Sigma_{\mathsf{ID}}}
    V(\sigma,\phi,\mu_0,\nu)
    <1.
    \]
\end{itemize}
\end{theorem}

Each accumulation policy stays silent during an accumulation stage of finite length, discloses the accumulated evidence at a deterministic date, and then fully discloses fresh evidence. The value of such policies approaches one as the accumulation horizon grows. Under \cref{as:generic-design}, static balance is also necessary for efficiency to be attainable: under strict imbalance, agents initially choose only the favored action, and Bayes plausibility bounds the probability of restoring balance.

\begin{figure}[H]
    \centering
      \caption{Accumulation policy} 
      \vspace{-0.8em}
    \subfloat[Accumulation under $\theta = (1,1)$]{\bslfigure{0.4\textwidth}{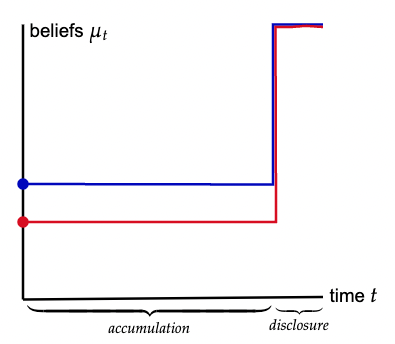}} 
    \subfloat[Accumulation under $\theta = (1,0)$]{\bslfigure{0.4\textwidth}{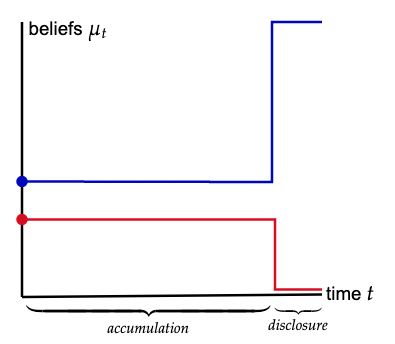}} 
    \label{fig:accumulation}
\end{figure}

The length of the accumulation stage does not depend on the realizations of interim signals: the designer commits to a period of silence. This commitment ensures that beliefs conditioned on no disclosure do not move. Both actions therefore continue to be chosen, which is the sense in which the policy \emph{accumulates} information. The fixed period of silence stands in contrast to the policies in \cite{che2018recommender,knoepfle2024dynamic}, who study how information policies should be designed in a setting with a single action of uncertain value. The difference arises because the designer in our setting must control public beliefs about \emph{both} actions: releasing evidence before both actions have been sampled sufficiently can make public beliefs diverge and undermine
subsequent exploration. A deterministic accumulation period avoids this risk by keeping the public belief fixed while
evidence is collected.

\textbf{How often can efficiency be achieved when static balance fails?} We have characterized when efficiency is attainable under design. We now characterize the \emph{second-best value} under adaptive information and study how the value of adaptive information can be \emph{approximated} by the value of adaptive evidence. For the remainder of this section, let $\overline\mu:=\max_{i\in\mathcal I}\mu_{i0}$ and $\underline\mu:=\min_{i\in\mathcal I}\mu_{i0}$.

\begin{theorem}\label{thrm:adaptiveinfo}
Suppose \cref{as:generic-design} holds, fix a (nontrivial) experiment $\phi$, and suppose the static
balance condition does not hold. Then
\[
\sup_{\sigma\in\Sigma_{\mathsf{ID}}}
V(\sigma,\phi,\mu_0,\nu)
=
1-\frac{\underline\mu(\overline\mu-(\underline\mu+\bar\lambda))}{1-(\underline\mu+\bar\lambda)}
\]
\end{theorem}

The starting point for \cref{thrm:adaptiveinfo} is to note that, when the static balance condition fails, the prior initially only allows for one action to be sampled. Without loss, call this action~$1$, so that $\mu_{10}=\overline\mu$ and $\mu_{20}=\underline\mu$. Define the hitting time $\tau := \inf \big\{t \in \mathcal{T}: \mu_{1t} \leq \underline\mu+\bar\lambda \big\}$. On the event $\{\tau=\infty,\theta=(1,1)\}$, action $1$ is chosen forever and efficiency fails. Hence, 
\[
V\leq 1-\overline\mu \underline\mu\cdot \mathbb{P}(\tau=\infty|\theta=(1,1))
\]
We then use a standard downcrossing argument to obtain the bound
\[
\mathbb{P}(\tau<\infty\mid\theta=(1,1))
\leq\frac{(\underline\mu+\bar\lambda)(1-\overline\mu)}{\overline\mu(1-(\underline\mu+\bar\lambda))}
\]
Combining these inequalities yields the $\leq$ direction of \cref{thrm:adaptiveinfo}. To show that this bound is tight, we consider policies in which the designer waits until her private posterior is sufficiently informed and then sends one of two messages, splitting her posterior so that the public belief about action~$1$ moves either strictly inside the balanced region or close to one. In the former case, the designer chains an accumulation policy as in \cref{thrm:evidence_iff}. In the latter case, she accepts the residual failure probability. Finally, we show that saturating the Bayes plausibility constraint and letting the two split points approach the boundary and one, respectively, approximates the upper bound.

\begin{figure}[H]
\vspace{0em}
\centering
{\bslfigure{0.55\textwidth}{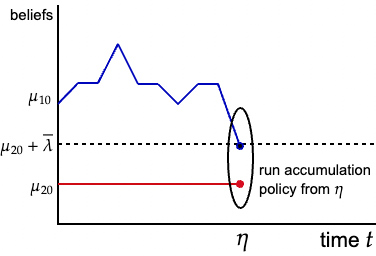}}
    \caption{`Chaining' accumulation policy after $\tau$}\label{fig:chain_accumulation}
\end{figure}

Adaptive information design may require more control than a designer has in some contexts. For instance, online marketplaces such as Amazon not only aggregate reviews (e.g., into star ratings) but also make some (though potentially not all) real and unedited reviews available to consumers. Within our taxonomy, such policies are \emph{adaptive evidence} policies.\footnote{A recent paper by \cite{knoepfle2024dynamic} studies such policies for a single innovation about which conclusive good/bad news arrives and the designer can choose to delay its disclosure.} How often can adaptive evidence policies achieve efficiency when static balance fails? We answer this question for the regime in which individual signals have a vanishing impact on beliefs, which the following definition formalizes. 

\begin{definition}[$\delta$-boundedness]
As in \cref{def:perfect-good-news}, let $\rho:=\phi^0+\phi^1$ and
$g_\vartheta:=d\phi^\vartheta/d\rho$ for $\vartheta\in\{0,1\}$. Define the posterior obtained from belief $m$ after directly observing signal $s$ as:
\[
B_\phi(m,s)
:=
\begin{cases}
\displaystyle
\frac{m g_1(s)}{m g_1(s)+(1-m)g_0(s)},
&\text{if }m g_1(s)+(1-m)g_0(s)>0,\\[1.1ex]
m,
&\text{if }m g_1(s)+(1-m)g_0(s)=0,
\end{cases}
\qquad m\in(0,1).
\]
The experiment $\phi$ is \emph{$\delta$-bounded} if, for each
$m\in(0,1)$,
\[
\bigl|B_\phi(m,s)-m\bigr|\leq\delta
\]
for $\bigl(m\phi^1+(1-m)\phi^0\bigr)$-almost every $s$. Let
$\Phi_\delta$ denote the set of experiments that satisfy this condition and
the maintained assumptions of \cref{sec:model}. This set is nonempty for every $\delta>0$: for example, it contains every nontrivial binary experiment whose two likelihood ratios are sufficiently close to one.
\end{definition}

$\delta$-boundedness limits the impact that each individual signal can have on the public belief about an action: it rules out one-step posterior jumps larger than $\delta$ from observing any given signal. After suitable time rescaling and under additional local-moment conditions, such sequences admit continuous-path limits.\footnote{Analogous diffusion limits have been studied in single-agent settings \citep{liang2021dynamically}.}

\begin{manualtheorem}{4B}\label{thrm:adaptiveinfoB}
Suppose \cref{as:generic-design} holds and the static balance condition does not hold. Then adaptive evidence approaches the
adaptive-information value of \cref{thrm:adaptiveinfo} uniformly over $\delta$-bounded experiments:
\[
\lim_{\delta\downarrow0}
\inf_{\phi\in\Phi_\delta}
\sup_{\sigma\in\Sigma_{\mathsf{EV}}}
V(\sigma,\phi,\mu_0,\nu)
=
1-\frac{\underline\mu(\overline\mu-(\underline\mu+\bar\lambda))}{1-(\underline\mu+\bar\lambda)}.
\]
\end{manualtheorem}

The key insight behind \cref{thrm:adaptiveinfoB} is that, when evidence arrives in sufficiently small pieces, selective evidence disclosure can stop strictly inside the balanced region with arbitrarily high precision and then run the accumulation policy. The limit value does not depend on the experiment and equals the value in \cref{thrm:adaptiveinfo}.

\section{Conclusion}
When preferences are heterogeneous, efficient social learning requires \emph{balance} such that every high-value action is discovered. This paper studies how the informational and preference environment affects the possibility of obtaining efficient outcomes. Under full transparency, we characterize when the efficiency probability can be made approximately one in terms of the \emph{noise} and \emph{drift} of experiments (\cref{thrm:fulltrans}). We then turn our attention to \emph{perfect good news} and \emph{perfect bad news}, documenting that the former lead to extreme cases of learning failure regardless of noise (\cref{prop:goodnews}), while the latter yield efficient outcomes at an \emph{exponential} rate in the noise parameter (\cref{thrm:exp_convergence}). Together, these results shed light on the key dimensions of the speed-efficiency tradeoff displayed in \cref{fig:diagram_outcomes}.

Next, we study the value of design in our setting. We find that efficiency can be achieved by adaptive evidence design (and more generally, adaptive information design) if and only if the prior is sufficiently balanced (\cref{thrm:evidence_iff}). Moreover, when the prior is imbalanced, we characterize the second-best adaptive-information value and show that adaptive evidence approaches it uniformly as the impact of individual signals vanishes (\cref{thrm:adaptiveinfo,thrm:adaptiveinfoB}). We conclude with a few remarks on the broader economic implications of our findings.

\textbf{1. Fast vs slow learning.} In environments with single-dimensional uncertainty (e.g., a single product or technology is either good or bad), a common prescription is to \emph{speed up} social learning. For instance, \cite{acemoglu2022learning} show how finer reviews in online marketplaces can facilitate quicker learning about product quality. More recent work by \cite{knoepfle2024dynamic} and \cite{board2025braeking} reinforces this view, showing that a designer would want to be fully transparent about bad news, disclosing it as soon as it arrives. We offer a complementary view of how and why a designer might instead want to slow learning down: slow learning balances exploration across actions even though each agent is myopic and does not internalize the learning externality on future agents. 

However, slow learning may also be costly: in the interim, agents risk taking a low-value action. We make three remarks about speed. First, if the designer is patient, the efficiency probability is what matters for welfare. Second, even if the designer is impatient, \cref{thrm:exp_convergence} suggests that the experiment may not need to be very noisy in practice: under binary bad news, the efficiency probability converges to $1$ at an exponential rate in the noise. A key takeaway is therefore that not all garblings are created equal, and slowing learning down \emph{in the right way} can improve welfare even away from the patient limit. Third, how a designer should resolve the tradeoff between speed and balance remains an open question. The tools we use to show convergence of the belief process to the solution of a stochastic differential equation apply only in the noisy-experiment limit. Understanding how often efficient outcomes arise away from this limit is an exciting question for future work.

\textbf{2. Coarsening in theory and practice.} In practice, the statistical experiment generated by an action is often fixed rather than a design choice. For instance, the probability that a new technology generates breakthroughs and/or breakdowns is technology-specific. Similarly, the informativeness of a consumer's experience is specific to the product in question. \cref{thrm:evidence_iff} offers a simple recipe for restoring balanced learning in such settings. Yet the associated policy is stark: it prescribes potentially long periods in which agents receive no information and public beliefs do not move. Such a policy might be undesirable for a designer (e.g., an e-commerce platform) who is patient and cares about long-run efficiency but, conditional on that, wishes to minimize the probability that low-value actions are chosen in the interim, i.e., at finite times. \cref{thrm:fulltrans} instead points to coarsenings that approximate noisy bad news. Along the sequences covered by the theorem, these coarsenings permit interim learning while the efficiency probability approaches one.

In product markets, where the exogenous experiment is typically not conclusive about quality (but might still be quite informative), one such coarsening is a simple \emph{upper-censorship} policy, in which favorable signals above a threshold are pooled while lower-tail signals are disclosed.\footnote{Such disclosure policies have been extensively studied in the literature on static persuasion. The reason they are useful here is distinct.} This policy might be operationalized by (i) offering consumers an option to report fraudulent products, which might then be taken off the platform---these reports correspond to the `bad news' signals; and (ii) coarsening positive reviews.\footnote{Casual introspection suggests that the marginal 5-star review on Amazon or Google Maps does not make us much more optimistic about the quality of the product or restaurant.} In technology adoption, where the exogenous experiment might be conclusive `good news' (breakthroughs) or `bad news' (breakdowns), one such coarsening is to disclose bad news stochastically (thereby slowing down the Poisson arrival rate) and never to disclose good news. 
    
\setstretch{0.8}
\small 
\setlength{\bibsep}{0pt}
\bibliography{WorksCited}

\setstretch{1.2}
\normalsize 
\appendix

\begin{center}
    \large \textbf{APPENDIX TO `BALANCED SOCIAL LEARNING'}
\end{center}

\setstretch{1.2}

\appendix

\section{Preliminaries}\label{appendix:prelim}

This appendix collects tools that are used repeatedly in the subsequent proofs: the link between sampling, learning, and efficiency (\cref{app:prelim-sampling}), bounds on the posterior belief process (\cref{app:prelim-trap}), and the efficiency of the accumulation policy (\cref{app:prelim-accumulation}). \cref{app:prelim-trap} additionally proves \cref{lem:sources}.

\paragraph{Notation.} We collect the conventions used in the appendix. Bold symbols denote $\mathbb R^2$-valued vectors, functions, and processes. The log-odds map $\mathsf L$ and its inverse $\ell$ are applied componentwise, with $\ell(+\infty)=1$ and $\ell(-\infty)=0$. For a belief vector $\mu=(\mu_1,\mu_2)\in[0,1]^2$ and $i\neq j$, let
\[
p_i(\mu):=\nu\bigl(\{\lambda\in\Lambda:\lambda_i+\mu_i\geq\lambda_j+\mu_j\}\bigr)
\]
denote the probability that action $i$ is chosen at beliefs $\mu$, so that $\Gamma_i(\bm x)=p_i(\ell(\bm x))$. Each $p_i$ is continuous (since $\nu$ has a density, indifference has probability zero), nondecreasing in $\mu_i$, and nonincreasing in $\mu_j$. Because regular horizontal demand implies that the density is bounded below on $\Lambda$,
\begin{equation}\label{eq:positive-limit-demand}
\mu_j-\mu_i<\bar\lambda
\implies
p_i(\mu)>0.
\end{equation}
Moreover, because $\lambda_t$ is independent of $(\theta,\mathscr H_t)$, in every environment and every state $\vartheta\in\Theta$,
\begin{equation}\label{eq:choice-prob}
\Pr^\vartheta(i_t=i\mid\mathscr H_t)=p_i(\mu_t)
\qquad\text{almost surely.}
\end{equation}
Recall that $N_i(t)$ is the number of times action $i$ has been chosen before date $t$ and that $(\mathscr H_t)_{t\in\mathcal T}$ is the public filtration. For a state $\vartheta\in\Theta$, $\Pr^\vartheta$ and $\Ex^\vartheta$ denote probability and expectation conditional on $\theta=\vartheta$, and we write $\Pr^{11}$ and $\Ex^{11}$ when $\vartheta=(1,1)$. Subscripts on $\Pr$ index components of the learning environment (such as the experiment $\phi_n$, the policy $\sigma$, or a horizon $T$). Along a sequence of experiments $(\phi_n)_n$, we write $\KL_n:=\KL[\phi_n]$, $\V_n:=\V[\phi_n]$, and $V_n:=V(\sigma_{\mathsf{TR}},\phi_n,\mu_0,\nu)$. We use the conditional Borel--Cantelli lemma \cite[Theorem 4.3.4]{durrett2019probability} repeatedly: if $A_t\in\mathscr H_{t+1}$ for each $t$, then $\{A_t\text{ infinitely often}\}=\{\sum_t\Pr(A_t\mid\mathscr H_t)=\infty\}$ almost surely.

In this appendix, unless stated otherwise, we assume that the prior $\mu_0\in(0,1)^2$ is independent across actions but need not satisfy $\max_i\mu_{i0}<1-\bar\lambda$, and the experiment $\phi$ need only satisfy $\phi^1\neq\phi^0$. This generality is needed because some results are applied to continuation environments that start from interim beliefs. For any such experiment $\phi$, let $\rho:=\phi^0+\phi^1$, $g_\vartheta:=d\phi^\vartheta/d\rho$, and $Z:=\log(g_1/g_0)$. Under $\phi^1$, the log-likelihood ratio $Z$ takes values in $(-\infty,+\infty]$, and it equals $\log\frac{d\phi^1}{d\phi^0}$ when $\phi^1\ll\phi^0$. Because $\Ex_{\phi^1}[e^{-Z}]=\phi^0(\{g_1>0\})\leq1$, the negative part of $Z$ is $\phi^1$-integrable, and Jensen's inequality gives
\begin{equation}\label{eq:positive-kl}
\Ex_{\phi^1}[Z]\geq-\log\Ex_{\phi^1}[e^{-Z}]\geq0,
\qquad\text{so that}\qquad
\Ex_{\phi^1}[Z]\in(0,+\infty].
\end{equation}
Positivity is strict because $\phi^1\neq\phi^0$.\footnote{If both inequalities held with equality, strict convexity of $-\log$ would give $Z=0$ $\phi^1$-almost surely and hence $g_0=g_1$ $\rho$-almost everywhere on $\{g_1>0\}$, and the second equality would give $\phi^0(\{g_1>0\})=1$. Together, these would imply $\phi^0=\phi^1$.}

\subsection{Sampling, learning, and efficiency}\label{app:prelim-sampling}
Our first lemma says that, when both actions are high value, abandoning an action rules out efficiency.

\begin{lemma}\label{lem:abandon}
In every environment,
\[
E\subseteq\{N_1(t)\to\infty\}\cap\{N_2(t)\to\infty\}
\qquad\Pr^{11}\text{-almost surely.}
\]
\end{lemma}

\begin{proof}[Proof of \cref{lem:abandon}]
For each ordered pair of distinct actions $(i,j)$, let
$\Lambda_{ij}:=\{\lambda\in\Lambda:\lambda_j-\lambda_i\geq\bar\lambda/2\}$.
This has Lebesgue measure $\bar\lambda^2/8$, so regular horizontal demand gives
$\nu(\Lambda_{ij})\geq\underline f\,\bar\lambda^2/8>0$. The types are i.i.d.\ and independent of $\theta$, so the second Borel--Cantelli lemma gives $\Pr^{11}(\lambda_t\in \Lambda_{ij}\text{ infinitely often})=1$ for both ordered pairs. Fix $(i,j)$. On $\{N_j(t)\not\to\infty\}$, action $i$ is chosen at every sufficiently late date. At each such date with $\lambda_t\in \Lambda_{ij}$, regret in state $(1,1)$ equals
\[
\max_{k\in\mathcal I}U(k,(1,1),\lambda_t)-U(i,(1,1),\lambda_t)
=\lambda_{jt}-\lambda_{it}\geq\frac{\bar\lambda}{2}.
\]
Hence, regret is at least $\bar\lambda/2$ infinitely often and $E$ fails.
\end{proof}

\begin{definition}[Sampled-action revelation]\label{def:sar}
An environment satisfies \emph{sampled-action revelation} if, for each action $i$,
\[
 \{N_i(t)\to\infty\}\subseteq\{\mu_{it}\to\theta_i\}
 \qquad\text{almost surely.}
\]
\end{definition}

Under an adaptive evidence policy, say that the record $x_t$ is \emph{eventually disclosed} if it is an entry of the public message $y_{t'}$ for some $t'\geq t$.

\begin{lemma}\label{lem:sar}
Let $\sigma\in\Sigma_{\mathsf{EV}}$ and fix an action $i$. If every record about action $i$ is eventually disclosed almost surely, then $\{N_i(t)\to\infty\}\subseteq\{\mu_{it}\to\theta_i\}$ almost surely. Consequently, adaptive evidence policies under which every record is eventually disclosed, such as full transparency and the accumulation policies of \cref{app:prelim-accumulation}, satisfy sampled-action revelation.
\end{lemma}

\begin{proof}[Proof of \cref{lem:sar}]
Choose a Borel set $A$ with $r_0:=\phi^0(A)\ne r_1:=\phi^1(A)$, such that the value of the action is identified. Applying the strong law of large numbers to the signal stream gives,
\[
 \frac1k\sum_{m=1}^k\mathbf 1\{s_{im}\in A\}\to r_{\theta_i}
 \quad\text{almost surely}.
\]

Define $B_i:=\{N_i(t)\to\infty\}\in\mathscr H_\infty$. By assumption, the sequence of eventually disclosed action-$i$ signals, ordered by date, is $\mathscr H_\infty$-measurable. On the event $B_i$, this sequence is the entire stream $(s_{im})_{m\geq1}$, so the limit $F_i$ of its empirical frequencies of $A$ exists and equals $r_{\theta_i}$ almost surely.\footnote{Off the event $B_i$, define $F_i:=0$ whenever the disclosed sequence is finite or its empirical-frequency limit does not exist.} Since $r_0\ne r_1$, $\mathbf 1_{B_i}\theta_i=\mathbf 1_{B_i}\mathbf 1\{F_i=r_1\}$ almost surely, which is an $\mathscr H_\infty$-measurable random variable. By L\'evy's upward theorem, the posterior martingale satisfies
\[
 \mu_{it}=\Ex[\theta_i\mid\mathscr H_t]
 \to\Ex[\theta_i\mid\mathscr H_\infty]
 \quad\text{almost surely},
\]
and hence
\begin{align*}
    \mathbf 1_{B_i}\lim_{t\to\infty}\mu_{it}
 =\Ex[\mathbf 1_{B_i}\theta_i\mid\mathscr H_\infty]
 \\
 =\Ex[\mathbf 1_{B_i}\mathbf 1\{F_i=r_1\}\mid\mathscr H_\infty]=\mathbf 1_{B_i}\mathbf 1\{F_i=r_1\}=\mathbf 1_{B_i}\theta_i
 \quad\text{almost surely}.
\end{align*}
Under full transparency, $x_t$ is an entry of $y_t=D_t$. Under $\sigma^{\mathsf{ACC},T}$, it is an entry of $y_{\max\{t,T\}}$.
\end{proof}

The next result generalizes \cref{lem:sources} to every environment with sampled-action revelation.

\begin{lemma}[Sources of inefficiency, general form]\label{lem:sources-general}
For any environment that satisfies sampled-action revelation,
\begin{itemize}
\item[(i)] $\Pr^\vartheta(E)=1$ for every state $\vartheta\neq(1,1)$.
\item[(ii)] $E=\{N_1(t)\to\infty\}\cap\{N_2(t)\to\infty\}=\{\mu_t\to(1,1)\}$ $\Pr^{11}$-almost surely. Moreover, in every state and for each action $i$, $\{p_i(\mu_\infty)>0\}\subseteq\{N_i(t)\to\infty\}$ almost surely, where $\mu_\infty:=\lim_t\mu_t$.
\item[(iii)] $V=1-\mu_{10}\mu_{20}\Pr^{11}(E^c)$.
\end{itemize}
\end{lemma}

\begin{proof}[Proof of \cref{lem:sources-general}]
For each $i$, $\mu_{it}=\Ex[\theta_i\mid\mathscr H_t]$ is a bounded martingale, so $\mu_{it}\to\mu_{i\infty}$ almost surely for some random variable $\mu_{i\infty}\in[0,1]$ by the martingale convergence theorem. Because every state has strictly positive prior probability, the same convergence holds $\Pr^\vartheta$-almost surely for each $\vartheta\in\Theta$.

Let $A_t^i:=\{i_t=i\}\in\mathscr H_{t+1}$. By \eqref{eq:choice-prob} and continuity of $p_i$, on $\{p_i(\mu_\infty)>0\}$,
\[
\sum_{t=0}^\infty
\Pr^\vartheta(A_t^i\mid\mathscr H_t)
=
\sum_{t=0}^\infty p_i(\mu_t)
=
\infty.
\]
The conditional Borel--Cantelli lemma therefore yields
\begin{equation}\label{eq:positive-demand-infinite-sampling}
\{p_i(\mu_\infty)>0\}
\subseteq
\{N_i(t)\to\infty\}
\qquad \Pr^\vartheta\text{-a.s.},
\end{equation}
which is the second claim in (ii).

We will also use the following fact. If
$\mu_t\to\vartheta$, let
$i_t^*\in\arg\max_i\{\lambda_{it}+\vartheta_i\}$. Because $i_t$
maximizes $\lambda_{it}+\mu_{it}$,
\begin{align*}
0
&\leq
U(i_t^*,\vartheta,\lambda_t)-U(i_t,\vartheta,\lambda_t)
\\
&\leq
|\vartheta_{i_t^*}-\mu_{i_t^*,t}|
+|\vartheta_{i_t}-\mu_{i_t,t}|
\\
&\leq
2\lVert\mu_t-\vartheta\rVert_\infty
\to 0.
\end{align*}
Thus, convergence of beliefs to the true state implies $E$.

\medskip
\noindent\emph{State $(0,0)$.}
Because $N_1(t)+N_2(t)=t$, at least one action is sampled infinitely
often on every path. On the event that action $i$ is sampled infinitely
often, sampled-action revelation gives $\mu_{i\infty}=0$. For $j\neq i$,
\[
\mu_{i\infty}-\mu_{j\infty}\leq0<\bar\lambda,
\]
so \eqref{eq:positive-limit-demand} and
\eqref{eq:positive-demand-infinite-sampling} imply that action $j$ is also
sampled infinitely often. Sampled-action revelation then gives
$\mu_t\to(0,0)$, and hence $E$, $\Pr^{00}$-almost surely. This gives (i) for the case $\vartheta=(0,0)$.

\medskip
\noindent\emph{Mixed states.}
Consider $\vartheta=(1,0)$ (the other mixed state is symmetric). On
$\{N_1(t)\not\to\infty\}$, action $2$ is sampled infinitely often, so
$\mu_{2\infty}=0$. Since
$\mu_{2\infty}-\mu_{1\infty}\leq0<\bar\lambda$,
\eqref{eq:positive-limit-demand}--\eqref{eq:positive-demand-infinite-sampling} imply
$N_1(t)\to\infty$, a contradiction. Thus $N_1(t)\to\infty$ and
$\mu_{1t}\to1$ almost surely. Sampled-action revelation then implies
\[
\{N_2(t)\to\infty\}
\subseteq
\{\mu_\infty=(1,0)\}
\qquad \Pr^\vartheta\text{-a.s.}
\]
On $\{\mu_\infty=(1,0)\}$, we have
$\mu_{1t}-\mu_{2t}\to1>\bar\lambda$, so
$\mu_{1t}-\mu_{2t}>\bar\lambda$ for all sufficiently large $t$.
Whenever this inequality holds, every $\lambda\in\Lambda$ satisfies
\[
(\lambda_1+\mu_{1t})-(\lambda_2+\mu_{2t})
\geq
\mu_{1t}-\mu_{2t}-\bar\lambda
>
0,
\]
and hence $p_2(\mu_t)=0$. Consequently,
\[
\{N_2(t)\to\infty\}
\subseteq
\left\{\sum_{t=0}^{\infty}p_2(\mu_t)<\infty\right\}
\qquad \Pr^\vartheta\text{-a.s.}
\]
On the other hand, the conditional Borel--Cantelli lemma and
\eqref{eq:choice-prob} give
\[
\{N_2(t)\to\infty\}
=
\{A_t^2\text{ infinitely often}\}
=
\left\{\sum_{t=0}^{\infty}p_2(\mu_t)=\infty\right\}
\qquad \Pr^\vartheta\text{-a.s.}
\]
Together, these imply
$\Pr^\vartheta(N_2(t)\to\infty)=0$, so action $2$ is chosen only
finitely often, $\Pr^\vartheta$-almost surely. Since, for each $\lambda\in\Lambda$,
\[
U(1,(1,0),\lambda)-U(2,(1,0),\lambda)
=
1+\lambda_1-\lambda_2
\geq
1-\bar\lambda
>
0,
\]
regret is eventually zero. This completes the proof of (i).

\medskip
\noindent\emph{State $(1,1)$.}
If both actions are sampled infinitely often, sampled-action revelation
gives $\mu_t\to(1,1)$, which implies $E$ by the bound above. Conversely, $E$ implies that both actions are sampled infinitely often by \cref{lem:abandon}. Finally, if $\mu_t\to(1,1)$, then $p_i(\mu_\infty)=p_i(1,1)>0$ for both $i$,
and \eqref{eq:positive-demand-infinite-sampling} implies that both actions
are sampled infinitely often. This proves (ii).

Because the prior is independent across actions and $E$ can fail only
in state $(1,1)$, $V=1-\Pr(\theta=(1,1))\Pr^{11}(E^c)=1-\mu_{10}\mu_{20}\Pr^{11}(E^c)$, which gives (iii).
\end{proof}

Several arguments stop the process at a random date and then use signals that have not yet been generated by agents. To handle this, we adapt the equivalence between the reward-stack and canonical models observed in the bandit literature \citep[Section~4.6]{lattimore2020bandit}. Formally, define an i.i.d.\ sequence $(u_t)_{t\in\mathcal T} \sim \mathrm{Unif}[0,1]$ that is independent of the state, types, and signal streams, where $u_t$ is used at date $t$ to specify the randomization of the policy (and of tie-breaking if necessary). Let $\upsilon_t:=(\lambda_t,u_t)$, with law $\bar\nu:=\nu\otimes\mathrm{Unif}[0,1]$. Let $\mathscr F_t$ be the $\sigma$-algebra generated by the draws $\upsilon_s$, the actions, and the signals of all dates $s<t$, so that $\mathscr H_t\subseteq\mathscr F_t$ and $N_1(t),N_2(t)$ are $\mathscr F_t$-measurable. 

The following lemma establishes that, conditional on the true state, stopping at a date chosen using past information does not change the joint distribution of the signals that have not yet been generated or future type and randomization draws.

\begin{lemma}\label{lem:fresh}
Fix a state $\vartheta$ and an $(\mathscr F_t)$-stopping time $\tau$.
\begin{itemize}
    \item[(i)] Under $\Pr^\vartheta$ and on the event $\{\tau<\infty\}$, the unused signal streams $(s_{1,N_1(\tau)+k})_{k\geq1}$ and $(s_{2,N_2(\tau)+k})_{k\geq1}$ and the future draws $(\upsilon_{\tau+k})_{k\geq0}$ are independent of $\mathscr F_\tau$ and of one another. The first two are i.i.d.\ with laws $\phi^{\vartheta_1}$ and $\phi^{\vartheta_2}$, and the third is i.i.d.\ with law $\bar\nu$. 
    \item[(ii)] Let $I$ be an $\mathscr F_\tau$-measurable label and let $J$ denote the other action. Conditional on $\mathscr F_\tau$, the unused streams of actions $I$ and $J$ and the future draws are mutually independent and i.i.d.\ with laws $\phi^{\vartheta_I}$, $\phi^{\vartheta_J}$, and $\bar\nu$. When $\vartheta_1=\vartheta_2$, they are in addition independent of $\mathscr F_\tau$.
\end{itemize}
\end{lemma}

\begin{proof}[Proof of \cref{lem:fresh}]
We apply the the strong Markov property of an i.i.d.\ sequence at a stopping time
\citep[Theorem~5.2.5]{durrett2019probability}. For stream $i$, let
$\mathscr G^i_k:=\sigma(s_{i,1},\ldots,s_{i,k})\vee\sigma\bigl((s_{j,k})_{k\ge1},(\upsilon_t)_{t\in\mathcal T}\bigr)$.
Because each action is selected before its next signal is observed, the history
through date $\tau$ is a function of the first $N_i(\tau)$ signals of stream $i$,
of stream $j$, and of the draws, so $N_i(\tau)$ is a $(\mathscr G^i_k)$-stopping time,
and $s_{i,k+1}$ is independent of $\mathscr G^i_k$ because the three components are
independent under $\Pr^\vartheta$. Theorem 5.2.5 of \cite{durrett2019probability} yields that $(s_{i,N_i(\tau)+k})_{k\ge1}$ is i.i.d.\ $\phi^{\vartheta_i}$
and independent of $\mathscr G^i_{N_i(\tau)}\supseteq\mathscr F_\tau$. Applying this to
$i=1,2$ and the same argument to the i.i.d.\ draws $(\upsilon_{\tau+k})_{k\ge0}$ gives part (i). Part (ii) follows by restricting to $\{I=1\}$ and $\{I=2\}$,
which lie in $\mathscr F_\tau$.
\end{proof}

Because $\mathscr H_\tau\subseteq\mathscr F_\tau$ when $\tau$ is an $(\mathscr H_t)$-stopping time, the conclusion of \cref{lem:fresh} also holds conditional on the public history $\mathscr H_\tau$. We also use the following unconditional version. Write $\Upsilon_\tau$ for the triple formed by the two unused streams and the future draws, and $Q^\vartheta:=(\phi^{\vartheta_1})^{\otimes\mathbb N}\otimes(\phi^{\vartheta_2})^{\otimes\mathbb N}\otimes\bar\nu^{\otimes\mathbb N}$ for its law under $\Pr^\vartheta$. For every $(\mathscr H_t)$-stopping time $\tau$, every measurable set $A$, and every state $\vartheta$, on $\{\tau<\infty\}$,
\begin{equation}\label{eq:fresh-unconditional}
\Pr\bigl(\theta=\vartheta,\ \Upsilon_\tau\in A\mid\mathscr H_\tau\bigr)
=\Ex\bigl[\mathbf 1\{\theta=\vartheta\}\,Q^\vartheta(A)\mid\mathscr H_\tau\bigr]
=\pi_\tau(\vartheta)\,Q^\vartheta(A),
\end{equation}
where the first equality conditions on $(\theta,\mathscr F_\tau)$ and uses \cref{lem:fresh}. Thus, conditional on $\mathscr H_\tau$, the continuation joint distribution over states, unused signals, types, and randomization draws is governed by the public posterior $\pi_\tau$ and by iid draws from $\phi,\bar\nu$.

\subsection{Posterior bounds}\label{app:prelim-trap}

Let $T_0$ be an almost surely finite $(\mathscr H_t)$-stopping time. Say that a policy is \emph{transparent from $T_0$ on} if the public posterior $\pi_{T_0}$ is a product measure and, at every date $t\geq T_0$, the public message discloses the record $x_t$ (and no new information besides this). Full transparency is transparent from $T_0=0$ on. By \cref{lem:fresh}, conditional on $(\theta,\mathscr H_t)$ and on $i_t=i$, the signal $s_t$ has law $\phi^{\theta_i}$, which does not depend on the other action's value. Hence, under a policy that is transparent from $T_0$ on, the public posterior is a product measure at every date $t\geq T_0$ and evolves by one-dimensional Bayesian updating: the belief about the chosen action updates from $m$ to the posterior $m'(s_t)$ after one signal, and the belief about the other action does not move.

\begin{lemma}[Posterior downcrossings]
\label{lem:nonsmooth-posterior}
Suppose the policy is transparent from $T_0$ on, fix an action $i$, and fix a state $\vartheta$ with $\vartheta_i=1$.
\begin{itemize}
\item[(i)] Under $\Pr^\vartheta$, the posterior $(\mu_{it})_{t\geq T_0}$ is an $(\mathscr H_t)$-submartingale, so $1-\mu_{it}$ is a nonnegative supermartingale.
\item[(ii)] For every almost surely finite $(\mathscr H_t)$-stopping time $T\geq T_0$ and every $\mathscr H_T$-measurable random variable $\alpha$ with values in $[0,1)$,
\[
\Pr^\vartheta\bigl(\mu_{it}\leq \alpha\text{ for some }t\geq T\mid\mathscr H_T\bigr)
\leq\frac{1-\mu_{iT}}{1-\alpha}
\qquad\text{almost surely.}
\]
\item[(iii)] Let $\mathscr G$ be a $\sigma$-algebra, let $m_0\in(0,1]$ and $\alpha\in[0,1)$ be $\mathscr G$-measurable random variables, and let $(s_k)_{k\geq1}$ be i.i.d.\ with law $\phi^1$ and independent of $\mathscr G$. For $k\geq1$, let $m_k$ be the posterior obtained from $m_{k-1}$ by Bayes' rule after observing $s_k$, with $m_k:=1$ if $m_{k-1}=1$. Then
\[
\Pr\bigl(m_k\leq \alpha\text{ for some }k\geq0\mid\mathscr G\bigr)
\leq\frac{1-m_0}{1-\alpha}
\qquad\text{almost surely.}
\]
\end{itemize}
\end{lemma}

\begin{proof}[Proof of \cref{lem:nonsmooth-posterior}]
(i) Fix a current posterior $m\in(0,1)$ for $\theta_i=1$. Conditional on
sampling action $i$, let
\[
P_m:=m\phi^1+(1-m)\phi^0
\]
be the predictive signal law and let $m'(s)$ be the posterior after signal
$s$. Bayes' rule gives
\[
\frac{d\phi^1}{dP_m}(s)=\frac{m'(s)}{m}.
\]
Therefore,
\[
\Ex_{\phi^1}[m']
=
\frac1m\Ex_{P_m}[(m')^2]
\geq
\frac1m\bigl(\Ex_{P_m}[m']\bigr)^2
=
m.
\]
The inequality is Jensen's inequality, and the last equality uses the
martingale property of beliefs under the predictive law. If $m=1$, the posterior remains equal to one almost surely. If action $i$ is not sampled, its posterior does not change. Note that this calculation does not use $\phi^1\ll\phi^0$.

(ii) By (i) and optional sampling \citep[Exercise~4.8.1]{durrett2019probability}, $(1-\mu_{i,T+k})_{k\geq0}$
is a bounded nonnegative supermartingale under $\Pr^\vartheta$
with respect to $(\mathscr H_{T+k})_{k\geq0}$.
The conditional version of Ville's maximal inequality
\citep[Lemma~1]{howard2020time},\footnote{For a nonnegative
$(\mathscr K_k)$-supermartingale $(Y_k)_{k\geq0}$ and a strictly
positive, finite, $\mathscr K_0$-measurable random variable $c$, define $d:=\max\{c,Y_0\}$. The process $L_k:=Y_k/d$ is a
nonnegative supermartingale, with
$\Ex[L_k]\leq\Ex[Y_0/d]\leq1$. The events
$\{\exists k\geq0:Y_k\geq c\}$ and
$\{\exists k\geq0:L_k\geq1\}$ coincide: on $\{Y_0<c\}$,
$d=c$, while on $\{Y_0\geq c\}$ both events occur at time zero.
The inequality at level $1$ therefore gives
$\Pr(\exists k\geq0:Y_k\geq c\mid\mathscr K_0)
\leq Y_0/d\leq Y_0/c$.} with initial information
$\mathscr H_T$ and the strictly positive
$\mathscr H_T$-measurable threshold $1-\alpha$, therefore gives
\[
\Pr^\vartheta\bigl(
\mu_{it}\leq\alpha\text{ for some }t\geq T
\mid\mathscr H_T
\bigr)
\leq
\frac{1-\mu_{iT}}{1-\alpha}
\qquad\text{almost surely}.
\]

(iii) Let
$\mathscr G_k:=\mathscr G\vee\sigma(s_1,\ldots,s_k)$.
Since $s_k$ has law $\phi^1$ and is independent of
$\mathscr G_{k-1}$, the calculation in (i) gives
$\Ex[m_k\mid\mathscr G_{k-1}]\geq m_{k-1}$.
Thus $(1-m_k)_{k\geq0}$ is a bounded nonnegative
$(\mathscr G_k)$-supermartingale. Applying the same conditional
maximal inequality with initial information $\mathscr G_0=\mathscr G$
and threshold $1-\alpha$ yields
\[
\Pr\bigl(
m_k\leq\alpha\text{ for some }k\geq0
\mid\mathscr G
\bigr)
\leq
\frac{1-m_0}{1-\alpha}
\qquad\text{almost surely}.
\]
\end{proof}

\begin{lemma}\label{lem:trap}
Suppose the policy is transparent from $T_0$ on. Let $T\geq T_0$ be an almost surely finite $(\mathscr H_t)$-stopping time, let $i\neq j$, and let $F\in\mathscr H_T$ be an event on which $\mu_{iT}-\mu_{jT}>\bar\lambda$. Then, $\Pr^{11}$-almost surely on $F$,
\[
\Pr^{11}\bigl(\text{action $j$ is not chosen at any date }t\geq T\mid\mathscr H_T\bigr)
\geq
\frac{\mu_{iT}-\mu_{jT}-\bar\lambda}{1-\mu_{jT}-\bar\lambda}
\geq
\mu_{iT}-\mu_{jT}-\bar\lambda.
\]
Consequently,
$\Pr^{11}(E^c)\geq\Ex^{11}\bigl[\mathbf 1_F\,(\mu_{iT}-\mu_{jT}-\bar\lambda)\bigr]$.
\end{lemma}

\begin{proof}[Proof of \cref{lem:trap}]
Define $\alpha:=\mu_{jT}+\bar\lambda$ on $F$ and $\alpha:=0$ on $F^c$. Then $\alpha$ is $\mathscr H_T$-measurable and, on $F$, $\bar\lambda\leq \alpha<\mu_{iT}\leq1$. Let $\tau_\alpha:=\inf\{t\geq T:\mu_{it}\leq \alpha\}$. By \cref{lem:nonsmooth-posterior}(ii), on $F$,
\[
\Pr^{11}(\tau_\alpha=\infty\mid\mathscr H_T)
\geq1-\frac{1-\mu_{iT}}{1-\alpha}
=\frac{\mu_{iT}-\alpha}{1-\alpha}
\geq\mu_{iT}-\alpha,
\]
where the last inequality uses $0<1-\alpha\leq1$. On $F\cap\{\tau_\alpha=\infty\}$, we show by induction on $t\geq T$ that action $j$ is not chosen at $t$ and $\mu_{j,t+1}=\mu_{jT}$. If $\mu_{jt}=\mu_{jT}$, then for every type $\lambda\in\Lambda$,
\[
\lambda_j+\mu_{jt}\leq\bar\lambda+\mu_{jT}=\alpha<\mu_{it}\leq\lambda_i+\mu_{it},
\]
so every type strictly prefers action $i$ and tie-breaking is irrelevant; because the policy is transparent, a signal about action $i$ leaves $\mu_j$ unchanged. This proves the displayed bound. On the event that action $j$ is not chosen after $T$, $N_j(t)\not\to\infty$, so $E$ fails $\Pr^{11}$-almost surely by \cref{lem:abandon}. Taking expectations over $F$ gives the last claim.
\end{proof}

The next lemma gives a lower bound on the probability that a run of favorable signals about one action induces herding on that action.

\begin{lemma}\label{lem:escape-prior}
Consider full transparency with a prior satisfying $\max_i\mu_{i0}<1-\bar\lambda$, and suppose $\phi^1(\{Z\geq z\})\geq q$ for some $z,q>0$. Let $I$ be an action with $\gamma:=p_I(\mu_0)>0$, let $J\neq I$, choose $\varepsilon>0$ with $\mu_{J0}+\bar\lambda+2\varepsilon<1$, and let $m\geq 0$ be an integer with
\[
\ell\bigl(\mathsf L(\mu_{I0})+mz\bigr)
>
\mu_{J0}+\bar\lambda+2\varepsilon.
\]
Then $\Pr^{11}(E^c)\geq2\varepsilon(\gamma q)^m$.
\end{lemma}

\begin{proof}[Proof of \cref{lem:escape-prior}]
Let $A_m\in\mathscr H_m$ be the event that the first $m$ agents choose action $I$ and all
$m$ resulting signals satisfy $Z\geq z$. Along this event, action $I$'s
belief weakly increases and action $J$'s belief remains at $\mu_{J0}$, so, by \eqref{eq:choice-prob} and monotonicity of $p_I$,
the conditional probability of choosing $I$ is at least $\gamma$ at each of these dates. By \cref{lem:fresh}, each resulting signal has law $\phi^1$ under $\Pr^{11}$ and is independent of the past. Thus
$\Pr^{11}(A_m)\geq(\gamma q)^m$. On $A_m$, $\mu_{Jm}=\mu_{J0}$ and
$\mu_{Im}\geq\ell(\mathsf L(\mu_{I0})+mz)>\mu_{J0}+\bar\lambda+2\varepsilon$. Applying \cref{lem:trap} with $T=m$ and $F=A_m$ gives
$\Pr^{11}(E^c)\geq2\varepsilon\Pr^{11}(A_m)\geq2\varepsilon(\gamma q)^m$.
\end{proof}

We may now prove \cref{lem:sources}.

\begin{proof}[Proof of \cref{lem:sources}]
Full transparency satisfies sampled-action revelation by \cref{lem:sar}. Part (i) of \cref{lem:sources-general} therefore gives $\Pr^\vartheta(E)=1$ for every $\vartheta\neq(1,1)$, and part (iii) gives the formula for $V$. It remains to show $\Pr^{11}(E^c)>0$. By \eqref{eq:positive-kl}, there are
$z,q>0$ with $\phi^1(\{Z\geq z\})\geq q$. Let $I$ be an action with $\mu_{I0}=\max_i\mu_{i0}$, so that $p_I(\mu_0)\geq\nu(\{\lambda:\lambda_I\geq\lambda_J\})>0$. Because $\mu_{J0}<1-\bar\lambda$, there are $\varepsilon>0$ and an integer $m$ as in \cref{lem:escape-prior}, which gives $\Pr^{11}(E^c)\geq2\varepsilon(\gamma q)^m>0$. Note that this argument uses only $\phi^1\neq\phi^0$ and $\max_i\mu_{i0}<1-\bar\lambda$, so the lemma also covers experiments with unbounded likelihood ratios, such as the Gaussian experiment in \cref{fig:sources}.
\end{proof}

\begin{lemma}\label{lem:hitting}
Fix any environment, let $a:=\mu_{10}$, let $0<c<a$, and let $\tau:=\inf\{t\in\mathcal T:\mu_{1t}\leq c\}$.
\begin{itemize}
\item[(i)] $\displaystyle\Pr(\tau<\infty\mid\theta_1=1)\leq\frac{c(1-a)}{a(1-c)}$.
\item[(ii)] If $\mu_{1\tau}\geq c'$ on $\{\tau<\infty\}$ for some $c'\in(0,c]$ and $\Pr(\tau<\infty\mid\theta_1=0)=1$, then $\displaystyle\Pr(\tau<\infty\mid\theta_1=1)\geq\frac{c'(1-a)}{a(1-c')}$.
\end{itemize}
\end{lemma}

\begin{proof}[Proof of \cref{lem:hitting}]
Because $\{\tau=t\}\in\mathscr H_t$ and $\mu_{1t}=\Pr(\theta_1=1\mid\mathscr H_t)$,
\[
\Pr(\theta_1=1,\tau=t)=\Ex\bigl[\mu_{1t}\mathbf 1_{\{\tau=t\}}\bigr]
\qquad\text{and}\qquad
\Pr(\theta_1=0,\tau=t)=\Ex\bigl[(1-\mu_{1t})\mathbf 1_{\{\tau=t\}}\bigr].
\]
On $\{\tau=t\}$, $\mu_{1t}\leq c$, which is equivalent to $\mu_{1t}\leq\frac{c}{1-c}(1-\mu_{1t})$. Summing over $t$ gives
\[
\Pr(\theta_1=1,\tau<\infty)
\leq\frac{c}{1-c}\Pr(\theta_1=0,\tau<\infty)
\leq\frac{c}{1-c}(1-a),
\]
and dividing by $\Pr(\theta_1=1)=a$ gives (i). For (ii), $\mu_{1t}\geq c'$ on $\{\tau=t\}$ is equivalent to $\mu_{1t}\geq\frac{c'}{1-c'}(1-\mu_{1t})$, so the same summation gives
$\Pr(\theta_1=1,\tau<\infty)\geq\frac{c'}{1-c'}\Pr(\theta_1=0,\tau<\infty)=\frac{c'}{1-c'}(1-a)$.
\end{proof}

Part (i) says that, conditional on $\theta_1=1$, the posterior odds in favor of $\theta_1=1$ rarely fall.\footnote{This also follows by applying Ville's maximal inequality for the inverse posterior odds $(1-\mu_{1t})/\mu_{1t}$, which form a nonnegative supermartingale conditional on $\theta_1=1$ under every information policy. We provide a direct argument which we use to prove both parts.}

\subsection{Accumulation}\label{app:prelim-accumulation}

For an integer $T\geq0$, the \emph{accumulation policy} $\sigma^{\mathsf{ACC},T}$ is defined by its cumulative
evidence stock:
\[
R_t^{\mathsf{ACC},T}
=
\begin{cases}
\varnothing, & t<T,\\
D_t, & t\geq T.
\end{cases}
\]
The policy is silent through date $T-1$, reveals the entire
stock $D_T$ at the end of date $T$, and thereafter reveals each record. It
is an element of $\Sigma_{\mathsf{EV}}$: disclosure is nested, no record is garbled or retracted, and the silence is deterministic and therefore uninformative.\footnote{The definition of $\Sigma_{\mathsf{EV}}$ requires nested disclosure at every evidence-consistent pair $(D_t,H_t)$, including pairs that the policy itself never reaches. At such a pair, let the message consist of the current stock $R(H_t)$ together with the records that the rule above discloses. This convention does not affect the law of the environment, and we adopt it for every evidence policy constructed below.}

\begin{lemma}[Accumulation]\label{lem:accumulation}
Fix an experiment with $\phi^1\neq\phi^0$, and let $\mathcal K$ be a compact subset of
$\{\mu\in(0,1)^2:|\mu_1-\mu_2|<\bar\lambda\}$.
\begin{itemize}
\item[(i)] $\displaystyle\lim_{T\to\infty}\ \sup_{\mu\in\mathcal K}\bigl[1-V(\sigma^{\mathsf{ACC},T},\phi,\mu,\nu)\bigr]=0$.
\item[(ii)] Let $\sigma$ be any information policy, $\tau$ an $(\mathscr H_t)$-stopping time, and $F\in\mathscr H_\tau$ an event with $F\subseteq\{\tau<\infty\}$ such that, on $F$, the public posterior $\pi_\tau$ is a product measure with marginals $\mu_\tau\in\mathcal K$ and, from date $\tau$ on, $\sigma$ applies the accumulation policy with horizon $T$ to the records generated from date $\tau$ on.\footnote{That is, beyond what $H_\tau$ already contains, its messages reveal nothing through date $\tau+T-1$, the records $x_\tau,\ldots,x_{\tau+T}$ at date $\tau+T$, and each record thereafter, and they do not depend on anything else the designer has observed.} Then, almost surely on $F$,
\[
\Pr(E^c\mid\mathscr H_\tau)
=1-V(\sigma^{\mathsf{ACC},T},\phi,\mu_\tau,\nu)
\leq\sup_{\mu\in\mathcal K}\bigl[1-V(\sigma^{\mathsf{ACC},T},\phi,\mu,\nu)\bigr].
\]
\end{itemize}
\end{lemma}

\begin{proof}[Proof of \cref{lem:accumulation}]
\underline{Part (i)}: Fix $\mu\in\mathcal K$ and write $\Pr_{T,\mu}$ for the law
of the environment
$(\sigma^{\mathsf{ACC},T},\phi,\mu,\nu)$.
Every record is eventually disclosed, so the environment satisfies
sampled-action revelation by \cref{lem:sar}. By
\cref{lem:sources-general}(iii),
\[
1-V(\sigma^{\mathsf{ACC},T},\phi,\mu,\nu)
=
\mu_1\mu_2\Pr^{11}_{T,\mu}(E^c)
\leq
\Pr^{11}_{T,\mu}(E^c).
\]
It therefore suffices to bound
$\Pr^{11}_{T,\mu}(E^c)$ uniformly over $\mu\in\mathcal K$.

First, note that before the date-$T$ disclosure, the null messages are deterministic. Since the publicly observed actions convey no information about
$\theta$, public beliefs remain equal to $\mu$ through date $T$.
The actions chosen at dates $0,\ldots,T$ are therefore i.i.d.\ with
choice probabilities $(p_1(\mu),p_2(\mu))$, and are independent of
$\theta$ and of the signal streams. By
\eqref{eq:positive-limit-demand}, continuity of $p_i$, and compactness
of $\mathcal K$,
\[
q_*:=\min_{\mu'\in\mathcal K}\min_{i\in\mathcal I}p_i(\mu')>0.
\]
Let $N:=T+1$ and $Q_i:=\sum_{t=0}^{T}\mathbf 1\{i_t=i\}$. For each $i$, the summands in $Q_i$ are independent,
take values in $[0,1]$, and have common mean $p_i(\mu)$ under
$\Pr^{11}_{T,\mu}$. Hoeffding's inequality therefore gives, for
every $u>0$,
\[
\Pr^{11}_{T,\mu}\left(
\frac{Q_i}{N}-p_i(\mu)\leq-u
\right)
\leq \exp(-2Nu^2).
\]
Since $p_i(\mu)\geq q_*$, applying this inequality with
$u=p_i(\mu)-q_*/2\geq q_*/2$ yields
\[
\Pr^{11}_{T,\mu}\left(Q_i<\frac{q_*N}{2}\right)
\leq
\exp\left[-2N\left(p_i(\mu)-\frac{q_*}{2}\right)^2\right]
\leq
\exp\left(-\frac{q_*^2N}{2}\right).
\]
Define the event that both actions have sufficiently many observations by
\[
C_N:=\left\{\min_{i\in\mathcal I}Q_i\geq\frac{q_*N}{2}\right\}.
\]
The union bound gives
\[
\Pr^{11}_{T,\mu}(C_N^c)
\leq
\sum_{i\in\mathcal I}
\Pr^{11}_{T,\mu}\left(Q_i<\frac{q_*N}{2}\right)
\leq
2\exp\left(-\frac{q_*^2N}{2}\right).
\]
This bound is uniform over $\mu\in\mathcal K$ because $q_*$
is a common lower bound on both choice probabilities.

Next, recall from \eqref{eq:positive-kl} that
$\Ex_{\phi^1}[Z^-]<\infty$, where $Z^-:=\max\{-Z,0\}$, and $\Ex_{\phi^1}[Z]\in(0,+\infty]$.
For $M>0$, define the upper truncation $Z^{(M)}:=Z\wedge M$. Its negative part is $Z^-$ and its positive part is bounded by $M$,
so $Z^{(M)}$ is integrable. Moreover,
\[
\Ex_{\phi^1}[Z^{(M)}]
=
\Ex_{\phi^1}[Z^+\wedge M]-\Ex_{\phi^1}[Z^-]
\uparrow
\Ex_{\phi^1}[Z]
\qquad\text{as }M\to\infty.
\]
We may therefore fix a finite $M>0$ such that
$d_M:=\Ex_{\phi^1}[Z^{(M)}]>0$.
If $(s_k)_{k\geq1}$ is i.i.d.\ with law $\phi^1$ and
$S_n:=\sum_{k=1}^n Z(s_k)$, the strong LLN applied to the
integrable variables $Z^{(M)}(s_k)$ gives
\[
\liminf_{n\to\infty}\frac{S_n}{n}
\geq
\lim_{n\to\infty}
\frac1n\sum_{k=1}^n Z^{(M)}(s_k)
=
d_M>0
\qquad\text{almost surely}.
\]
Thus $S_n\to+\infty$ almost surely, including when
$\Ex_{\phi^1}[Z]=+\infty$ or $Z=+\infty$ with positive probability.

Next, fix $\zeta\in(0,1)$. Compactness of $\mathcal K$ inside
$(0,1)^2$ gives the common finite threshold
\[
A_\zeta
:=
\mathsf L(1-\zeta)
-
\min_{\mu\in\mathcal K}\min_{i\in\mathcal I}\mathsf L(\mu_i)
<\infty.
\]
Let
\[
w_N(\zeta)
:=
\sup_{\substack{n\in\mathbb N\\ n\geq q_*N/2}}
(\phi^1)^{\otimes n}(S_n<A_\zeta).
\]
Because $S_n\to+\infty$ almost surely under the infinite product law,
\[
w_N(\zeta)
\leq
(\phi^1)^{\otimes\mathbb N}
\left(
S_n<A_\zeta
\text{ for some integer }n\geq q_*N/2
\right)
\longrightarrow0.
\]
Indeed, the events on the right decrease to the event that
$S_n<A_\zeta$ infinitely often, which has probability zero.

The action sequence before date $T$ is independent of the signal
streams. Consequently, conditional on the counts $(Q_1,Q_2)$
and on $\theta=(1,1)$, the date-$T$ disclosure contains $Q_i$ i.i.d.\ signals
with law $\phi^1$ about each action $i$. The resulting posterior is therefore a product measure, with marginals
\[
\mu_{i,T+1}
=
\ell\bigl(\mathsf L(\mu_i)+S^i_{Q_i}\bigr),
\qquad
S^i_n:=\sum_{k=1}^nZ(s_{ik}),
\qquad S^i_0:=0.
\]
For every $\mu\in\mathcal K$,
\[
\{\mu_{i,T+1}<1-\zeta\}
=
\left\{
S^i_{Q_i}
<
\mathsf L(1-\zeta)-\mathsf L(\mu_i)
\right\}
\subseteq
\{S^i_{Q_i}<A_\zeta\}.
\]
Using this inclusion and independence of $Q_i$ and the signal
stream under $\Pr^{11}_{T,\mu}$, we obtain
\[
\begin{aligned}
&\Pr^{11}_{T,\mu}
\left(C_N\cap\{\mu_{i,T+1}<1-\zeta\}\right)\\
&\quad\leq
\sum_{\substack{n\in\mathbb N\\q_*N/2\leq n\leq N}}
\Pr^{11}_{T,\mu}(Q_i=n)\,
(\phi^1)^{\otimes n}(S_n<A_\zeta)
\leq w_N(\zeta).
\end{aligned}
\]
Define
\[
B_T:=\left\{\min_{i\in\mathcal I}\mu_{i,T+1}\geq1-\zeta\right\}.
\]
Since
\[
B_T^c
\subseteq
C_N^c
\cup
\bigcup_{i\in\mathcal I}
\left(C_N\cap\{\mu_{i,T+1}<1-\zeta\}\right),
\]
another union bound gives
\[
\Pr^{11}_{T,\mu}(B_T^c)
\leq
2\exp\left(-\frac{q_*^2N}{2}\right)+2w_N(\zeta).
\]
Note that both terms on the right are independent of $\mu$. Thus,
\[
r_T(\zeta)
:=
\sup_{\mu\in\mathcal K}
\Pr^{11}_{T,\mu}(B_T^c)
\leq
2\exp\left(-\frac{q_*^2N}{2}\right)+2w_N(\zeta)
\longrightarrow0
\qquad\text{as }T\to\infty.
\]

From date $T+1$ on, the public history contains all evidence,
so the policy is transparent from $T+1$ on. Define $\beta:=1-\bar\lambda/2$ and now fix
$\zeta\in(0,\bar\lambda/2)$.
For each action $i$, let
\[
A_{i,T}
:=
\{\mu_{it}\leq\beta\text{ for some }t\geq T+1\}.
\]
By \cref{lem:nonsmooth-posterior}(ii), on $B_T$,
\[
\Pr^{11}_{T,\mu}(A_{i,T}\mid\mathscr H_{T+1})
\leq
\frac{1-\mu_{i,T+1}}{1-\beta}
\leq
\frac{2\zeta}{\bar\lambda}.
\]
Let
\[
G_T
:=
\{\mu_{it}>\beta
\text{ for every }t\geq T+1
\text{ and every }i\in\mathcal I\}.
\]
Then $G_T^c=\bigcup_{i\in\mathcal I}A_{i,T}$.
Because $B_T\in\mathscr H_{T+1}$, the union bound and the
tower property give
\[
\begin{aligned}
\Pr^{11}_{T,\mu}(G_T^c)
&\leq
\Pr^{11}_{T,\mu}(B_T^c)
+
\sum_{i\in\mathcal I}
\Pr^{11}_{T,\mu}(B_T\cap A_{i,T})\\
&=
\Pr^{11}_{T,\mu}(B_T^c)
+
\sum_{i\in\mathcal I}
\Ex^{11}_{T,\mu}\left[
\mathbf1_{B_T}
\Pr^{11}_{T,\mu}(A_{i,T}\mid\mathscr H_{T+1})
\right]\\
&\leq
\Pr^{11}_{T,\mu}(B_T^c)
+\frac{4\zeta}{\bar\lambda}.
\end{aligned}
\]
Taking the supremum over $\mu\in\mathcal K$ yields
\begin{equation}\label{eq:acc-safe-region}
\sup_{\mu\in\mathcal K}\Pr^{11}_{T,\mu}(G_T^c)
\leq
r_T(\zeta)+\frac{4\zeta}{\bar\lambda}.
\end{equation}

On $G_T$, both limit beliefs lie in $[\beta,1]$, almost surely.
Hence
\[
|\mu_{1\infty}-\mu_{2\infty}|
\leq
1-\beta
=
\bar\lambda/2
<
\bar\lambda,
\]
so \eqref{eq:positive-limit-demand} gives
$p_i(\mu_\infty)>0$ for both actions.
By \cref{lem:sources-general}(ii), both actions are sampled
infinitely often and $E$ occurs, $\Pr^{11}_{T,\mu}$-almost surely
on $G_T$. Therefore,
\[
\sup_{\mu\in\mathcal K}
\bigl[1-V(\sigma^{\mathsf{ACC},T},\phi,\mu,\nu)\bigr]
\leq
\sup_{\mu\in\mathcal K}\Pr^{11}_{T,\mu}(E^c)
\leq
r_T(\zeta)+\frac{4\zeta}{\bar\lambda}.
\]
For each fixed $\zeta\in(0,\bar\lambda/2)$,
$r_T(\zeta)\to0$ as $T\to\infty$. Taking the $\limsup$
as $T\to\infty$ and then letting $\zeta\downarrow0$ proves (i).

\underline{Part (ii)}: On $F$, the conditional law of $\theta$ given $\mathscr H_\tau$ is the product measure with marginals $\mu_\tau$. By \eqref{eq:fresh-unconditional}, conditional on $\mathscr H_\tau$ the unused signal streams and the future draws are fresh given the state. From date $\tau$ on, the messages of $\sigma$ reveal only records generated from date $\tau$ on, according to the accumulation rule. Hence, conditional on $\mathscr H_\tau$, the continuation of the environment has the law of $(\sigma^{\mathsf{ACC},T},\phi,\mu_\tau,\nu)$: public beliefs start from $\mu_\tau$ and are updated using the same records. Because $\tau<\infty$ on $F$, deleting the finitely many dates
before $\tau$ does not change whether per-period regret converges
to zero. This gives the displayed identity. The inequality follows
from $\mu_\tau\in\mathcal K$ on $F$ and the definition of the
supremum.
\end{proof}

\section{Proof of Theorem \ref{thrm:fulltrans}}\label{appendix:proofs1}

\paragraph{Outline.} The proof of \cref{thrm:fulltrans} is organized into the following blocks.
\begin{itemize}[leftmargin = 1.5em]
    \item \textbf{Housekeeping.} \cref{app:existence_ode} verifies existence and uniqueness for the path balance ODE in the main text. \cref{app:sde_existence} verifies that the stochastic differential equation \eqref{eqn:sde} has a unique solution and that the associated martingale problem is well posed.
    \item \textbf{Weak convergence.}
     \cref{app:weakconvergence} shows weak convergence in Skorokhod space to the solution of \eqref{eqn:sde} when $\KL_n \to 0$ and the stochastic smoothness condition holds. The condition holds automatically when drift dominates noise $(\V_n/\KL_n\to0)$.
    \item \textbf{`If' direction of \cref{thrm:fulltrans}.}
    \cref{app:T1_sufficiency} combines convergence to the path-balance ODE with the maximal inequality of \cref{lem:nonsmooth-posterior}.
    \item \textbf{`Only if' direction of \cref{thrm:fulltrans}.} \cref{app:T1_necessity} considers three cases. First, if $\KL_n \to 0$ and $\V_n/\KL_n \to 0$ but path balance fails, the ODE limit enters a herding region. Second, if $\KL_n\not \to 0$, finitely many signals reach the herding region. Third, if $\KL_n\to0$ but $\V_n/\KL_n\not\to0$, the stochastically smooth subcase has a diffusion limit with $C>0$, while the nonsmooth subcase is handled by a direct finite-$n$ rare-signal construction.
\end{itemize}

\subsection{Existence and uniqueness for the path-balance ODE}
\label{app:existence_ode}
Let $(\lambda_1,\lambda_2)\sim\nu$, define
$D:=\lambda_2-\lambda_1$, and let $F_D$ be its CDF on
$\mathbb R$. To derive the density of $D$, consider the change
of variables
\[
(u,z)=(\lambda_1,\lambda_2-\lambda_1).
\]
The inverse transformation is
$(\lambda_1,\lambda_2)=(u,u+z)$ and has Jacobian determinant
equal to one. Consequently, the joint density of
$(\lambda_1,D)$ is $f_\nu(u,u+z)$ whenever
$0\leq u\leq \bar\lambda$ and $0\leq u+z\leq \bar\lambda$, and is zero otherwise.
For a fixed $z\in[-\bar\lambda,\bar\lambda]$, these inequalities are equivalent to $\max\{0,-z\}\leq u\leq\min\{\bar\lambda,\bar\lambda-z\}$. Integrating out $u$ therefore gives the following density of $D$:
\[
f_D(z)
:=
\begin{cases}
\displaystyle
\int_{\max\{0,-z\}}^{\min\{\bar\lambda,\bar\lambda-z\}}
f_\nu(u,u+z)\,du,
& |z|\leq \bar\lambda,\\[1.2em]
0,
& |z|>\bar\lambda.
\end{cases}
\]
For every $z\in[-\bar\lambda,\bar\lambda]$, the interval of integration has
length $\bar\lambda-|z|$. Thus, integrating the lower and upper bounds
$\underline f\leq f_\nu\leq\overline f$ from regular horizontal
demand over this interval yields
\begin{equation}
\underline f\,(\bar\lambda-|z|)
\leq f_D(z)
\leq \overline f\,(\bar\lambda-|z|)
\qquad\text{for almost every }z\in[-\bar\lambda,\bar\lambda].
\label{eq:difference-density-bounds-ode}
\end{equation}
In particular,
$\|f_D\|_{L^\infty(\mathbb R)}\leq\overline f\bar\lambda$.
Since
\[
F_D(y)-F_D(x)=\int_x^y f_D(z)\,dz
\qquad\text{for every }x<y,
\]
$F_D$ is absolutely continuous and globally
$(\overline f\bar\lambda)$-Lipschitz.

For $\bm x=(x_1,x_2)\in\mathbb R^2$, define $M(\bm x):=\ell(x_1)-\ell(x_2)$. Since indifference has probability zero,
\[
\Gamma_1(\bm x)
=
\Pr(D\leq M(\bm x))
=
F_D(M(\bm x)),
\qquad
\Gamma_2(\bm x)=1-\Gamma_1(\bm x).
\]

\begin{lemma}\label{lem:F-Lip}
For all $\bm x,\bm y\in\mathbb R^2$,
\[
|M(\bm x)-M(\bm y)|\leq\frac{\sqrt2}{4}\|\bm x-\bm y\|_2
\qquad\text{and}\qquad
\|\bm\Gamma(\bm x)-\bm\Gamma(\bm y)\|_2
\leq
\frac{\overline f\bar\lambda}{2}\|\bm x-\bm y\|_2.
\]
In particular, $\bm\Gamma=(\Gamma_1,\Gamma_2)^\top$ is globally Lipschitz on $\mathbb R^2$.
\end{lemma}

\begin{proof}[Proof of \cref{lem:F-Lip}]
Because
\[
\ell'(x)=\frac{e^x}{(1+e^x)^2}\leq\frac14,
\]
the mean-value theorem implies $|\ell(x)-\ell(y)|\leq\frac14|x-y|$.
Consequently, by the Cauchy--Schwarz inequality,
\[
|M(\bm x)-M(\bm y)|
\leq
\frac14\bigl(|x_1-y_1|+|x_2-y_2|\bigr)
\leq
\frac{\sqrt2}{4}\|\bm x-\bm y\|_2.
\]
Since $F_D$ is globally $(\overline f\bar\lambda)$-Lipschitz,
\[
|\Gamma_1(\bm x)-\Gamma_1(\bm y)|
\leq
\frac{\sqrt2\,\overline f\bar\lambda}{4}
\|\bm x-\bm y\|_2.
\]
Finally, because $\Gamma_2=1-\Gamma_1$,
\[
\|\bm\Gamma(\bm x)-\bm\Gamma(\bm y)\|_2
=
\sqrt2\,|\Gamma_1(\bm x)-\Gamma_1(\bm y)| \leq \frac{\overline f\bar\lambda}{2} \|\bm x-\bm y\|_2.
\]
\end{proof}

\begin{lemma}\label{lem:path-ode-wellposed}
For each $\bm x_0\in\mathbb R^2$, the initial-value problem
\[
\dot{\bm X}(t)=\bm\Gamma(\bm X(t)),
\qquad
\bm X(0)=\bm x_0,
\]
has a unique solution
$\bm X\in C^1([0,\infty);\mathbb R^2)$. In particular,
\eqref{eqn:path} has a unique global solution
$\bm X^{\mathrm{PB}}$.
\end{lemma}

\begin{proof}[Proof of \cref{lem:path-ode-wellposed}]
By \cref{lem:F-Lip}, $\bm\Gamma: \mathbb{R}^2 \rightarrow \mathbb{R}^2$ is globally Lipschitz. The Picard--Lindel\"of theorem \citep[Theorem 2.2]{teschl2012ordinary} therefore gives a unique local solution. Moreover,
\[
\|\bm\Gamma(\bm x)\|_2^2
=
\Gamma_1(\bm x)^2+\Gamma_2(\bm x)^2
\leq1.
\]
Hence, every solution satisfies $\|\bm X(t)\|_2
\leq
\|\bm x_0\|_2+t$ on its interval of existence. The unique local solution may therefore be uniquely extended to $[0,\infty)$. See, for example,
\cite[Corollary 2.6]{teschl2012ordinary}.
\end{proof}

\subsection{Existence and uniqueness for the limit SDE}
\label{app:sde_existence}
Fix $C\geq0$ and define the drift and diffusion coefficients
\[
b(\bm x):=\bm\Gamma(\bm x),
\qquad
a_C^{1/2}(\bm x)
:=
C\operatorname{diag}(\bm\Gamma(\bm x))^{1/2}
=
C
\begin{pmatrix}
\sqrt{\Gamma_1(\bm x)}&0\\
0&\sqrt{\Gamma_2(\bm x)}
\end{pmatrix},
\]
so that $a_C(\bm x):=a_C^{1/2}(\bm x)a_C^{1/2}(\bm x)^\top=C^2\operatorname{diag}(\bm\Gamma(\bm x))$. In this notation, \eqref{eqn:sde} is the stochastic integral equation
\begin{equation}
\bm X_t
=
\bm x
+\int_0^t b(\bm X_s)\,ds
+\int_0^t a_C^{1/2}(\bm X_s)\,d\bm B_s.
\label{eq:sde-integral-form}
\end{equation}
We use the standard notions of weak and strong solutions, pathwise uniqueness, and uniqueness in law \citep[Chapter 8]{le2016brownian}. Filtered probability spaces are assumed to satisfy the usual conditions, and an $(\mathcal F_t)$-Brownian motion is a standard two-dimensional Brownian motion $\bm B$ started from zero that is $(\mathcal F_t)$-adapted and such that $\bm B_t-\bm B_s$ is independent of $\mathcal F_s$ for all $s<t$.

\begin{lemma}\label{lem:ex-unique-SDE}
\begin{itemize}
    \item[(i)] The coefficients $b$ and $a_C^{1/2}$ are bounded and globally Lipschitz. Consequently, for every filtered probability space carrying an $(\mathcal F_t)$-Brownian motion $\bm B$ and for each $\bm x\in\mathbb R^2$, \eqref{eqn:sde} has a pathwise unique strong solution started from $\bm x$, and weak solutions of \eqref{eqn:sde} are unique in law.
    \item[(ii)] For each $\bm x\in\mathbb R^2$, the martingale problem of \cref{def:main-martingale-problem} for
    \[
    \mathcal A_C f(\bm x)
    =
    \bm\Gamma(\bm x)\mathbin{\cdot}\nabla f(\bm x)
    +\frac{C^2}{2}
    \left[
    \Gamma_1(\bm x)\,\partial_{11}f(\bm x)
    +\Gamma_2(\bm x)\,\partial_{22}f(\bm x)
    \right]
    \]
    is well posed, and its solution started from $\bm x$ is the law of the solution of \eqref{eqn:sde} started from $\bm x$.
\end{itemize}
\end{lemma}

\begin{proof}[Proof of \cref{lem:ex-unique-SDE}]
\underline{Part (i)}: First, we show that the density bounds ensure that $F_D$ vanishes quadratically at the edge of its support, so that its square root is Lipschitz. Recall that $D=\lambda_2-\lambda_1$ has distribution function
$F_D$ and density $f_D$ satisfying
\eqref{eq:difference-density-bounds-ode}. Define
\[
g_0(z):=\sqrt{F_D(z)},
\qquad
g_1(z):=\sqrt{1-F_D(z)},
\qquad
K_0:=\frac{\overline f}{\sqrt{2\underline f}}.
\]
We show that $g_0$ and $g_1$ are globally
$K_0$-Lipschitz.

First, we bound $F_D$ and $f_D$. For $z\in[-\bar\lambda,0]$, \eqref{eq:difference-density-bounds-ode} gives
$\underline f(\bar\lambda+u)\leq f_D(u)\leq\overline f(\bar\lambda+u)$ for almost every $u\in[-\bar\lambda,z]$, and integrating the lower bound from $-\bar\lambda$
to $z$ gives
\[
F_D(z)
\geq
\frac{\underline f}{2}(\bar\lambda+z)^2
\qquad\text{for every }z\in[-\bar\lambda,0].
\]
For $z\in[0,\bar\lambda]$, monotonicity of $F_D$ gives
$F_D(z)\geq F_D(0)\geq\frac{\underline f}{2}\bar\lambda^2$, and
\eqref{eq:difference-density-bounds-ode} gives
$f_D(z)\leq\overline f\bar\lambda$ almost everywhere.

Second, we show that $g_0$ is $K_0$-Lipschitz on $[-\bar\lambda,\bar\lambda]$. Fix $-\bar\lambda\leq x<y\leq\bar\lambda$. Because $F_D(y)>0$,
\[
g_0(y)-g_0(x)
=
\frac{F_D(y)-F_D(x)}{g_0(y)+g_0(x)}
\geq0.
\]
If $y\leq0$, Step 1 bounds the numerator and the denominator:
\begin{align*}
F_D(y)-F_D(x)&\leq\overline f\int_x^y(\bar\lambda+u)\,du
=\frac{\overline f}{2}(y-x)\bigl[(\bar\lambda+y)+(\bar\lambda+x)\bigr],
\\
g_0(y)+g_0(x)&\geq\sqrt{\frac{\underline f}{2}}\bigl[(\bar\lambda+y)+(\bar\lambda+x)\bigr],
\end{align*}
so $g_0(y)-g_0(x)\leq K_0(y-x)$. If $x\geq0$, then $F_D(y)-F_D(x)\leq\overline f\bar\lambda(y-x)$ and $g_0(y)+g_0(x)\geq2\sqrt{\underline f/2}\,\bar\lambda=\sqrt{2\underline f}\,\bar\lambda$, so again $g_0(y)-g_0(x)\leq K_0(y-x)$. If $x<0<y$, apply the two previous cases to $(x,0)$ and $(0,y)$ and add.

Third, we show that $g_1$ is $K_0$-Lipschitz on $[-\bar\lambda,\bar\lambda]$. Because $D$ has no atoms, $g_1(z)=\sqrt{\Pr(-D<-z)}=\sqrt{F_{-D}(-z)}$. The random variable $-D=\lambda_1-\lambda_2$ is the taste difference for the type distribution obtained from $\nu$ by exchanging the two coordinates, which satisfies regular horizontal demand with the same bounds $\underline f$ and $\overline f$. Step 2 applied to that distribution shows that $z\mapsto\sqrt{F_{-D}(z)}$, and hence $g_1$, is $K_0$-Lipschitz.

Fourth, we extend these facts from $[-\bar\lambda,\bar\lambda]$ to all of $\mathbb{R}$. Both $g_0$ and $g_1$ are constant on each of
$(-\infty,-\bar\lambda]$ and $[\bar\lambda,\infty)$. With the projection
$\pi(z):=\max\{-\bar\lambda,\min\{z,\bar\lambda\}\}$, we have $g_i(z)=g_i(\pi(z))$ and
$|\pi(y)-\pi(x)|\leq|y-x|$, so
$|g_i(y)-g_i(x)|
\leq
K_0|y-x|$ for every
$x,y\in\mathbb R$ and $i\in\{0,1\}$.

\medskip
By \cref{lem:F-Lip},
$\|b(\bm x)-b(\bm y)\|_2
\leq
\frac{\overline f\bar\lambda}{2}\|\bm x-\bm y\|_2$ and $|M(\bm x)-M(\bm y)|\leq\frac{\sqrt2}{4}\|\bm x-\bm y\|_2$.
Since $\sqrt{\Gamma_1}=g_0\circ M$ and $\sqrt{\Gamma_2}=g_1\circ M$, the Frobenius norm satisfies
\begin{align*}
\|a_C^{1/2}(\bm x)-a_C^{1/2}(\bm y)\|_{\mathrm F}^2
&=
C^2\Bigl(
 |g_0(M(\bm x))-g_0(M(\bm y))|^2+|g_1(M(\bm x))-g_1(M(\bm y))|^2
\Bigr)\\
&\leq
\frac{C^2K_0^2}{4}\|\bm x-\bm y\|_2^2.
\end{align*}
The coefficients are also bounded:
$\|b(\bm x)\|_2^2
=
\Gamma_1(\bm x)^2+\Gamma_2(\bm x)^2
\leq1$
and
$\|a_C^{1/2}(\bm x)\|_{\mathrm F}^2
=
C^2\bigl(\Gamma_1(\bm x)+\Gamma_2(\bm x)\bigr)
=
C^2$.
Pathwise uniqueness and the existence of a strong solution follow from the standard theorem for SDEs with globally Lipschitz coefficients \citep[Theorem 8.3]{le2016brownian}, and pathwise uniqueness implies uniqueness in law by the Yamada--Watanabe theorem \citep[Proposition 5.3.20]{karatzas1991brownian}.

\underline{Part (ii)}:
Fix $\bm x\in\mathbb R^2$.
By It\^o's formula, boundedness of $a_C^{1/2}$, and the
SDE--martingale-problem correspondence
\citep[Proposition~5.4.6 and Remark~5.4.12]{karatzas1991brownian},
solutions of the martingale problem in
\cref{def:main-martingale-problem} started from $\bm x$
are precisely the laws of weak solutions of
\eqref{eq:sde-integral-form}.
The compactly supported cutoffs in \citep[Remark~5.4.12]{karatzas1991brownian} may be chosen
in $C_c^\infty(\mathbb R^2)$.
Existence and uniqueness in law from Part~(i) therefore prove
the claim.
\end{proof}

\subsection{Weak convergence} \label{app:weakconvergence}
Let $(\phi_n)_n$ be a sequence of experiments. For each $n\geq 1$, define
\[
Z_n
:=
\log\frac{d\phi_n^1}{d\phi_n^0}(s),
\qquad s\sim\phi_n^1
\]
and let $F_{Z,n}$ denote its law. $Z_n$ is the log-likelihood-ratio \emph{increment} to the log-odds for the sampled action. Unless otherwise stated, all probabilities and expectations in this subsection concern the full-transparency process conditional on $\theta=(1,1)$. Let $\KL_n=\Ex[Z_n]$ and $\V_n=\operatorname{Var}(Z_n)$. Throughout this subsection, we maintain the following assumption.

\begin{assumption}\label{as:weak convergence}
For all sufficiently large $n$, $0<\KL_n<\infty$ and $\V_n<\infty$. Moreover,
\[
\KL_n\to0,
\qquad
\sqrt{\frac{\V_n}{\KL_n}}\to C
\quad\text{for some }C\in[0,\infty).
\]
\end{assumption}

Under \cref{as:weak convergence}, after discarding finitely many terms we may assume without loss that $\KL_n$ and $\V_n$ are finite for each $n$. Define the initial condition
\[
\bm X_0
:=
\begin{pmatrix}
\mathsf L(\mu_{10})\\
\mathsf L(\mu_{20})
\end{pmatrix}
=:
\begin{pmatrix}
X_{10}\\X_{20}
\end{pmatrix}.
\]
For each $n\geq 1$, define the independent random variables $\{Z_{i,k}^{(n)}:i\in\{1,2\},\ k\in\mathcal T\}$ and $\{U_k^{(n)}:k\in\mathcal T\}$, where the $Z_{i,k}^{(n)}$ are i.i.d.\ with law $F_{Z,n}$ and the
$U_k^{(n)}$ are i.i.d.\ uniform on $[0,1]$. Recursively, given
$\bm X_k^{(n)}$, define
\[
B_{1,k}^{(n)}
:=
\mathbf 1\!\left\{
U_k^{(n)}\leq\Gamma_1(\bm X_k^{(n)})
\right\},
\qquad
B_{2,k}^{(n)}:=1-B_{1,k}^{(n)}
\]
and, for $t\in\mathcal T$,
\[
\bm X_t^{(n)}
=
\bm X_0+
\sum_{k=0}^{t-1}
\begin{pmatrix}
B_{1,k}^{(n)}Z_{1,k}^{(n)}\\
B_{2,k}^{(n)}Z_{2,k}^{(n)}
\end{pmatrix}.
\]
If $\mathcal G_k$ is the $\sigma$-algebra generated by the variables dated
strictly before $k$, then $\bm X_k^{(n)}$ is
$\mathcal G_k$-measurable, while $U_k^{(n)}$ and the two current
potential signals are independent of $\mathcal G_k$. This construction
therefore has the same law as the full-transparency log-odds process
conditional on $\theta=(1,1)$. For each $n\geq 1$, define the continuous-time, c\`adl\`ag process
\[
\bm X^{(n)}(t)
:=
\bm X_{\lfloor t/\KL_n\rfloor}^{(n)}
=
\bm X_0+
\sum_{k=0}^{\lfloor t/\KL_n\rfloor-1}
\begin{pmatrix}
B_{1,k}^{(n)}Z_{1,k}^{(n)}\\
B_{2,k}^{(n)}Z_{2,k}^{(n)}
\end{pmatrix}
\qquad \forall t\geq0.
\]
$\bm X^{(n)}$ is the right-continuous, piecewise-constant interpolation of the above construction, where each discrete date $k$ is mapped to continuous time $k\KL_n$. For each $i\in\{1,2\},\ t\in\mathcal T$, we also define
\[
X_{it}^{(n)}:=\bigl[\bm X_t^{(n)}\bigr]_i,
\qquad
\mu_{it}^{(n)}:=\ell\!\left(X_{it}^{(n)}\right).
\]
$\mu_{it}^{(n)}$ is the posterior belief corresponding to the action-$i$ log-odds $X_{it}^{(n)}$.

Fix a filtered probability space $(\Omega,\mathcal F,(\mathcal F_t)_{t\geq0},\mathbb P)$ satisfying the usual conditions and an $(\mathcal F_t)_{t\geq0}$-Brownian motion $\bm B=(B_1,B_2)$ started from zero. By \cref{lem:ex-unique-SDE}, there is a unique strong solution of \eqref{eqn:sde} with $\bm X^*_0=\bm X_0$, which we denote as $(\bm X^*_t)_{t\geq0}$.

\begin{definition}
    Say that a sequence of experiments $(\phi_n)_n$ satisfies the \emph{stochastic smoothness condition} if
    \[
    \frac{\mathbb{E}\big[Z_n^2\mathbf 1\{|Z_n|\geq 1\}\big]}{\mathbb{E}[Z_n]} \to 0 \quad \text{and} \quad \frac{\mathbb{P}(|Z_n|\geq \beta)}{\mathbb{E}[Z_n]} \to 0 \text{ for each $\beta > 0$} \tag{SS} \label{eqn:stochsmoothness}
    \]
\end{definition}

We record an inequality that is used several times. The function $z\mapsto e^{-z}-1+z$ is nonnegative, and absolute continuity gives $\Ex[e^{-Z_n}]=\phi_n^0(\{d\phi_n^1/d\phi_n^0>0\})\leq1$. Hence, whenever $\phi_n^1\ll\phi_n^0$ and $\KL_n<\infty$,
\begin{equation}\label{eq:exp-moment}
\Ex\!\left[e^{-Z_n}-1+Z_n\right]\leq\KL_n.
\end{equation}

The two conditions in \eqref{eqn:stochsmoothness} are used to establish weak convergence. The first part of the next lemma shows that they hold automatically when drift dominates noise. The second part shows that, when log-likelihood ratios are uniformly bounded above, the second condition implies the first, so that a failure of stochastic smoothness is always a failure of the second condition.

\begin{lemma}\label{lem:smoothness-housekeeping}
\begin{enumerate}[label=(\roman*)]
\item If \cref{as:weak convergence} holds with $C=0$, then $(\phi_n)_n$ satisfies \eqref{eqn:stochsmoothness}.

\item Suppose $\KL_n<\infty$ for each $n$ and, for some $K<\infty$, $Z_n\leq K$ $\phi_n^1$-almost surely for each $n$. If
\[
\frac{\Pr(|Z_n|\geq1)}{\KL_n}\to0,
\qquad\text{then}\qquad
\frac{\Ex\!\left[
Z_n^2\mathbf 1\{|Z_n|\geq1\}
\right]}{\KL_n}\to0.
\]
Consequently, under these assumptions, the second
condition in \eqref{eqn:stochsmoothness} implies the first.
\end{enumerate}
\end{lemma}

\begin{proof}[Proof of \cref{lem:smoothness-housekeeping}]
For part (i), the first condition holds because
\[
\frac{\Ex\!\left[Z_n^2\mathbf 1\{|Z_n|\geq1\}\right]}{\KL_n}
\leq
\frac{\Ex[Z_n^2]}{\KL_n}
=
\frac{\V_n}{\KL_n}+\KL_n
\to0.
\]
For the second condition, fix $\beta>0$. For all sufficiently large $n$,
$\KL_n<\beta/2$, and hence
$\{|Z_n|\geq\beta\}
\subseteq
\{|Z_n-\KL_n|\geq\beta/2\}$.
Chebyshev's inequality therefore gives
\[
\frac{\Pr(|Z_n|\geq\beta)}{\KL_n}
\leq
\frac{4}{\beta^2}\frac{\V_n}{\KL_n}
\to0.
\]

For part (ii), fix $M\geq1$. For $z\leq-M$, writing $w=-z\geq M$,
\[
e^{-z}-1+z
=e^{w}-1-w
\geq\frac{w^2}{2}+\frac{w^3}{6}
\geq z^2\,\frac{3+M}{6}.
\]
Because $e^{-z}-1+z\geq0$ everywhere, \eqref{eq:exp-moment} gives $\Ex[Z_n^2\mathbf 1\{Z_n<-M\}]\leq\frac{6}{3+M}\KL_n$. On $\{|Z_n|\geq1,\,Z_n\geq-M\}$ we have $Z_n^2\leq\max\{K,M\}^2$. Therefore
\[
\frac{\Ex\!\left[Z_n^2\mathbf 1\{|Z_n|\geq1\}\right]}{\KL_n}
\leq
\max\{K,M\}^2\,\frac{\Pr(|Z_n|\geq1)}{\KL_n}
+\frac{6}{3+M}.
\]
Take $n\to\infty$ and then $M\to\infty$.
\end{proof}

We now state the weak convergence of the rescaled log-odds
processes. We use the martingale problem of \cref{def:main-martingale-problem} with
\[
b=\bm\Gamma,
\qquad
a=a_C=C^2\operatorname{diag}(\bm\Gamma),
\]
so that $\mathcal A=\mathcal A_C$, where $\mathcal A_C$ is the
operator in \cref{lem:ex-unique-SDE}(ii).
That lemma establishes well-posedness for every deterministic
initial state. In particular, the unique solution law with initial
state $\bm X_0$ is the law of $\bm X^*$.

\begin{lemma}[Weak convergence]\label{lemma:convergence}
Suppose \cref{as:weak convergence} holds and
$(\phi_n)_n$ satisfies \eqref{eqn:stochsmoothness}. Conditional
on $\theta=(1,1)$, for each $T\in(0,\infty)$,
\[
\bm X^{(n)}
\Rightarrow
\bm X^*
\qquad
\text{in }\mathbb D([0,T];\mathbb R^2)
\text{ equipped with the Skorokhod $J_1$ topology},
\]
where $\bm X^*$ is the unique solution of \eqref{eqn:sde}
with $\bm X^*_0=\bm X_0$.
\end{lemma}

\begin{proof}[Proof of \cref{lemma:convergence}]
We use the notation of \citep[Chapter 11]{stroock2007multidimensional}. Let $h_n:=\KL_n$. For $\bm x\in\mathbb R^2$, let $\Delta^{(n)}(\bm x)$ be a random vector with the law of the one-step increment of the chain from $\bm x$:
\[
\Delta^{(n)}(\bm x)
:=
\bigl(\mathbf 1\{U\leq\Gamma_1(\bm x)\}Z_1,\ \mathbf 1\{U>\Gamma_1(\bm x)\}Z_2\bigr),
\]
where $U$, $Z_1$, and $Z_2$ are independent, $U$ is uniform on $[0,1]$, and $Z_1,Z_2$ have law $F_{Z,n}$; write $\Ex_{\bm x}$ and $\Pr_{\bm x}$ for expectation and probability with respect to this vector.
Exactly one coordinate of $\Delta^{(n)}(\bm x)$ is potentially nonzero,
and its increment has law $F_{Z,n}$.

Using the closed-unit-ball truncation, define
\begin{align*}
a_{ij}^n(\bm x)
&:=
\frac1{h_n}
\Ex_{\bm x}\!\left[
\Delta_i^{(n)}(\bm x)\Delta_j^{(n)}(\bm x)
\mathbf 1\{\|\Delta^{(n)}(\bm x)\|_2\leq1\}
\right],\\
b_i^n(\bm x)
&:=
\frac1{h_n}
\Ex_{\bm x}\!\left[
\Delta_i^{(n)}(\bm x)
\mathbf 1\{\|\Delta^{(n)}(\bm x)\|_2\leq1\}
\right],\\
\Delta_\varepsilon^n(\bm x)
&:=
\frac1{h_n}
\Pr_{\bm x}\!\left(
\|\Delta^{(n)}(\bm x)\|_2>\varepsilon
\right).
\end{align*}

For $i\neq j$, exactly one coordinate moves at each date, and hence
\[
a_{ij}^n(\bm x)=0.
\]
For $i=j$,
\begin{align*}
a_{ii}^n(\bm x)
&=
\Gamma_i(\bm x)
\frac{\Ex[Z_n^2\mathbf 1\{|Z_n|\leq1\}]}{h_n}\\
&=
\Gamma_i(\bm x)
\left[
\frac{\V_n}{\KL_n}+\KL_n
-
\frac{\Ex[Z_n^2\mathbf 1\{|Z_n|>1\}]}{\KL_n}
\right].
\end{align*}
The last term converges to zero by the first condition in \eqref{eqn:stochsmoothness}, because
$\{|Z_n|>1\}\subseteq\{|Z_n|\geq1\}$. Consequently, uniformly in
$\bm x$,
\[
a_{11}^n(\bm x)\to C^2\Gamma_1(\bm x),
\qquad
a_{22}^n(\bm x)\to C^2\Gamma_2(\bm x),
\qquad
a_{12}^n(\bm x)=a_{21}^n(\bm x)=0.
\]

Similarly,
\[
b_i^n(\bm x)
=
\Gamma_i(\bm x)
\left[
1-
\frac{\Ex[Z_n\mathbf 1\{|Z_n|>1\}]}{\KL_n}
\right].
\]
Because $|z|\leq z^2$ when $|z|>1$, the last term is bounded in absolute value by the ratio $\Ex[Z_n^2\mathbf 1\{|Z_n|>1\}]/\KL_n$, which converges to zero as just shown.
It follows, uniformly in $\bm x$, that
\[
b_1^n(\bm x)\to\Gamma_1(\bm x),
\qquad
b_2^n(\bm x)\to\Gamma_2(\bm x).
\]

Finally, for each $\varepsilon>0$,
\[
\Delta_\varepsilon^n(\bm x)
=
\frac{\Pr(|Z_n|>\varepsilon)}{\KL_n}
\leq
\frac{\Pr(|Z_n|\geq\varepsilon)}{\KL_n}
\to0,
\]
uniformly in $\bm x$. Fix any subsequence, and pass if necessary to a further subsequence along which $h_n$ is strictly decreasing.\footnote{A sequence of laws converges weakly if every subsequence has a further subsequence that converges to the same limit \citep[Theorem 2.6]{billingsley1999convergence}.} We now match these definitions to the diffusion-approximation
criterion in \cite[Theorem 11.2.3]{stroock2007multidimensional},\footnote{Note that the criterion is stated for a family of kernels indexed by all mesh sizes $h>0$, but its proof uses the family only along sequences $h_k\downarrow0$.} with dimension $d=2$, the kernels $\Pi_n$ defined below, and mesh sizes $h=h_n=\KL_n\to0$.

Let $\bm e_1,\bm e_2$ be the standard basis vectors of
$\mathbb R^2$. The transition kernel corresponding to the
\cite[Theorem 11.2.3]{stroock2007multidimensional}'s $\Pi_h$ is, in our notation,
\[
\Pi_n(\bm x,A)
:=
\sum_{i=1}^2 \Gamma_i(\bm x)
\int_{\mathbb R}
\mathbf 1\{\bm x+z\bm e_i\in A\}\,F_{Z,n}(dz),
\qquad A\in\mathcal B(\mathbb R^2).
\]
This is a Borel probability kernel because $\Gamma_1,\Gamma_2$ are
continuous, nonnegative, and sum to one. The construction of
the discrete process gives
\[
\Pr\!\left(
\bm X_{k+1}^{(n)}\in A
\,\middle|\,\mathcal G_k
\right)
=
\Pi_n(\bm X_k^{(n)},A)
\qquad\text{almost surely}.
\]
Hence, one transition of $\Pi_n$ corresponds to elapsed time
$h_n$ in the rescaled process.

The truncated coefficients $a_h^{ij}$ and $b_h^i$ in \cite[Theorem 11.2.3]{stroock2007multidimensional} are our $a_{ij}^n$ and $b_i^n$: writing
$\bm y-\bm x=\Delta^{(n)}(\bm x)$ in their defining
integrals gives
\begin{align*}
\frac1{h_n}
\int_{\|\bm y-\bm x\|_2\leq1}
(y_i-x_i)(y_j-x_j)\,\Pi_n(\bm x,d\bm y)
&=a_{ij}^n(\bm x),\\
\frac1{h_n}
\int_{\|\bm y-\bm x\|_2\leq1}
(y_i-x_i)\,\Pi_n(\bm x,d\bm y)
&=b_i^n(\bm x).
\end{align*}
In particular, these are truncated raw second moments, not
centered covariances. The limiting coefficients are
\[
a(\bm x)=C^2\operatorname{diag}(\bm\Gamma(\bm x)),
\qquad
b(\bm x)=\bm\Gamma(\bm x),
\]
so \cite[Theorem 11.2.3]{stroock2007multidimensional}'s diffusion operator is our $\mathcal A_C$.
The calculations above show, for every $R<\infty$,
\[
\sup_{\|\bm x\|_2\leq R}
\|a^n(\bm x)-a(\bm x)\|_{\mathrm F}\to0,
\qquad
\sup_{\|\bm x\|_2\leq R}
\|b^n(\bm x)-b(\bm x)\|_2\to0.
\]

The large-jump condition also holds when the boundary of the
ball is included. Specifically, for every $\varepsilon>0$,
\[
\sup_{\bm x\in\mathbb R^2}
\frac1{h_n}
\Pi_n\!\left(
\bm x,\{\bm y:\|\bm y-\bm x\|_2\geq\varepsilon\}
\right)
=
\frac{\Pr(|Z_n|\geq\varepsilon)}{\KL_n}
\longrightarrow0.
\]
This is the second condition in \eqref{eqn:stochsmoothness}. Thus the reference's
large-jump requirement is satisfied whether it is written
with a strict or a nonstrict inequality.\footnote{These probabilities may differ since $F_{Z,n}$ may have atoms.}

We also verify the uniform boundedness condition on the
approximating coefficients. By \cref{as:weak convergence},
\[
K_2
:=
\sup_{n\geq1}\frac{\Ex[Z_n^2]}{h_n}
=
\sup_{n\geq1}
\left(\frac{\V_n}{\KL_n}+\KL_n\right)
<\infty.
\]
The supremum is finite because the displayed sequence
converges and every term is finite. Since
$\Gamma_1(\bm x)^2+\Gamma_2(\bm x)^2\leq1$, the formulas above give
\[
\sup_{\bm x\in\mathbb R^2}
\|a^n(\bm x)\|_{\mathrm F}
\leq K_2.
\]
Moreover,
\[
\frac{
\left|\Ex[Z_n\mathbf 1\{|Z_n|\leq1\}]\right|
}{h_n}
\leq
\frac{|\Ex[Z_n]|}{h_n}
+
\frac{\Ex[|Z_n|\mathbf 1\{|Z_n|>1\}]}{h_n}
\leq
1+K_2,
\]
where the last inequality uses
$|z|\mathbf 1\{|z|>1\}\leq z^2$.
Consequently,
\[
\sup_{n\geq1}\sup_{\bm x\in\mathbb R^2}
\left(
\|a^n(\bm x)\|_{\mathrm F}
+
\|b^n(\bm x)\|_2
\right)
\leq1+2K_2<\infty.
\]

The limiting coefficients $a$ and $b$ are bounded and
continuous, and \cref{lem:ex-unique-SDE}(ii) establishes
well-posedness of their martingale problem for every
deterministic initial state. Finally, \cite[Theorem 11.2.3]{stroock2007multidimensional}'s measure $P_{\bm x}^h$ is the law on
continuous path space obtained by linearly interpolating the
Markov chain with kernel $\Pi_h$ and initial state $\bm x$.
Accordingly, define
\[
\widehat{\bm X}^{(n)}(t)
:=
\bm X_k^{(n)}
+
\frac{t-kh_n}{h_n}
\left(\bm X_{k+1}^{(n)}-\bm X_k^{(n)}\right),
\qquad
k=\lfloor t/h_n\rfloor,\quad t\geq0.
\]
With $h=h_n$ and initial state $\bm x=\bm X_0$,
$P_{\bm x}^h$ therefore corresponds to the law of
$\widehat{\bm X}^{(n)}$. The reference's limiting measure $P_{\bm X_0}$ is the
unique solution law of the martingale problem for
$\mathcal A_C$, hence the law of $\bm X^*$.

All hypotheses of the diffusion-approximation criterion
are therefore satisfied. Restricting its conclusion to
any finite time interval gives
\[
\widehat{\bm X}^{(n)}
\Rightarrow
\bm X^*
\qquad
\text{in }C([0,T];\mathbb R^2)
\]
equipped with the uniform topology.

It remains to transfer this convergence to the piecewise-constant
interpolation $\bm X^{(n)}$. By construction,
\[
\sup_{0\leq t\leq T}
\left\|
\widehat{\bm X}^{(n)}(t)-\bm X^{(n)}(t)
\right\|_2
\leq
\max_{0\leq k\leq\lfloor T/h_n\rfloor}
\left\|\bm X_{k+1}^{(n)}-\bm X_k^{(n)}\right\|_2.
\]
Conditional on $\mathcal G_k$, exactly one action is sampled, and
the selected signal increment has law $F_{Z,n}$. Consequently, for
every $\varepsilon>0$,
\[
\Pr\!\left(
\left\|\bm X_{k+1}^{(n)}-\bm X_k^{(n)}\right\|_2>\varepsilon
\,\middle|\,\mathcal G_k
\right)
=
\Pr(|Z_n|>\varepsilon)
\quad\text{a.s.}
\]
A union bound therefore gives
\begin{align*}
&\Pr\!\left(
\sup_{0\leq t\leq T}
\left\|
\widehat{\bm X}^{(n)}(t)-\bm X^{(n)}(t)
\right\|_2>\varepsilon
\right)\\
&\qquad\leq
\left(\lfloor T/h_n\rfloor+1\right)
\Pr(|Z_n|>\varepsilon)\\
&\qquad\leq
(T+h_n)\frac{\Pr(|Z_n|>\varepsilon)}{h_n}
\longrightarrow0.
\end{align*}
The last limit follows from the vanishing-large-jump bound
established above.

The inclusion of $C([0,T];\mathbb R^2)$ with the uniform topology
into $\mathbb D([0,T];\mathbb R^2)$ with the Skorokhod $J_1$
topology is continuous. Moreover, choosing the identity time change
shows that the $J_1$ distance between two paths is bounded above by
their uniform distance. Hence $\widehat{\bm X}^{(n)}\Rightarrow\bm X^*$ in $\mathbb D([0,T];\mathbb R^2)$, the $J_1$ distance between $\widehat{\bm X}^{(n)}$ and $\bm X^{(n)}$ converges to zero in probability, and \citet[Theorem 3.1]{billingsley1999convergence}
implies
\[
\bm X^{(n)}
\Rightarrow
\bm X^*
\qquad
\text{in }\mathbb D([0,T];\mathbb R^2)
\]
equipped with the Skorokhod $J_1$ topology, as desired.
\end{proof}

\subsection{The `if' direction of \cref{thrm:fulltrans}} \label{app:T1_sufficiency} We now show that the conditions in the statement of \cref{thrm:fulltrans} are sufficient for $V_n\to 1$. The argument has two steps. First, convergence to the path-balance ODE implies that both beliefs are close to one by a deterministic rescaled time with high probability. Second, the maximal inequality of \cref{lem:nonsmooth-posterior} keeps both beliefs above $1-\bar\lambda/2$ forever with high probability, so that both actions remain active and both values are learned.

\begin{lemma}\label{lem:T1-sufficiency}
Suppose
\[
\KL_n\to0,
\qquad
\frac{\V_n}{\KL_n}\to0,
\]
and suppose $(\mu_0,\nu)$ satisfies path balance. Then
$V_n\to1$.
\end{lemma}

\begin{proof}[Proof of \cref{lem:T1-sufficiency}]
Let $\Pr_n^{11}$ denote probability conditional on $\theta=(1,1)$ under experiment $\phi_n$. Fix $\varepsilon\in(0,1)$ and choose
$\bar\mu\in(1-\bar\lambda/2,1)$ such that
\begin{equation}\label{eq:suff-mubar}
\frac{4(1-\bar\mu)}{\bar\lambda}<\frac{\varepsilon}{2}.
\end{equation}

\medskip
\noindent
\emph{Step 1: Both posteriors are close to one at a
deterministic rescaled time.}
Path balance gives a finite $T>0$ such that
$m_*:=
\min_{i\in\mathcal I}\ell(X_i^{\mathrm{PB}}(T))
>\bar\mu$.
Fix this $T$, and define
\[
T_n:=\left\lfloor\frac{T}{\KL_n}\right\rfloor,
\qquad
m_n:=\min_{i\in\mathcal I}\mu_{i,T_n}^{(n)},
\qquad
A_n:=\{m_n\geq\bar\mu\}\in\mathscr H_{T_n}.
\]
By \cref{lemma:convergence} with $C=0$, the rescaled
log-odds process converges weakly on $\mathbb D([0,T];\mathbb R^2)$ to the deterministic
continuous path $\bm X^{\mathrm{PB}}$. Evaluation at the endpoint $T$
is continuous in the Skorokhod
$J_1$ topology, and $\bm X^{(n)}(T)=\bm X^{(n)}_{T_n}$. Since $\ell$ and the minimum of two
coordinates are continuous, $m_n\to m_*$ in probability under $\Pr_n^{11}$. Because $m_*>\bar\mu$, there is an integer $N$ such
that $\Pr_n^{11}(A_n^c)\leq\varepsilon/2$ for every $n\geq N$.

\medskip
\noindent
\emph{Step 2: Both posteriors stay above $1-\bar\lambda/2$.}
Fix any $n$ and let
\[
G_n:=\bigl\{\mu_{it}^{(n)}>1-\bar\lambda/2\text{ for all }t\geq T_n\text{ and }i\in\mathcal I\bigr\}.
\]
Full transparency is transparent from date $0$ on. Applying \cref{lem:nonsmooth-posterior}(ii) to each action with the stopping time $T_n$ and $\alpha=1-\bar\lambda/2$, and using the union bound, gives on $A_n$
\[
\Pr_n^{11}(G_n^c\mid\mathscr H_{T_n})
\leq
\sum_{i\in\mathcal I}
\frac{1-\mu_{i,T_n}^{(n)}}{\bar\lambda/2}
\leq
\frac{4(1-\bar\mu)}{\bar\lambda}
<
\frac{\varepsilon}{2}.
\]

\medskip
\noindent
\emph{Step 3: On $G_n$, learning is efficient.}
On $G_n$, the limit beliefs satisfy $\mu^{(n)}_{i\infty}\geq1-\bar\lambda/2$ for both $i$, so $|\mu^{(n)}_{1\infty}-\mu^{(n)}_{2\infty}|\leq\bar\lambda/2<\bar\lambda$ and $p_i(\mu^{(n)}_\infty)>0$ for both $i$ by \eqref{eq:positive-limit-demand}. Full transparency satisfies sampled-action revelation (\cref{lem:sar}), so \cref{lem:sources-general}(ii) implies that both actions are sampled infinitely often and that $E$ occurs. Thus
$\Pr_n^{11}(G_n\cap E^c)=0$ for every $n$.

\medskip
\noindent
\emph{Step 4: Combining the bounds.}
For every $n\geq N$,
\[
\Pr_n^{11}(E^c)
\leq
\Pr_n^{11}(G_n^c)
\leq
\Pr_n^{11}(A_n^c)
+
\Ex_n^{11}\!\left[
\mathbf 1_{A_n}
\Pr_n^{11}(G_n^c\mid\mathscr H_{T_n})
\right]
<
\varepsilon,
\]
where the second inequality uses $A_n\in\mathscr H_{T_n}$ and the tower
property. Since $\varepsilon\in(0,1)$ was arbitrary,
$\Pr_n^{11}(E^c)\to0$, and \cref{lem:sources} gives
$V_n
=
1-\mu_{10}\mu_{20}\Pr_n^{11}(E^c)
\to1$.
\end{proof}

\subsection{The `only if' direction of \cref{thrm:fulltrans}}
\label{app:T1_necessity}

Throughout this subsection, write $\Pr_n^{11}$ and $\Ex_n^{11}$ for
probability and expectation under experiment $\phi_n$ conditional on
$\theta=(1,1)$, abbreviated to $\Pr^{11}$ and $\Ex^{11}$ when $n$ is clear. By \cref{lem:sources}, $V_n=1-\mu_{10}\mu_{20}\Pr_n^{11}(E^c)$. To prove necessity, it therefore suffices to show that whenever one of
\[
\KL_n\to0,
\qquad
\frac{\V_n}{\KL_n}\to0,
\qquad
\text{path balance}
\]
fails, there exists a subsequence along which $\Pr_n^{11}(E^c)$ is bounded away from zero. The argument considers three exhaustive cases. Case A handles the failure of path balance when the first two conditions hold, Case B the failure of $\KL_n\to0$, and Case C the failure of $\V_n/\KL_n\to0$ when $\KL_n\to0$. In each case we show that, with probability bounded away from zero, some high-value action is abandoned. With one exception (the negative tail in Subcase C.II, which argues directly through \cref{lem:abandon}), the argument ends by invoking \cref{lem:trap}. Cases A and C.I do so from the limit process, through the following lemma. Throughout, events such as $F_n$ below are defined through the log-odds process constructed in \cref{app:weakconvergence}. That process has the law of the log-odds of the public beliefs under full transparency in state $(1,1)$, so we identify the two. The events then belong to the public filtration, so \cref{lem:trap} applies.

\begin{lemma}[Escape in the limit]\label{lem:escape-limit}
Suppose that, along a sequence of experiments, \cref{as:weak convergence} and \eqref{eqn:stochsmoothness} hold, and let $\bm X^*$ be the solution of \eqref{eqn:sde} started from $\bm X_0$. Suppose that, for some $T\in(0,\infty)$, some labeling $i\neq j$ of the actions, and some $\kappa>0$,
\[
p:=\mathbb P\bigl(\ell(X^*_{iT})-\ell(X^*_{jT})>\bar\lambda+\kappa\bigr)>0.
\]
Then $\liminf_{n\to\infty}\Pr_n^{11}(E^c)\geq\kappa p$.
\end{lemma}

\begin{proof}[Proof of \cref{lem:escape-limit}]
Let $k_n:=\lfloor T/\KL_n\rfloor$, so that $\bm X^{(n)}(T)=\bm X^{(n)}_{k_n}$. By \cref{lemma:convergence}, $\bm X^{(n)}\Rightarrow\bm X^*$ in $\mathbb D([0,T];\mathbb R^2)$, and evaluation at the endpoint $T$ is continuous in the Skorokhod $J_1$ topology, so $\bm X^{(n)}_{k_n}\Rightarrow\bm X^*_T$. The set
$\{\bm x:\ell(x_i)-\ell(x_j)>\bar\lambda+\kappa\}$
is open, so the portmanteau theorem gives $\liminf_n\Pr_n^{11}(F_n)\geq p$ for
\[
F_n:=
\left\{
\mu_{i,k_n}^{(n)}-\mu_{j,k_n}^{(n)}>
\bar\lambda+\kappa
\right\}\in\mathscr H_{k_n}.
\]
\cref{lem:trap}, applied with the deterministic stopping time $k_n$ and the event $F_n$, gives $\Pr_n^{11}(E^c)\geq\kappa\Pr_n^{11}(F_n)$.
\end{proof}

\medskip
\noindent
\underline{Case A: $\KL_n\to0$ and $\V_n/\KL_n\to0$, but path balance
fails.}
\cref{as:weak convergence} holds with $C=0$, stochastic smoothness holds by \cref{lem:smoothness-housekeeping}(i), and the solution of \eqref{eqn:sde} with $C=0$ is the deterministic path $\bm X^{\mathrm{PB}}$. By Assumption \ref{as:no-tangential-pb} and continuity of $\bm X^{\mathrm{PB}}$, after relabeling the actions there
exists $T_0\in(0,\infty)$ such that
$2\kappa:=\ell(X_i^{\mathrm{PB}}(T_0))
-
\ell(X_j^{\mathrm{PB}}(T_0))
-\bar\lambda>0$.
\cref{lem:escape-limit} with $T=T_0$ and $p=1$ gives
$\liminf_{n}\Pr_n^{11}(E^c)\geq\kappa>0$.

\medskip
\noindent
\underline{Case B: $\KL_n \not\to 0$.} Let $K<\infty$ be such that $Z_n\leq K$ $\phi_n^1$-almost surely for every $n$. Choose a subsequence and a constant $\kappa_0>0$ such that $\KL_n\geq\kappa_0$ along it, fix $\alpha\in(0,1)$, and put $z:=\alpha\kappa_0$. Because $\Ex_{\phi_n^1}[Z_n]=\KL_n\leq K$ and $Z_n\leq K$,
\[
\KL_n\leq z\,\phi_n^1(\{Z_n<z\})+K\,\phi_n^1(\{Z_n\geq z\})
\leq z+(K-z)\,\phi_n^1(\{Z_n\geq z\}),
\]
so that
\[
 \phi_n^1(\{Z_n\geq z\})\geq
 \frac{\KL_n-z}{K-z}
 \geq \frac{(1-\alpha)\kappa_0}{K}=:q>0.
\]
Let $I$ be an action with $\mu_{I0}=\max_i\mu_{i0}$, so that $\gamma:=p_I(\mu_0)>0$, and choose $\varepsilon>0$ and an integer $m$ as in \cref{lem:escape-prior}; these constants depend on $(z,\mu_0,\bar\lambda)$ but not on $n$. \cref{lem:escape-prior} gives
$\Pr_n^{11}(E^c)\geq2\varepsilon(\gamma q)^m>0$ for every $n$ along the subsequence. Note that only the upper bound on $Z_n$ is used in this case.

\medskip
\noindent
\underline{Case C: $\KL_n\to0$ and $\V_n/\KL_n\not\to0$.}
Let $K\in(0,\infty)$ be such that $Z_n\leq K$ $\phi_n^1$-almost surely for every $n$. We first establish a uniform moment bound on the log-likelihood ratio.

\begin{lemma}[Likelihood-ratio moment bound]\label{lem:nonsmooth-second-moment}
If $Z_n\leq K$ $\phi_n^1$-almost surely for some $K\geq0$, then
\begin{equation}
\label{eq:nonsmooth-second-moment}
 \Ex[Z_n^2]\leq A\KL_n
 \qquad\text{with }A:=2e^K.
\end{equation}
\end{lemma}

\begin{proof}
For $z\leq K$, Taylor's formula with integral remainder gives
\[
e^{-z}-1+z
=
z^2\int_0^1(1-s)e^{-sz}\,ds
\geq
\frac{e^{-K}}{2}z^2,
\]
because $e^{-sz}\geq e^{-K}$ for every $s\in[0,1]$: if $z\geq0$ this follows from $sz\leq K$, and if $z<0$ then $e^{-sz}\geq1\geq e^{-K}$. Taking expectations and using \eqref{eq:exp-moment} yields
$\frac{e^{-K}}{2}\Ex[Z_n^2]
\leq
\KL_n$.
\end{proof}

It follows that $\V_n/\KL_n\leq A$ for all $n$. Pass to a subsequence, retained throughout Case C, on which
\[
 \sqrt{\frac{\V_n}{\KL_n}}\to C
 \qquad\text{for some }C\in(0,\sqrt A].
\]
This subsequence satisfies \cref{as:weak convergence}. We split according to whether it satisfies \eqref{eqn:stochsmoothness}.

\underline{Subcase C.I: stochastic smoothness holds.}
By \cref{lemma:convergence}, conditional on $\theta=(1,1)$,
the rescaled log-odds processes converge on every finite horizon
to the unique solution of
\[
\begin{aligned}
dX_{1t}^*
&=
\Gamma_1(\bm X_t^*)\,dt
+C\sqrt{\Gamma_1(\bm X_t^*)}\,dB_{1t},\\
dX_{2t}^*
&=
\Gamma_2(\bm X_t^*)\,dt
+C\sqrt{\Gamma_2(\bm X_t^*)}\,dB_{2t}.
\end{aligned}
\]
where $C>0$. Let $\mathbb P$ be the law of this diffusion. Relabel the actions so that $\mu_{10}\geq\mu_{20}$, and choose $\varepsilon>0$ such that
\[
\mu_{20}+\bar\lambda+4\varepsilon<1.
\]
By continuity of $\ell$, and then by $\ell(H)\to1$ as
$H\to+\infty$, we can choose $\delta>0$ and $H>X_{10}$ such that
\begin{equation}
\ell(X_{10}-\delta)-\ell(X_{20}+\delta)>-\bar\lambda,
\qquad
\ell(H-\delta)-\ell(X_{20}+\delta)>
\bar\lambda+3\varepsilon.
\label{eq:smooth-escape-target}
\end{equation}
Define
\[
\mathcal D
=
[X_{10}-\delta,H+\delta]
\times[X_{20}-\delta,X_{20}+\delta],
\qquad
\tau
=
\inf\{t\geq0:\bm X_t^*\notin\operatorname{int}\mathcal D\}.
\]
The first inequality in \eqref{eq:smooth-escape-target},
regular horizontal demand, and compactness imply
\[
g:=\inf_{\bm x\in\mathcal D}\Gamma_1(\bm x)>0.
\]

Choose $T>0$ such that $C^2T/\delta^2<1/2$, and set
$v=(H-X_{10})/T$. For $0\leq t\leq T$, define
\[
\bm u_t
=
\mathbf 1\{t<\tau\}
\begin{pmatrix}
\displaystyle
\frac{v-\Gamma_1(\bm X_{t\wedge\tau}^*)}
     {C\sqrt{\Gamma_1(\bm X_{t\wedge\tau}^*)}}\\[1.2em]
\displaystyle
-\frac{\sqrt{\Gamma_2(\bm X_{t\wedge\tau}^*)}}{C}
\end{pmatrix}.
\]
This progressively measurable process is bounded because
$\Gamma_1\geq g>0$ on $\mathcal D$. In particular, its stochastic
exponential satisfies Novikov's criterion. By
\cite[Theorems~5.22--5.23]{le2016brownian}, the probability measure
defined on $\mathcal F_T$ by
\[
\left.\frac{d\mathbb Q}{d\mathbb P}\right|_{\mathcal F_T}
=
\exp\left(
\int_0^T\bm u_s\cdot d\bm B_s
-\frac12\int_0^T\|\bm u_s\|_2^2\,ds
\right)
\]
is equivalent to $\mathbb P$, and
\[
\bm W_t
=
\bm B_t-\int_0^t\bm u_s\,ds,
\qquad 0\leq t\leq T,
\]
is a standard two-dimensional Brownian motion under $\mathbb Q$.

The choice of $\bm u$ changes the drift to $(v,0)$ before
$\tau$. Thus, for $0\leq t\leq T$,
\[
\bm X_{t\wedge\tau}^*
=
\bm X_0+
\begin{pmatrix}
v(t\wedge\tau)\\0
\end{pmatrix}
+\bm M_t,
\qquad
\bm M_t
:=
\int_0^{t\wedge\tau}
a_C^{1/2}(\bm X_s^*)\,d\bm W_s.
\]
Since $\operatorname{tr}(a_C)=C^2$,
the It\^o isometry gives
\[
\Ex_{\mathbb Q}\|\bm M_T\|_2^2
=
C^2\Ex_{\mathbb Q}[T\wedge\tau]
\leq C^2T.
\]
Applying Doob's maximal inequality to the nonnegative
submartingale $\|\bm M_t\|_2^2$ yields
\[
\mathbb Q\left(
\sup_{0\leq t\leq T}\|\bm M_t\|_\infty\geq\delta
\right)
\leq
\frac{\Ex_{\mathbb Q}\|\bm M_T\|_2^2}{\delta^2}
<
\frac12.
\]

On the complementary event, $\tau>T$. Indeed, if $\tau\leq T$,
the stopped representation at $\tau$ would place the first
coordinate strictly between $X_{10}-\delta$ and $H+\delta$,
and the second strictly between $X_{20}-\delta$ and
$X_{20}+\delta$, contradicting the definition of $\tau$.
Consequently, on this event,
\[
X_{1T}^*>H-\delta,
\qquad
X_{2T}^*<X_{20}+\delta.
\]
By \eqref{eq:smooth-escape-target} and equivalence of the two
measures,
\begin{equation}
p:=
\mathbb P\left(
\ell(X_{1T}^*)-\ell(X_{2T}^*)>
\bar\lambda+3\varepsilon
\right)>0.
\label{eq:smooth-terminal-escape}
\end{equation}
Since \cref{as:weak convergence} and \eqref{eqn:stochsmoothness} hold along the retained subsequence, \cref{lem:escape-limit} with $i=1$, $j=2$, and $\kappa=3\varepsilon$ gives
$\liminf_n\Pr_n^{11}(E^c)\geq3\varepsilon p>0$ along it.
This completes Subcase C.I.

\medskip
\noindent

\underline{Subcase C.II: stochastic smoothness fails.}
By \cref{lem:smoothness-housekeeping}(ii), the second condition in \eqref{eqn:stochsmoothness} fails along the retained subsequence. Hence, there are constants $\beta,c>0$ and a further subsequence along which
\begin{equation}
    \phi_n^1(|Z_n|\geq\beta)\geq c\KL_n.
    \label{eq:nonsmooth-rare-lower}
\end{equation}
For each $n$, at least one of $\phi_n^1(Z_n\geq\beta)\geq\tfrac c2\KL_n$ and $\phi_n^1(Z_n\leq-\beta)\geq\tfrac c2\KL_n$ holds. Passing to one more subsequence, the same alternative holds for every $n$:
\begin{align}
 \phi_n^1(Z_n\geq\beta)&\geq\tfrac c2\KL_n\quad\text{for every }n,
 \tag{P}\label{eq:nonsmooth-positive-tail}\\
 \intertext{or}
 \phi_n^1(Z_n\leq-\beta)&\geq\tfrac c2\KL_n\quad\text{for every }n.
 \tag{N}\label{eq:nonsmooth-negative-tail}
\end{align}
We prove failure under either sign.

\smallskip
\noindent\textit{Proof sketch of Subcase C.II.}
Call a signal \emph{rare} if $|Z_n|\geq\beta$ and \emph{small} otherwise. A rare signal moves the log-odds of the sampled action by at least $\beta$ in absolute value, an amount that does not shrink with $n$, and it occurs with probability of order $\KL_n$: the lower bound is \eqref{eq:nonsmooth-rare-lower}, and the matching upper bound follows from $\Ex[Z_n^2]\leq A\KL_n$. A small signal has conditional mean and variance bounded by a multiple of $\KL_n$. Consider a block of $L_n=\lfloor r/\KL_n\rfloor$ consecutive signals for a fixed $r>0$. The event that the block contains exactly one rare signal, and that this signal has the sign in \eqref{eq:nonsmooth-positive-tail} or \eqref{eq:nonsmooth-negative-tail}, has probability bounded below by a constant $q_r>0$ that does not depend on $n$, while the small signals in the block move the log-odds by $O(r)$ in mean and $O(\sqrt r)$ in standard deviation. Therefore, over $m$ consecutive blocks, $m$ rare signals of one sign move one action's log-odds by at least $m\beta$, whereas with probability at least one half, all small signals together move either action's log-odds by at most $2\delta$ once $mr$ is small relative to $\delta^2$. Choosing $m$ large and then $r$ small, with every constant fixed independently of $n$, opens a belief gap larger than $\bar\lambda$ with probability bounded away from zero.\footnote{The blocks are needed because requiring $m$ consecutive rare signals instead would give probability of order $\KL_n^m$, so \cref{lem:escape-prior} would yield only a bound that vanishes with $n$}. Once the gap exceeds $\bar\lambda$, every type strictly prefers the favored action, and \cref{lem:trap,lem:abandon} ensure that abandonment of the other action holds with positive probability.

The two signs require different constructions because rare signals move beliefs in opposite directions. Under \eqref{eq:nonsmooth-positive-tail}, rare signals are good news, and we deliver $m$ of them to the action with the higher prior over the first $mL_n$ calendar dates. On an event of conditional probability at least one half, the small signals move neither action's log-odds by more than $2\delta$, so the favored action is chosen with probability at least $g>0$ at every date, and the uniform variables of the coupling in \cref{app:weakconvergence} let us force its selection at the $m$ scheduled dates at a cost of $g^m$. At date $mL_n$ the gap exceeds $\bar\lambda+\eta$, and \cref{lem:trap} applies. Under \eqref{eq:nonsmooth-negative-tail}, rare signals are bad news about a high-value action. We wait for the first date at which some action $I$ has belief at least $a>\bar\lambda$, which arrives almost surely, and then work on the unused signal streams, which by \cref{lem:fresh} are independent of the past and of each other. With probability bounded below, action $I$'s belief stays above $a_0<a$ (the maximal inequality of \cref{lem:nonsmooth-posterior}), and the next $mL_n$ signals in the stream of the other action $J$ contain $m$ rare negative signals with small signals totaling at most $2\delta$. On both events, action $J$ is either sampled fewer than $mL_n$ more times or, at its $mL_n$th further sample, its belief falls below $b$ while action $I$'s exceeds $a_0$, where $a_0-b>\bar\lambda$, so that no type chooses $J$ again. Either way $J$ is abandoned, and \cref{lem:abandon} gives $E^c$. Because the bad news waits on action $J$'s own stream whether or not $J$ is sampled, this construction needs no control over which action is selected, which is why it avoids the coupling.

\smallskip
\noindent\textit{Rare and small signals.}
Throughout this subcase we suppress the superscript $(n)$ on the variables of the coupling in \cref{app:weakconvergence} and index dates by $t$. Let $p_n:=\phi_n^1(|Z_n|\geq\beta)$ be the probability of a rare signal, and let $Q_n$ be the law of $Z_n$ conditional on $|Z_n|<\beta$, the law of a small signal. There is a constant $\bar c<\infty$, depending only on $A$ and $\beta$, such that for all large $n$
\begin{equation}
 p_n\leq\bar c\KL_n,
 \qquad
 \bigl|\Ex_{Q_n}Z_n\bigr|\leq\bar c\KL_n,
 \qquad
 \Ex_{Q_n}\bigl[Z_n^2\bigr]\leq\bar c\KL_n.
 \label{eq:nonsmooth-small-moments}
\end{equation}
Indeed, Markov's inequality and \eqref{eq:nonsmooth-second-moment} give $p_n\leq(A/\beta^2)\KL_n$, so $p_n\leq\tfrac12$ for large $n$, and $\Ex[|Z_n|\mathbf 1\{|Z_n|\geq\beta\}]\leq\Ex[Z_n^2]/\beta\leq(A/\beta)\KL_n$, so $|\Ex[Z_n\mathbf 1\{|Z_n|<\beta\}]|\leq(1+A/\beta)\KL_n$. Since also $\Ex[Z_n^2\mathbf 1\{|Z_n|<\beta\}]\leq A\KL_n$, dividing both bounds by $1-p_n\geq\tfrac12$ yields the last two inequalities with $\bar c:=2\max\{A/\beta^2,\,1+A/\beta,\,A\}$.

\smallskip
\noindent\textit{Constants.}
The constants chosen here depend on $(\mu_0,\nu,\bar\lambda,\beta,\bar c)$ but not on $n$. Relabel the actions so that $\mu_{10}\geq\mu_{20}$. Since $\mu_{20}<1-\bar\lambda$ and $\mu_{10}-\mu_{20}\geq0>-\bar\lambda$, continuity of $\ell$ allows us to choose $\delta>0$ such that
\begin{equation}
 \ell(X_{20}+2\delta)<1-\bar\lambda
 \quad\text{and}\quad
 \ell(X_{10}-2\delta)-\ell(X_{20}+2\delta)>-\bar\lambda,
\label{eq:nonsmooth-positive-room}
\end{equation}
and we set $\eta:=\tfrac12\bigl(1-\bar\lambda-\ell(X_{20}+2\delta)\bigr)>0$. Fix $a\in(\max\{\bar\lambda,\mu_{10},\mu_{20}\},1)$ and set $a_0:=(a+\bar\lambda)/2$ and $b:=(a_0-\bar\lambda)/2$, so that $a>a_0=\bar\lambda+2b>\bar\lambda$. Choose an integer $m$ large enough that
\begin{equation}
 \ell(X_{10}+m\beta-2\delta)-\ell(X_{20}+2\delta)>\bar\lambda+\eta
 \quad\text{and}\quad
 \mathsf L(a)-m\beta+2\delta<\mathsf L(b);
\label{eq:nonsmooth-targets}
\end{equation}
the first inequality is attainable because $\ell(x)\to1$ as $x\to\infty$ and $1-\ell(X_{20}+2\delta)=\bar\lambda+2\eta$. Then choose $r>0$ such that
\begin{equation}
 \bar c\,mr\leq\min\{\delta,\ \delta^2/4\},
\label{eq:nonsmooth-r}
\end{equation}
and let $L_n:=\lfloor r/\KL_n\rfloor$, so that $L_n\KL_n\leq r$ for every $n$ and $L_n\KL_n\geq r/2$ for all large $n$. Finally, the second inequality in \eqref{eq:nonsmooth-positive-room} and regular horizontal demand give
\begin{equation}
 g:=\Gamma_1\bigl(X_{10}-2\delta,\,X_{20}+2\delta\bigr)>0.
\label{eq:nonsmooth-positive-demand}
\end{equation}

\smallskip
\noindent\textit{Unbalanced blocks.}
Let $\sigma\in\{+,-\}$ be the sign for which \eqref{eq:nonsmooth-positive-tail} or \eqref{eq:nonsmooth-negative-tail} holds. Consider $L_n$ \emph{designated} draws together with at most $L_n$ further draws, all independent with the law of $Z_n$. Call the collection \emph{unbalanced} if exactly one designated draw $Z$ satisfies $\sigma Z\geq\beta$ and every other draw in the collection is small. For all large $n$, the collection is unbalanced with probability at least
\begin{equation}
 q_r:=\frac{cr}{4}\,e^{-4\bar cr}>0,
\label{eq:nonsmooth-block}
\end{equation}
and, conditional on the collection being unbalanced and on the position of the exceptional draw, the remaining draws are independent with law $Q_n$. The conditional statement holds because the conditioning event is a product over independent coordinates. For the bound, note that $p_n\leq\bar c\KL_n\leq\tfrac12$ and $\log(1-p)\geq-2p$ on $[0,\tfrac12]$, so any set of at most $2L_n-1$ draws is entirely small with probability at least $(1-p_n)^{2L_n-1}\geq e^{-4p_nL_n}\geq e^{-4\bar cr}$. The exceptional draw can occupy any of $L_n$ positions, and these events are disjoint; by \eqref{eq:nonsmooth-positive-tail} or \eqref{eq:nonsmooth-negative-tail}, their total probability is at least $L_n\cdot\tfrac c2\KL_n\cdot e^{-4\bar cr}\geq q_r$.

\smallskip
\noindent\textit{Positive rare signals.}
Suppose \eqref{eq:nonsmooth-positive-tail} holds. Recall the coupling of \cref{app:weakconvergence}: at each date $t$, two independent potential signals $(Z_{1,t},Z_{2,t})$ with the law of $Z_n$ and an independent uniform variable $U_t$ are drawn, action $1$ is selected when $U_t\leq\Gamma_1(\bm X_t)$, and only the selected action's signal is observed. Divide the first $mL_n$ dates into $m$ blocks of $L_n$ consecutive dates. Let $\mathcal C_n^+$ be the event that in every block exactly one potential signal of action $1$ satisfies $Z_{1,t}\geq\beta$, every other potential signal of either action in the block is small, and $U_t\leq g$ at the date of the exceptional signal. Call the set of $m$ exceptional dates the \emph{schedule} $\mathcal T_n^+$. The $m$ blocks involve disjoint sets of independent coordinates, and $U_t$ is independent of the signals, so the unbalanced-block bound \eqref{eq:nonsmooth-block}, with action $1$'s $L_n$ potential signals in the block designated, gives
\begin{equation}
 \Pr^{11}(\mathcal C_n^+)\geq(q_rg)^m.
\label{eq:nonsmooth-positive-block}
\end{equation}

Fix a value of the schedule and condition on $\mathcal C_n^+$ and on this value. The conditioning event is a product over the independent coordinates $(Z_{1,t},Z_{2,t},U_t)_t$, so under the conditional law the coordinates remain independent, every small potential signal has law $Q_n$, and every unscheduled $U_t$ remains uniform on $[0,1]$. Let $\mathscr A_t$ be the $\sigma$-algebra generated by $(U_s)_{s\leq t}$ and $(Z_{1,s},Z_{2,s})_{s<t}$. The action $i_t$ is $\mathscr A_t$-measurable because it is chosen before the date-$t$ signal is observed. For $t\leq mL_n$, define the centered sums of the observed small signals,
\[
 M_{1t}:=\sum_{s<t,\ s\notin\mathcal T_n^+}\mathbf 1\{i_s=1\}\bigl(Z_{1,s}-\Ex_{Q_n}Z_n\bigr),
 \qquad
 M_{2t}:=\sum_{s<t}\mathbf 1\{i_s=2\}\bigl(Z_{2,s}-\Ex_{Q_n}Z_n\bigr).
\]
Each signal entering these sums is independent of $\mathscr A_s$ with law $Q_n$, so $M_1$ and $M_2$ are square-integrable $(\mathscr A_t)$-martingales with $\Ex[M_{i,mL_n}^2]\leq mL_n\Ex_{Q_n}[Z_n^2]\leq\bar cmr$ by \eqref{eq:nonsmooth-small-moments}. Doob's maximal inequality \citep[Theorem~4.4.2]{durrett2019probability} applied to the submartingale $M_{it}^2$ gives, under the conditional law, $\Pr(\max_{t\leq mL_n}|M_{it}|>\delta)\leq\bar cmr/\delta^2\leq\tfrac14$ by \eqref{eq:nonsmooth-r}, and the centering removed from each sum has absolute value at most $mL_n|\Ex_{Q_n}Z_n|\leq\bar cmr\leq\delta$. Hence, with conditional probability at least $\tfrac12$, the sum of the observed small signals of each action lies in $[-2\delta,2\delta]$ at every date $t\leq mL_n$. Call this the \emph{standard signal event}.

On the standard signal event, at every date $t\leq mL_n$,
\[
 X_{1t}\geq X_{10}-2\delta
 \qquad\text{and}\qquad
 X_{2t}\leq X_{20}+2\delta,
\]
because the exceptional signals observed by date $t$ contribute nonnegative increments to $X_1$ and action $2$ has no exceptional signals. Since $\Gamma_1$ is nondecreasing in $x_1$ and nonincreasing in $x_2$, $\Gamma_1(\bm X_t)\geq g$ at every such date. At each scheduled date, $U_t\leq g$ therefore forces action $1$ to be selected, and all $m$ exceptional signals are observed. The first inequality in \eqref{eq:nonsmooth-targets} then gives $\mu_{1,mL_n}-\mu_{2,mL_n}>\bar\lambda+\eta$. The conditional bound $\tfrac12$ holds for every value of the schedule, so summing over the schedule and using \eqref{eq:nonsmooth-positive-block},
\begin{equation}
 \Pr^{11}(F_n^+)\geq\tfrac12(q_rg)^m,
 \qquad
 F_n^+:=\bigl\{\mu_{1,mL_n}-\mu_{2,mL_n}>\bar\lambda+\eta\bigr\}\in\mathscr H_{mL_n}.
\label{eq:nonsmooth-positive-escape}
\end{equation}
\cref{lem:trap}, applied with the deterministic stopping time $mL_n$ and the event $F_n^+$, gives $\Pr^{11}(E^c)\geq\eta\Pr^{11}(F_n^+)\geq\tfrac\eta2(q_rg)^m>0$ uniformly along the positive-tail subsequence.

\smallskip
\noindent\textit{Negative rare signals.}
Suppose instead that \eqref{eq:nonsmooth-negative-tail} holds. Let $\tau_n$ be the first date at which some posterior is at least $a$. This $(\mathscr H_t)$-stopping time is almost surely finite: some action is sampled infinitely often, and its log-odds equal its initial log-odds plus the partial sums of i.i.d.\ log-likelihood ratios with mean $\KL_n\in(0,\infty)$ (integrable because $Z_n\leq K$), which diverge to $+\infty$ by the strong law. Let $I$ be the action whose posterior is at least $a$ at $\tau_n$ and $J$ the other action. Only one posterior moves per date and $\tau_n$ is the first crossing, so $\mu_{I\tau_n}\geq a>\mu_{J\tau_n}$, and the labels are $\mathscr H_{\tau_n}$-measurable. By \cref{lem:fresh}(ii), in state $(1,1)$ the unused signal streams of $I$ and $J$ are independent of $\mathscr F_{\tau_n}$ and of each other, and each is i.i.d.\ with law $\phi_n^1$.

Let $(m_k)_{k\geq0}$ be the posterior sequence of \cref{lem:nonsmooth-posterior}(iii) started at $m_0=\mu_{I\tau_n}$ and driven by action $I$'s unused stream, and let $\mathcal G_n^I:=\{m_k>a_0\text{ for all }k\geq0\}$. Under full transparency, the posterior of action $I$ updates one-dimensionally on its own signals (see the discussion preceding \cref{lem:nonsmooth-posterior}), so $\mu_{It}=m_{N_I(t)-N_I(\tau_n)}$ for every $t\geq\tau_n$, and on $\mathcal G_n^I$ the posterior of action $I$ exceeds $a_0$ at every date from $\tau_n$ on. \cref{lem:nonsmooth-posterior}(iii), applied with $\mathscr G=\mathscr F_{\tau_n}$ and $\alpha=a_0$, gives
\begin{equation}
 \Pr^{11}\bigl(\mathcal G_n^I\mid\mathscr F_{\tau_n}\bigr)
 \geq1-\frac{1-\mu_{I\tau_n}}{1-a_0}
 \geq\frac{a-a_0}{1-a_0}=:q_I>0.
\label{eq:nonsmooth-high-stays-high}
\end{equation}

Let $Y_k$ be the log-likelihood ratio of the $k$th unused signal of action $J$, so that $(Y_k)_{k\geq1}$ is i.i.d.\ with the law of $Z_n$ and independent of $\mathscr F_{\tau_n}$ and of action $I$'s unused stream. Divide $Y_1,\dots,Y_{mL_n}$ into $m$ blocks of $L_n$ consecutive draws, and let $\mathcal B_n^J$ be the event that every block is unbalanced (exactly one draw with $Y_k\leq-\beta$, all other draws small) and that the sum of the $mL_n-m$ small draws lies in $[-2\delta,2\delta]$. By \eqref{eq:nonsmooth-block}, the block events have probability at least $q_r^m$, and conditional on them and on the exceptional positions the small draws are i.i.d.\ with law $Q_n$. Their sum then has mean of absolute value at most $mL_n\bar c\KL_n\leq\bar cmr\leq\delta$ and variance at most $\bar cmr\leq\delta^2/4$, by \eqref{eq:nonsmooth-small-moments} and \eqref{eq:nonsmooth-r}, so Chebyshev's inequality bounds the conditional probability that the sum leaves $[-2\delta,2\delta]$ by $\tfrac14$. Hence
\begin{equation}
 \Pr^{11}\bigl(\mathcal B_n^J\mid\mathscr F_{\tau_n}\bigr)\geq\tfrac34\,q_r^m=:q_J>0,
 \qquad
 \sum_{k=1}^{mL_n}Y_k\leq-m\beta+2\delta\quad\text{on }\mathcal B_n^J.
\label{eq:nonsmooth-negative-block}
\end{equation}
Conditional on $\mathscr F_{\tau_n}$, the events $\mathcal G_n^I$ and $\mathcal B_n^J$ are determined by independent streams and are therefore independent, so $\Pr^{11}(\mathcal G_n^I\cap\mathcal B_n^J\mid\mathscr F_{\tau_n})\geq q_Iq_J$.

On $\mathcal G_n^I\cap\mathcal B_n^J$, either action $J$ receives fewer than $mL_n$ further samples, so that it is sampled only finitely often, or it receives $mL_n$ further samples. In the latter case, immediately after its $mL_n$-th further sample its log-odds are at most $\mathsf L(\mu_{J\tau_n})-m\beta+2\delta<\mathsf L(a)-m\beta+2\delta<\mathsf L(b)$ by \eqref{eq:nonsmooth-negative-block} and the second inequality in \eqref{eq:nonsmooth-targets}, so $\mu_J<b$, while $\mu_I>a_0=\bar\lambda+2b$ on $\mathcal G_n^I$. For every type $\lambda\in\Lambda$, $\lambda_J+\mu_J<\bar\lambda+b<\mu_I\leq\lambda_I+\mu_I$, so the next agent chooses $I$; this leaves $\mu_J$ unchanged and, on $\mathcal G_n^I$, keeps $\mu_I$ above $a_0$. By induction, action $J$ is never chosen again. In either case $N_J(t)\not\to\infty$, and \cref{lem:abandon} gives $E^c$. Since $\tau_n<\infty$ almost surely,
\begin{equation}
 \Pr^{11}(E^c)
 \geq\Ex^{11}\bigl[\Pr^{11}(\mathcal G_n^I\cap\mathcal B_n^J\mid\mathscr F_{\tau_n})\bigr]
 \geq q_Iq_J>0
\label{eq:nonsmooth-negative-failure}
\end{equation}
uniformly along the negative-tail subsequence. This completes Subcase C.II and Case C.

\textbf{Taking stock.} Sufficiency follows from ODE convergence plus the maximal inequality of \cref{lem:nonsmooth-posterior}. Necessity has three cases: non-tangential path-balance failure (\cref{as:no-tangential-pb}) ensures that beliefs can separate along the ODE limit, nonvanishing drift ensures that beliefs can separate after finitely many signals, and nonvanishing noise leads to the abandonment of a high-value action either through a diffusion in the stochastically smooth case or through fixed-size rare-signal blocks in the nonsmooth case.

\section{Proofs of remaining results} \label{appendix:proof2}
\begin{proposition}[Local robustness of path balance]
\label{prop:path-balance-robustness}
Fix $\bar\lambda\in(0,1)$ and $(\mu_0^*,\nu^*)$ satisfying
the maintained assumptions of \cref{sec:model} and
path balance. There exists $\delta>0$ such that every
$(\mu_0,\nu)$ satisfying the maintained assumptions of \cref{sec:model} and
\[
\|\mu_0-\mu_0^*\|_2
+
\|\nu-\nu^*\|_{\mathrm{TV}}
\leq \delta
\]
also satisfies path balance.
\end{proposition}

\begin{proof}[Proof of \cref{prop:path-balance-robustness}]
Let $\bm\Gamma^*$ and $\bm\Gamma$ denote the choice probability mappings associated with $\nu^*$ and $\nu$, respectively. Let $\bm X^{\mathrm{PB},*}$ and $\bm X^{\mathrm{PB}}$ denote
the corresponding solutions of \eqref{eqn:path}, with initial
conditions $\mathsf L(\mu_0^*)$ and $\mathsf L(\mu_0)$,
respectively. Both solutions exist uniquely on $[0,\infty)$
by \cref{lem:path-ode-wellposed}.

We first bound the effect of perturbing the environment on any finite time interval. By \cref{lem:F-Lip}, there exists $K>0$, depending only on the reference environment, such that
\[
\|\bm\Gamma^*(\bm x)-\bm\Gamma^*(\bm y)\|_2
\leq K\|\bm x-\bm y\|_2
\qquad
\text{for all }\bm x,\bm y\in\mathbb R^2.
\]
Let $\eta:=\|\nu-\nu^*\|_{\mathrm{TV}}$. For each
$\bm x\in\mathbb R^2$, define the Borel set
\[
A_{\bm x}
:=
\left\{
\lambda\in\Lambda:
\lambda_1+\ell(x_1)\geq\lambda_2+\ell(x_2)
\right\}.
\]
By the definitions of the choice probabilities and total
variation distance,
\[
|\Gamma_1(\bm x)-\Gamma_1^*(\bm x)|
=
|\nu(A_{\bm x})-\nu^*(A_{\bm x})|
\leq\eta.
\]
Since $\Gamma_2=1-\Gamma_1$ and
$\Gamma_2^*=1-\Gamma_1^*$, it follows that
\[
\sup_{\bm x\in\mathbb R^2}
\|\bm\Gamma(\bm x)-\bm\Gamma^*(\bm x)\|_2
\leq\sqrt{2}\eta.
\]
Consequently, writing
\[
E(t)
:=
\|\bm X^{\mathrm{PB}}(t)-\bm X^{\mathrm{PB},*}(t)\|_2,
\qquad
d_0
:=
\|\mathsf L(\mu_0)-\mathsf L(\mu_0^*)\|_2,
\]
and subtracting the integral forms of the two ODEs gives
\[
E(t)
\leq
d_0+\int_0^t\bigl(KE(s)+\sqrt{2}\eta\bigr)\,ds.
\]
Gronwall's inequality therefore yields
\begin{equation}
E(t)
\leq
e^{Kt}d_0
+
\frac{\sqrt{2}\eta}{K}\bigl(e^{Kt}-1\bigr),
\qquad t\geq0.
\label{eq:pb-robustness-perturbation}
\end{equation}
See also \cite[Theorem~2.8]{teschl2012ordinary}.
Note that this bound uses the Lipschitz constant for
$\bm\Gamma^*$ only.

Since $(\mu_0^*,\nu^*)$ satisfies path balance, we can choose
$T>0$ such that
\[
r
:=
\min_{i\in\mathcal I}X_i^{\mathrm{PB},*}(T)
-
\mathsf L(1-\bar\lambda)
>0.
\]
The componentwise log-odds map $\mathsf L$ is continuous
at $\mu_0^*\in(0,1)^2$. Hence, there exists $\delta_0>0$
such that, for every $\mu_0\in(0,1)^2$,
\[
\|\mu_0-\mu_0^*\|_2\leq\delta_0
\quad\Longrightarrow\quad
\|\mathsf L(\mu_0)-\mathsf L(\mu_0^*)\|_2
\leq \frac{r}{4}e^{-KT}.
\]
Choose
\[
\delta
:=
\min\left\{
\delta_0,\,
\frac{rK}{4\sqrt{2}(e^{KT}-1)}
\right\}
>0.
\]
If
$\|\mu_0-\mu_0^*\|_2+\|\nu-\nu^*\|_{\mathrm{TV}}
\leq\delta$, then
\eqref{eq:pb-robustness-perturbation} implies
\[
E(T)\leq\frac r4+\frac r4=\frac r2.
\]
Therefore
\[
\min_{i\in\mathcal I}X_i^{\mathrm{PB}}(T)
\geq
\min_{i\in\mathcal I}X_i^{\mathrm{PB},*}(T)-E(T)
\geq
\mathsf L(1-\bar\lambda)+\frac r2,
\]
and thus
\[
c
:=
\min_{i\in\mathcal I}\ell(X_i^{\mathrm{PB}}(T))
>
1-\bar\lambda.
\]

It remains to show that this finite-time inequality
implies path balance. Both coordinates of
$\bm X^{\mathrm{PB}}$ are nondecreasing, because their
derivatives are choice probabilities. Hence, for every
$t\geq T$ and $i\in\mathcal I$,
\[
c\leq\ell(X_i^{\mathrm{PB}}(t))<1.
\]
Fix $i\in\mathcal I$ and let $j\neq i$. Every type satisfying
$\lambda_i-\lambda_j>1-c$ strictly prefers action $i$
at every $t\geq T$. It follows that
\[
\Gamma_i(\bm X^{\mathrm{PB}}(t))
\geq
q_i
:=
\nu\!\left(
\{\lambda\in\Lambda:\lambda_i-\lambda_j>1-c\}
\right),
\qquad t\geq T.
\]
Because $0<1-c<\bar\lambda$, the above set of types has Lebesgue measure
$(\bar\lambda-1+c)^2/2>0$. Since regular horizontal demand implies the lower bound $\underline f>0$ for the density of $\nu$,
\[
q_i
\geq
\frac{\underline f}{2}(\bar\lambda-1+c)^2
>0.
\]
Consequently,
\[
X_i^{\mathrm{PB}}(t)
\geq
X_i^{\mathrm{PB}}(T)+q_i(t-T)
\longrightarrow+\infty
\qquad\text{as }t\to\infty.
\]
This holds for both actions, so $(\mu_0,\nu)$ satisfies
path balance.
\end{proof}

\subsection{Proof of \cref{prop:goodnews}}

\begin{proof}[Proof of \cref{prop:goodnews}]
Let
\[
\Pr^{11}(\,\cdot\,)
:=
\Pr_{\sigma_{\mathsf{TR}},\phi,\mu_0,\nu}
\bigl(\,\cdot\mid\theta=(1,1)\bigr),
\]
and let $G$ and $q>0$ be as in \cref{def:perfect-good-news}. If the
current belief that a sampled action is high is $m\in(0,1)$ and the realized
signal $s$ lies in $G^c$, Bayes' rule gives
\[
m'(s)
=
\frac{m g_1(s)}{m g_1(s)+(1-m)g_0(s)}
\leq
m
\]
for $\rho$-almost every signal that can occur. A signal in $G$, by contrast,
raises that action's posterior to one. Let
\[
\tau
:=
\inf\{t\geq1:\max_i\mu_{it}=1\}.
\]
Before $\tau$, one fresh signal from a high-value action is generated in every
period. Conditional on every past history and on $\theta=(1,1)$, that signal
belongs to $G$ with probability $q$, irrespective of which action is chosen.
Hence
\[
\Pr^{11}(\tau>t)
=
(1-q)^t
\to
0,
\]
so $\tau<\infty$ $\Pr^{11}$-almost surely.

On $\{\tau<\infty\}$, let $I$ be the action whose conclusive signal produces
$\mu_{I\tau}=1$, and let $J\neq I$. Every signal observed about action $J$
before $\tau$ was nonconclusive and therefore weakly lowered its belief.
Consequently,
\[
\mu_{J\tau}
\leq
\mu_{J0}
<
1-\bar\lambda.
\]
For each $\lambda\in\Lambda$,
\[
(\lambda_I+\mu_{I\tau})-(\lambda_J+\mu_{J\tau})
\geq
1-\bar\lambda-\mu_{J\tau}
>
0.
\]
Thus every type strictly prefers action $I$ at date $\tau$. By induction, the same holds at every date $t\geq\tau$: action $I$'s
posterior remains one, while the posterior of the unsampled action
$J$ remains equal to $\mu_{J\tau}$. Hence, action $J$ is never chosen after $\tau$, so $N_J(t)\not\to\infty$, and $E$ fails $\Pr^{11}$-almost surely by \cref{lem:abandon}. We conclude that
$\Pr^{11}(E)=0$.
\end{proof}

\subsection{Proof of \cref{thrm:exp_convergence}}

We first record the two-sided maximal Azuma--Hoeffding
inequality.\footnote{We state this standard inequality without proof. See, e.g., \citet[Theorem~3.2.1]{roch2024modern}.}

\begin{lemma}[Maximal Azuma--Hoeffding inequality]\label{thm:azuma}
Let $(W_t,\mathcal G_t)_{t\geq0}$ be a martingale with $W_0=0$, and let
$(c_t)_{t\geq1}$ be a deterministic sequence of nonnegative constants such
that
\[
|W_t-W_{t-1}|\leq c_t
\qquad\text{almost surely.}
\]
Then, for each integer $T\geq1$ and every $x>0$,
\[
\Pr\!\left(\max_{1\leq t\leq T}|W_t|\geq x\right)
\leq
2\exp\!\left(-\frac{x^2}{2\sum_{s=1}^T c_s^2}\right),
\]
with the right-hand side interpreted as zero when
$\sum_{s=1}^T c_s^2=0$.
\end{lemma}

\noindent\textit{Proof sketch.}
We condition on $\theta=(1,1)$, the only state in which efficiency can fail (\cref{lem:sources}). Under binary bad news, a high-value action's log-odds rise by the fixed amount $a_p=\log\frac1{1-p}$ each time it is chosen, so the public log-odds equal $\mathsf L(\mu_0)+a_p(N_1(t),N_2(t))$: beliefs are a deterministic function of the sampling counts, and the only randomness is in which action each agent chooses. Because exactly one action is chosen per period, the sum of the two log-odds is deterministic, and it coincides with the coordinate sum of the path-balance solution $\bm X^{\mathrm{PB}}$ once time is measured in units of $a_p$. The deviation of beliefs from the path is therefore the scalar $e_t=X_{1t}-X_1^{\mathrm{PB}}(ta_p)$. It is the sum of three terms: a martingale $W_t$ (the compensated count of action-$1$ choices whose increments are bounded by $a_p$), a feedback term (which the Lipschitz constant $C_\Gamma$ of the choice probability $\Gamma_1$ bounds by $C_\Gamma a_p\sum_{s<t}|e_s|$), and a discretization error of order $a_p^2$ per period.
 
The exponential rate comes from the martingale term. Path balance gives a finite time $S$ at which both coordinates of $\bm X^{\mathrm{PB}}$ exceed a target $\mathsf L(\beta)+\delta$, where $\beta>1-\bar\lambda$. The horizon $S$ is fixed in ODE time but takes $N_p\approx S/a_p$ periods in calendar time, which is the source of the $1/p$ in the exponent. Over this horizon, the martingale accumulates $N_pa_p^2\approx Sa_p\approx Sp$ in squared increments, so the Azuma--Hoeffding inequality bounds the probability that it exceeds a fixed level at some date $t\leq N_p$ by $2\exp(-\text{const}/p)$. A discrete Gronwall inequality absorbs the feedback term over the fixed horizon with the constant $C_0=e^{C_\Gamma(S+1)}$, and the accumulated discretization error, of order $Sa_p$, falls below the tolerance once $p<p_0$. Hence, with probability at least $1-2e^{-K/p}$, the log-odds stay within $\delta$ of the path, in sup norm, through date $N_p$, at which both beliefs are at least $\beta$. Beliefs never decrease in state $(1,1)$, so both then remain in $[\beta,1]$: the belief gap never exceeds $1-\beta<\bar\lambda$, both actions keep being chosen with positive probability, both are sampled infinitely often, and both beliefs converge to one (\cref{lem:sources-general}). The constants $K$ and $p_0$ depend on $(\mu_0,\bar\lambda,\nu)$ only through $S$, $C_\Gamma$, and the fixed choice of $(\beta,\delta)$.

\begin{proof}[Proof of \cref{thrm:exp_convergence}]
Throughout, $\Pr^{11}$ and $\Ex^{11}$ refer to the environment $(\sigma_{\mathsf{TR}},\phi^p,\mu_0,\nu)$ conditional on $\theta=(1,1)$. In this state, binary bad news never arrives, so each time an action is chosen its log-odds increase by
\[
a_p:=\log\frac{1}{1-p}>0,
\]
and its posterior never decreases. Writing $\bm X_t:=\mathsf L(\mu_t)$ for the log-odds of the public beliefs,
\begin{equation}\label{eq:exp-counts}
 X_{it}=\mathsf L(\mu_{i0})+a_pN_i(t),\qquad i=1,2,
\end{equation}
so the belief path is a deterministic function of the sampling counts, and the only randomness is in which action each agent chooses.
 
\smallskip
\noindent\textit{Constants.}
Define the level $\beta:=1-\bar\lambda/2\in(1-\bar\lambda,1)$ and the tolerance $\delta:=1$. Each coordinate of $\bm\Gamma$ is nonnegative, so $\bm X^{\mathrm{PB}}$ is coordinatewise nondecreasing, and by path balance both coordinates diverge. Hence
\[
S:=\inf\bigl\{s\geq0:\min_{i}X_i^{\mathrm{PB}}(s)\geq\mathsf L(\beta)+\delta\bigr\}
\]
is finite, the infimum is attained because $\bm X^{\mathrm{PB}}$ is continuous, and $X_i^{\mathrm{PB}}(s)\geq\mathsf L(\beta)+\delta$ for both $i$ and all $s\geq S$. Moreover, $S>0$ because $\mu_{i0}<1-\bar\lambda<\beta$ for both $i$. By \cref{lem:F-Lip} and $\|\cdot\|_2\leq\sqrt2\,\|\cdot\|_\infty$ on $\mathbb R^2$, the choice probability $\Gamma_1$ is Lipschitz with respect to $\|\cdot\|_\infty$ with constant $C_\Gamma:=\overline f\bar\lambda$, where $\overline f$ is the essential supremum of $f_\nu$ (any larger constant also works). Let
\[
N_p:=\left\lceil\frac{S}{a_p}\right\rceil,
\qquad
C_0:=e^{C_\Gamma(S+1)},
\qquad
K:=\frac{\delta^2}{16C_0^2(S+1)},
\]
and choose $p_0\in(0,\tfrac12]$ such that $C_\Gamma(S+1)a_p\leq\delta/C_0$ for all $p<p_0$, which is possible because $a_p\to0$ as $p\to0$. The constants $S$, $C_\Gamma$, $C_0$, $K$, and $p_0$ depend only on $(\mu_0,\bar\lambda,\nu)$. Fix $p<p_0$. Since $p\leq\tfrac12$, $a_p\leq p/(1-p)\leq2p\leq1$. Hence, $N_p\geq1$, $S\leq N_pa_p<S+1$, and $N_pa_p^2\leq2(S+1)p$.
 
\smallskip
\noindent\textit{Step 1: the deviation from the path is a martingale plus a Lipschitz drift.}
Because $\Gamma_1+\Gamma_2=1$, the path satisfies $X_1^{\mathrm{PB}}(s)+X_2^{\mathrm{PB}}(s)=\mathsf L(\mu_{10})+\mathsf L(\mu_{20})+s$, while \eqref{eq:exp-counts} and $N_1(t)+N_2(t)=t$ give $X_{1t}+X_{2t}=\mathsf L(\mu_{10})+\mathsf L(\mu_{20})+ta_p$. The deviation of the beliefs from the path at matched times therefore has zero coordinate sum: with
\[
e_t:=X_{1t}-X_1^{\mathrm{PB}}(ta_p),
\qquad
\bm X_t-\bm X^{\mathrm{PB}}(ta_p)=e_t(1,-1)
\quad\text{and}\quad
\|\bm X_t-\bm X^{\mathrm{PB}}(ta_p)\|_\infty=|e_t|.
\]
Subtracting the integral form of \eqref{eqn:path}, $X_1^{\mathrm{PB}}(ta_p)=\mathsf L(\mu_{10})+\int_0^{ta_p}\Gamma_1(\bm X^{\mathrm{PB}}(u))\,du$, from \eqref{eq:exp-counts} and splitting the integral over periods,
\begin{equation}\label{eq:exp-decomposition}
 e_t
 =W_t
 +a_p\sum_{s<t}\bigl[\Gamma_1(\bm X_s)-\Gamma_1(\bm X^{\mathrm{PB}}(sa_p))\bigr]
 +\sum_{s<t}\int_{sa_p}^{(s+1)a_p}\bigl[\Gamma_1(\bm X^{\mathrm{PB}}(sa_p))-\Gamma_1(\bm X^{\mathrm{PB}}(u))\bigr]\,du,
\end{equation}
where
\[
 W_t:=a_p\sum_{s<t}\bigl(\mathbf 1\{i_s=1\}-\Gamma_1(\bm X_s)\bigr),
 \qquad W_0=0 .
\]
The three terms are a martingale, a feedback term, and a discretization error. First, $W_t$ is a function of $(i_s,\mu_s)_{s<t}$ and hence $\mathscr H_t$-measurable, and by \eqref{eq:choice-prob} $\Ex^{11}[\mathbf 1\{i_s=1\}\mid\mathscr H_s]=p_1(\mu_s)=\Gamma_1(\bm X_s)$, so $(W_t,\mathscr H_t)$ is a martingale with increments in $[-a_p,a_p]$. Second, the Lipschitz bound and $\|\bm X_s-\bm X^{\mathrm{PB}}(sa_p)\|_\infty=|e_s|$ bound the feedback term by $C_\Gamma a_p\sum_{s<t}|e_s|$. Third, since $\|\dot{\bm X}^{\mathrm{PB}}\|_\infty\leq1$, each integrand is at most $C_\Gamma(u-sa_p)$, so each summand is at most $C_\Gamma a_p^2/2$. Hence
\begin{equation}\label{eq:exp-recursion}
 |e_t|\leq|W_t|+C_\Gamma a_p\sum_{s<t}|e_s|+\frac{C_\Gamma}{2}\,ta_p^2 .
\end{equation}
 
\smallskip
\noindent\textit{Step 2: concentration over the horizon $N_p$.}
Let $\mathcal A_p:=\{\max_{t\leq N_p}|W_t|<\delta/(2C_0)\}$. \cref{thm:azuma} with $T=N_p$, $c=a_p$, and $x=\delta/(2C_0)$ gives
\begin{equation}\label{eq:exp-concentration-bound}
 \Pr^{11}(\mathcal A_p^c)
 \leq2\exp\!\left(-\frac{\delta^2}{8C_0^2N_pa_p^2}\right)
 \leq2\exp\!\left(-\frac Kp\right),
\end{equation}
using $N_pa_p^2\leq2(S+1)p$. On $\mathcal A_p$, for every $t\leq N_p$, the last term in \eqref{eq:exp-recursion} is at most $\tfrac{C_\Gamma}{2}(S+1)a_p\leq\delta/(2C_0)$ by the choice of $p_0$, so $|e_t|\leq\delta/C_0+C_\Gamma a_p\sum_{s<t}|e_s|$. By induction on $t$ (the discrete Gronwall inequality), $|e_t|\leq(\delta/C_0)(1+C_\Gamma a_p)^t\leq(\delta/C_0)e^{C_\Gamma ta_p}\leq\delta$ for all $t\leq N_p$, since $ta_p<S+1$.
 
\smallskip
\noindent\textit{Step 3: the target region is absorbing.}
On $\mathcal A_p$, since $N_pa_p\geq S$ and $\bm X^{\mathrm{PB}}$ is coordinatewise nondecreasing, for each action $i$,
\[
X_{i,N_p}\geq X_i^{\mathrm{PB}}(N_pa_p)-|e_{N_p}|\geq X_i^{\mathrm{PB}}(S)-\delta\geq\mathsf L(\beta),
\]
so $\mu_{i,N_p}\geq\beta$. Beliefs are nondecreasing in state $(1,1)$, so $\mu_{it}\in[\beta,1]$ for both $i$ and all $t\geq N_p$, and the limit beliefs satisfy $|\mu_{1\infty}-\mu_{2\infty}|\leq1-\beta<\bar\lambda$. By \eqref{eq:positive-limit-demand}, $p_i(\mu_\infty)>0$ for both $i$. Full transparency satisfies sampled-action revelation (\cref{lem:sar}), so \cref{lem:sources-general}(ii) implies that both actions are sampled infinitely often and $E$ occurs, $\Pr^{11}$-almost surely on $\mathcal A_p$. Hence $\Pr^{11}(E^c)\leq\Pr^{11}(\mathcal A_p^c)\leq2e^{-K/p}$ by \eqref{eq:exp-concentration-bound}, and \cref{lem:sources} gives
\[
1-V(\sigma_{\mathsf{TR}},\phi^p,\mu_0,\nu)=\mu_{10}\mu_{20}\Pr^{11}(E^c)\leq2\exp\!\left(-\frac Kp\right).
\]
\end{proof}

\subsection{Proof of \cref{thrm:evidence_iff}}

Part (ii) of \cref{thrm:evidence_iff} and the upper bound in \cref{thrm:adaptiveinfo} both follow from the next lemma.

\begin{lemma}[Upper bound under strict imbalance]\label{lem:ID-upper}
Fix an experiment with $\phi^1\neq\phi^0$, and suppose $a>c$, where $a=\max_i\mu_{i0}$, $b=\min_i\mu_{i0}$, and $c=b+\bar\lambda$. Then every $\sigma\in\Sigma_{\mathsf{ID}}$ satisfies
\[
V(\sigma,\phi,\mu_0,\nu)\leq1-\frac{b(a-c)}{1-c}<1.
\]
\end{lemma}

\begin{proof}[Proof of \cref{lem:ID-upper}]
Relabel the actions so that $\mu_{10}=a$ and $\mu_{20}=b$, fix $\sigma\in\Sigma_{\mathsf{ID}}$, and let
$\tau:=\inf\{t\in\mathcal T:\mu_{1t}\leq c\}$, which is an $(\mathscr H_t)$-stopping time. The maintained restriction on the prior gives $0<b<c<a<1$.

First, consider the auxiliary process in which action~$1$ is chosen at every date regardless of types, with the same signal streams and the same policy randomization. Let $\widetilde H_t$ be its public history, $\widetilde\mu_t$ the associated posterior marginals, and $\widetilde\tau:=\inf\{t:\widetilde\mu_{1t}\leq c\}$. The auxiliary history is a function of action~$1$'s signal stream and of the randomization, so it is independent of $\theta_2$ conditional on $\theta_1$. Hence $\widetilde\mu_{2t}=b$ for every $t$ and
\[
\Pr(\widetilde\tau<\infty\mid\theta=(1,1))
=\Pr(\widetilde\tau<\infty\mid\theta_1=1).
\]
We claim that the true and the auxiliary process coincide until $\widetilde\tau$. Let $\mathsf C_t\subseteq\mathsf H_t$ be the set of date-$t$ public histories in which every action equals~$1$ and $\widetilde\mu_{1s}>c$ at every prefix of length $s<t$, so that $\{\widetilde\tau\geq t\}=\{\widetilde H_t\in\mathsf C_t\}$. We show by induction on $t$ that, up to null sets, $\{H_t\in\mathsf C_t\}=\{\widetilde H_t\in\mathsf C_t\}$ and $H_t=\widetilde H_t$ on this event. This is trivial for $t=0$. Suppose it holds for $t$. Then the laws of $(\theta,H_t)$ and $(\theta,\widetilde H_t)$ coincide on $\Theta\times\mathsf C_t$, so the posteriors computed from the two histories agree there: $\mu_t=\widetilde\mu_t=(\widetilde\mu_{1t},b)$ on $\{H_t\in\mathsf C_t\}$. On the event $\{H_t\in\mathsf C_t,\ \widetilde\mu_{1t}>c\}$ we therefore have $\mu_{1t}-\mu_{2t}>c-b=\bar\lambda$, so every type strictly prefers action~$1$; thus $i_t=1$, and the same record and the same message are generated in both processes, so that $H_{t+1}=\widetilde H_{t+1}\in\mathsf C_{t+1}$. Conversely, every history in $\mathsf C_{t+1}$ has its date-$t$ prefix in $\mathsf C_t$ with $\widetilde\mu_{1t}>c$, which gives the claim for $t+1$. Consequently $\tau=\widetilde\tau$ almost surely, $\mu_{2t}=b$ and $i_t=1$ for every $t<\tau$, and
\[
\Pr(\tau<\infty\mid\theta=(1,1))
=\Pr(\tau<\infty\mid\theta_1=1)
\leq\frac{c(1-a)}{a(1-c)},
\]
where the inequality is \cref{lem:hitting}(i).

Finally, observe that on $\{\tau=\infty\}$, action~$2$ is never chosen, so $N_2(t)\not\to\infty$ and $E$ fails $\Pr^{11}$-almost surely by \cref{lem:abandon}. Because $\Pr(\theta=(1,1))=ab$,
\[
1-V(\sigma,\phi,\mu_0,\nu)
\geq ab\,\Pr(\tau=\infty\mid\theta=(1,1))
\geq ab\left(1-\frac{c(1-a)}{a(1-c)}\right)
=\frac{b(a-c)}{1-c}>0.
\]
\end{proof}

\begin{proof}[Proof of \cref{thrm:evidence_iff}]
(i) If $\mu_0$ satisfies static balance, then $\mathcal K:=\{\mu_0\}$ is a compact subset of $\{\mu\in(0,1)^2:|\mu_1-\mu_2|<\bar\lambda\}$, and \cref{lem:accumulation}(i) gives $V(\sigma^{\mathsf{ACC},T},\phi,\mu_0,\nu)\to1$ as $T\to\infty$. Each accumulation policy belongs to $\Sigma_{\mathsf{EV}}$.

(ii) If $\mu_0$ does not satisfy static balance, then $|\mu_{10}-\mu_{20}|>\bar\lambda$ by \cref{as:generic-design}, that is, $a>c$. The first inequality in (ii) holds because $\Sigma_{\mathsf{EV}}\subseteq\Sigma_{\mathsf{ID}}$, and the second follows from \cref{lem:ID-upper}, whose bound is uniform over $\sigma\in\Sigma_{\mathsf{ID}}$.
\end{proof}

\subsection{Proof of \cref{thrm:adaptiveinfo}}

\begin{proof}[Proof of \cref{thrm:adaptiveinfo}]
Relabel the actions so that $\mu_{10}=a$ and $\mu_{20}=b$. \cref{lem:ID-upper} gives the upper bound. For the lower bound, we construct policies whose value approaches it. The obstacle is that the designer must lower the public belief about action~$1$ to a point strictly inside the balanced region, with the largest probability that Bayes plausibility allows, even though she observes only a finite sample of signals about action~$1$. She does so by waiting until her posterior is nearly degenerate and then garbling it into two messages.

Fix $\varepsilon>0$ and choose
\[
 \underline m\in(\max\{0,b-\bar\lambda\},c),
 \qquad \overline m\in(a,1),
 \qquad q:=\frac{\overline m-a}{\overline m-\underline m}\in(0,1),
\]
so that $q\underline m+(1-q)\overline m=a$.

For an integer $n\geq1$, the designer sends the null message $\varnothing$ at dates $0,\ldots,n-2$. Null messages are deterministic, so the public belief stays at $\mu_0$, and because $a-b>\bar\lambda$ every type chooses action~$1$ at each date $t\leq n-1$. Let $P_n:=\Pr(\theta_1=1\mid s_{11},\ldots,s_{1n})$ be the designer's posterior after the first $n$ action-$1$ signals, and let
\[
v_n:=\Ex[P_n(1-P_n)],
\qquad\text{so that}\qquad
\Ex[P_n]=a
\quad\text{and}\quad
\Ex[P_n^2]=a-v_n.
\]
Choose a Borel set $A$ with $\phi^0(A)\neq\phi^1(A)$. By the strong law of large numbers, the empirical frequency of $A$ along action~$1$'s stream reveals $\theta_1$ almost surely, so L\'evy's upward theorem gives $P_n\to\theta_1$ almost surely, and bounded convergence gives $v_n\to\Ex[\theta_1(1-\theta_1)]=0$. Fix $n$ large enough that
\begin{equation}\label{eq:t4-feasible}
v_n\leq\min\bigl\{\underline m(1-a),\,a(1-\overline m)\bigr\}.
\end{equation}

At date $n-1$, conditional on $P_n$ and using independent randomization, the designer sends the message $M=\underline m$ with probability $\alpha(P_n)$ and the message $M=\overline m$ otherwise, where
\[
\alpha(P):=xP+y(1-P),
\qquad
x:=q\,\frac{\underline m(1-a)-v_n}{a(1-a)-v_n},
\qquad
y:=q\,\frac{a(1-\underline m)-v_n}{a(1-a)-v_n}.
\]
The denominators are strictly positive because $v_n\leq\underline m(1-a)<a(1-a)$. Because $\underline m<a$, we have $x\leq y$. Condition \eqref{eq:t4-feasible} gives $x\geq0$; and $y\leq1$ is equivalent to $v_n(1-q)\leq a\bigl[(1-a)-q(1-\underline m)\bigr]=a(1-q)(1-\overline m)$, which is again \eqref{eq:t4-feasible}. Hence $\alpha(P)\in[x,y]\subseteq[0,1]$ for every $P\in[0,1]$. Since $M$ is independent of $\theta$ conditional on the signals,
\begin{align*}
\Pr(M=\underline m)
&=\Ex[\alpha(P_n)]
=xa+y(1-a)
=q,\\
\Pr(\theta_1=1,M=\underline m)
&=\Ex[P_n\alpha(P_n)]
=x(a-v_n)+yv_n
=q\underline m,
\end{align*}
where the last equality in each line is a direct computation. Thus $\Pr(\theta_1=1\mid M=\underline m)=\underline m$, and Bayes plausibility gives $\Pr(\theta_1=1\mid M=\overline m)=(a-q\underline m)/(1-q)=\overline m$. The public history at date $n$ consists of $n$ choices of action~$1$, null messages, and $M$, and $M$ depends only on action~$1$'s signals and the randomization. Hence, the public posterior at date $n$ is the product measure with marginals $(M,b)$.

After the message $\underline m$, the designer applies the accumulation policy with horizon $T$ to the records generated from date $n$ on. Because $\underline m\in(0,1)$ and $|\underline m-b|<\bar\lambda$, \cref{lem:accumulation}(i) with $\mathcal K=\{(\underline m,b)\}$ yields a horizon $T$ with $1-V(\sigma^{\mathsf{ACC},T},\phi,(\underline m,b),\nu)\leq\varepsilon$, and \cref{lem:accumulation}(ii) with $\tau=n$ and $F=\{M=\underline m\}$ gives $\Pr(E^c\mid\mathscr H_n)\leq\varepsilon$ on $F$. After the message $\overline m$, the designer sends null messages forever. Because $\overline m-b>\bar\lambda$, action~$1$ is then chosen at every date, which is efficient in state $(1,0)$ because $1+\lambda_1>\lambda_2$ for every type. We count every realization with $\theta_1=0$ or $\theta=(1,1)$ on this branch as a failure. Since $\theta_2$ is independent of $(\theta_1,M)$,
\[
 \Pr(\theta_1=0,M=\overline m)=(1-\overline m)(1-q),
 \qquad
 \Pr(\theta=(1,1),M=\overline m)=b\,\overline m(1-q)=b(a-\underline mq).
\]
Writing $\widehat\sigma$ for the resulting policy,
\[
 V(\widehat\sigma,\phi,\mu_0,\nu)
 \geq1-b(a-\underline m q)-(1-\overline m)(1-q)-q\varepsilon.
\]
Letting $\underline m\uparrow c$ and $\overline m\uparrow1$ gives $q\to(1-a)/(1-c)$, so that
\[
 \underline m q\to\frac{c(1-a)}{1-c},
 \qquad
 (1-\overline m)(1-q)\to0,
 \qquad
 b(a-\underline mq)\to\frac{b(a-c)}{1-c}.
\]
Thus, the supremum is at least
$1-b(a-c)/(1-c)-\varepsilon$. Letting $\varepsilon\downarrow0$ and
combining with the upper bound proves the result.
\end{proof}

\subsection{Proof of \texorpdfstring{\cref{thrm:adaptiveinfoB}}{Theorem 4B}}

\begin{proof}[Proof of \texorpdfstring{\cref{thrm:adaptiveinfoB}}{Theorem 4B}]
Relabel the actions so that $\mu_{10}=a$ and $\mu_{20}=b$. The upper bound is immediate: for every $\delta$ and every $\phi\in\Phi_\delta$,
\begin{equation}\label{eq:t4b-upper}
\sup_{\sigma\in\Sigma_{\mathsf{EV}}}
V(\sigma,\phi,\mu_0,\nu)
\leq
\sup_{\sigma\in\Sigma_{\mathsf{ID}}}
V(\sigma,\phi,\mu_0,\nu)
=
1-\frac{b(a-c)}{1-c}
\end{equation}
by \cref{thrm:adaptiveinfo}. For the lower bound, an adaptive evidence policy cannot garble, so it cannot stop the public belief exactly where it wishes. Instead, it discloses evidence about action~$1$ until the public belief first falls below $c-\delta$. $\delta$-boundedness limits the overshoot, so the belief stops strictly inside the balanced region, and the designer then accumulates.

Fix
$0<\delta<\frac12\min\{c,\bar\lambda\}$, an experiment $\phi\in\Phi_\delta$, and
$\varepsilon>0$. Let $I_\delta:=[c-2\delta,c-\delta]$ and $\mathcal K:=I_\delta\times\{b\}$. Every $\mu\in\mathcal K$ is interior and satisfies $-\bar\lambda<\bar\lambda-2\delta\leq\mu_1-\mu_2\leq\bar\lambda-\delta<\bar\lambda$, so \cref{lem:accumulation}(i) yields a horizon $T=T(\phi,\delta,\varepsilon)$ with
\begin{equation}\label{eq:t4b-horizon}
\sup_{\mu\in\mathcal K}\bigl[1-V(\sigma^{\mathsf{ACC},T},\phi,\mu,\nu)\bigr]\leq\varepsilon.
\end{equation}

Let $m_0:=a$ and, for $t\geq0$,
\[
m_{t+1}:=
\begin{cases}
B_\phi(m_t,s_t), & \text{if } i_t=1 \text{ and } m_s>c-\delta \text{ for every } s\leq t,\\
m_t, & \text{otherwise,}
\end{cases}
\]
and let $\tau_\delta:=\inf\{t\in\mathcal T:m_t\leq c-\delta\}$, where $B_\phi$ is the one-signal Bayesian update from the definition of $\delta$-boundedness, extended by $B_\phi(m,s):=m$ for $m\in\{0,1\}$. Because $a>c$, $\tau_\delta\geq1$. At each date $t<\tau_\delta$, the designer discloses the record $x_t$ if $i_t=1$ and withholds it if $i_t=2$, where records withheld in this first stage are never disclosed. From date $\tau_\delta$ on, she applies the accumulation policy with horizon $T$ to the records generated from date $\tau_\delta$ on, carrying forward the action-$1$ records already disclosed. The index $m_t$ depends only on the actions and on the action-$1$ signals disclosed before date $t$, so it is a function of the public history $H_t$, $\tau_\delta$ is an $(\mathscr H_t)$-stopping time, and the policy is a well-defined sequence of kernels. Disclosure is nested and no record is garbled, so this policy, denoted $\sigma^\star$, belongs to $\Sigma_{\mathsf{EV}}$. Write $\Pr^\star$ for the law of the environment $(\sigma^\star,\phi,\mu_0,\nu)$.

Through date $\tau_\delta$, choices depend on the public history and on types, which are independent of $\theta$, and the only signals disclosed are action-$1$ signals. The likelihood of a first-stage public history therefore depends on $\theta$ only through $\theta_1$. Consequently, by induction on $t$, for each $t\leq\tau_\delta$ the public posterior is the product measure with marginals $(m_t,b)$: the belief about action~$1$ evolves by Bayesian updating from directly observed action-$1$ signals, and the belief about action~$2$ does not move. In particular, $\mu_{1t}=m_t$ for $t\leq\tau_\delta$, so $\tau_\delta$ is the first date at which the public belief about action~$1$ is at most $c-\delta$. Moreover, the first-stage history is a function of the exogenous draws and of action~$1$'s stream, so that
\begin{equation}\label{eq:t4b-state-independence}
\Pr^\star(\tau_\delta<\infty\mid\theta=(1,1))
=
\Pr^\star(\tau_\delta<\infty\mid\theta_1=1).
\end{equation}

On $\{\tau_\delta=\infty\}$, $\mu_{1t}>c-\delta$ at every date, so monotonicity of $p_1$ and \eqref{eq:positive-limit-demand} give
$p_1(\mu_t)\geq p_1(c-\delta,b)>0$. The conditional Borel--Cantelli lemma implies that action~$1$ is sampled infinitely often on $\{\tau_\delta=\infty\}$. Every action-$1$ record is eventually disclosed under $\sigma^\star$, so \cref{lem:sar} gives $\mu_{1t}\to\theta_1$ on this event. Since $\mu_{1t}>c-\delta>0$ there,
\begin{equation}\label{eq:t4b-no-hit-limit}
\Pr^\star(\tau_\delta=\infty,\theta_1=0)=0
\qquad\text{and}\qquad
\mu_{1t}\to1
\quad\text{almost surely on }\{\tau_\delta=\infty\}.
\end{equation}

On $\{\tau_\delta<\infty\}$, the belief $\mu_1$ exceeds $c-\delta$ at date $\tau_\delta-1$, and the update that crosses the threshold is the
Bayesian update from one directly observed signal. By
$\delta$-boundedness,
\begin{equation}\label{eq:t4b-overshoot}
c-2\delta
<
\mu_{1\tau_\delta}
\leq
c-\delta.
\end{equation}
By \eqref{eq:t4b-no-hit-limit},
$\Pr^\star(\tau_\delta<\infty\mid\theta_1=0)=1$. \cref{lem:hitting}(ii), applied with the threshold $c-\delta\in(0,a)$ and with $c'=c-2\delta>0$, therefore gives
\begin{equation}\label{eq:t4b-high-hit}
\Pr^\star(\tau_\delta<\infty\mid\theta_1=1)
\geq
\frac{(c-2\delta)(1-a)}
     {a(1-c+2\delta)}.
\end{equation}

On $F:=\{\tau_\delta<\infty\}\in\mathscr H_{\tau_\delta}$, the public posterior is a product measure with marginals $\mu_{\tau_\delta}\in\mathcal K$ by \eqref{eq:t4b-overshoot}, and the continuation messages depend only on records generated from date $\tau_\delta$ on. \cref{lem:accumulation}(ii) and \eqref{eq:t4b-horizon} give
\begin{equation}\label{eq:t4b-hit-branch}
\Pr^\star(E^c\mid\mathscr H_{\tau_\delta})\leq\varepsilon
\qquad\text{almost surely on }\{\tau_\delta<\infty\}.
\end{equation}
This bounds the total continuation failure probability,
averaged over states using the public posterior.

On $\{\tau_\delta=\infty\}$, \eqref{eq:t4b-no-hit-limit} gives $\theta_1=1$ and $\mu_{1t}\to1$ almost surely, while $\mu_{2t}=b$ at every date. Because $1-b>\bar\lambda$, the belief gap eventually exceeds $\bar\lambda$, after which every type chooses action~$1$. In state $(1,0)$ this choice is optimal for every type, because $1+\lambda_1>\lambda_2$, so regret is eventually zero and $E$ occurs. Hence, up to a null set, $E^c\cap\{\tau_\delta=\infty\}\subseteq\{\theta=(1,1),\tau_\delta=\infty\}$.

Combining the two branches with \eqref{eq:t4b-state-independence} and \eqref{eq:t4b-high-hit},
\begin{equation}\label{eq:t4b-value-lower}
\begin{aligned}
1-V(\sigma^\star,\phi,\mu_0,\nu)
&\leq
ab\left[
 1-\Pr^\star(\tau_\delta<\infty\mid\theta_1=1)
\right]+\varepsilon \\
&\leq
ab\left[
 1-
 \frac{(c-2\delta)(1-a)}
      {a(1-c+2\delta)}
\right]+\varepsilon.
\end{aligned}
\end{equation}
The last bound does not
depend on $\phi\in\Phi_\delta$. Taking the supremum over adaptive evidence policies, the
infimum over $\phi\in\Phi_\delta$, and then letting $\varepsilon\downarrow0$ and
$\delta\downarrow0$ gives
\[
\liminf_{\delta\downarrow0}
\inf_{\phi\in\Phi_\delta}
\sup_{\sigma\in\Sigma_{\mathsf{EV}}}
V(\sigma,\phi,\mu_0,\nu)
\geq
1-ab\left[1-\frac{c(1-a)}{a(1-c)}\right]
=
1-\frac{b(a-c)}{1-c}.
\]
Together with \eqref{eq:t4b-upper}, this proves the result.
\end{proof}

\end{document}